\documentclass[%
preprint,
 amsmath,amssymb,
 aps,
]{revtex4-1}

\usepackage{graphicx}% Include figure files
\usepackage[caption=false]{subfig}
\usepackage{dcolumn}% Align table columns on decimal point
\usepackage{bm}% bold math
\usepackage{xcolor}
\usepackage{soul}
\usepackage[colorlinks=true,linkcolor=blue,filecolor=blue,menucolor=yellow,urlcolor=blue,
citecolor=blue,anchorcolor=blue]{hyperref}
\usepackage[export]{adjustbox}
\usepackage{floatrow}

\begin{document}

\title{Cell growth rate dictates the onset of glass to fluid-like transition and long time super-diffusion in an evolving cell colony}

% Use letters for affiliations, numbers to show equal authorship (if applicable) and to indicate the corresponding author
\author{Abdul N Malmi-Kakkada$^1$, Xin Li$^1$, Himadri S. Samanta$^1$, Sumit Sinha$^2$, D. Thirumalai$^1$}
\email{dave.thirumalai@gmail.com}
\affiliation{$^1$Department of Chemistry, University of Texas, Austin, TX 78712, USA.}
%\author{Xin Li$^1$}
%\affiliation{$^1$Department of Chemistry, University of Texas, Austin, TX 78712, USA.} 
%\author{Himadri S. Samanta$^1$}
%\affiliation{$^1$Department of Chemistry, University of Texas, Austin, TX 78712, USA.}
%\author{Sumit Sinha$^2$}
\affiliation{$^2$Department of Physics, University of Texas, Austin, TX 78712, USA.} 
%\author{D. Thirumalai$^1$}
%\affiliation{$^1$Department of Chemistry, University of Texas, Austin, TX 78712, USA.}

\date{\today}

\begin{abstract}
Collective migration dominates many phenomena, from cell movement in living systems to 
abiotic self-propelling particles. 
Focusing on the early stages of tumor evolution,  
we enunciate the principles involved in cell dynamics and highlight their implications in understanding similar behavior in seemingly unrelated soft glassy materials and possibly chemokine-induced migration of CD8$^{+}$ T cells.  
We performed simulations of tumor invasion using a minimal three dimensional model, accounting for cell elasticity and adhesive cell-cell interactions as well as cell birth and death to establish that cell growth rate-dependent 
tumor expansion results in the emergence of distinct topological niches.  Cells at the periphery move with higher velocity perpendicular to the tumor boundary,  
while motion of interior cells is slower and isotropic. The mean square displacement, $\Delta(t)$, of cells exhibits 
glassy behavior at times comparable to the cell cycle time, while exhibiting super-diffusive behavior, $\Delta (t) \approx t^{\alpha}$ ($\alpha > 1$), at longer times. 
We derive the value of $\alpha \approx 1.33$ using a field theoretic approach based on stochastic quantization. In the process we establish the universality of 
super-diffusion in a class of seemingly unrelated non-equilibrium systems. Super diffusion at long times arises only if there is an imbalance between cell birth and death rates. 
Our findings for the collective migration, which also 
suggests that tumor evolution occurs in a polarized manner, are in quantitative agreement with {\it in vitro} experiments. Although set in the context of tumor invasion 
the findings should also hold in describing collective motion in growing cells and in active systems where creation and annihilation of particles play a role. 
\end{abstract}

%\dates{This manuscript was compiled on \today}
%\doi{\url{www.pnas.org/cgi/doi/10.1073/pnas.XXXXXXXXXX}}

%\begin{document}

% Optional adjustment to line up main text (after abstract) of first page with line numbers, when using both lineno and twocolumn options.
% You should only change this length when you've finalised the article contents.
%\verticaladjustment{-2pt}
%
%\maketitle
%\thispagestyle{firststyle}
%\ifthenelse{\boolean{shortarticle}}{\ifthenelse{\boolean{singlecolumn}}{\abscontentformatted}{\abscontent}}{}
\maketitle

% If your first paragraph (i.e. with the \dropcap) contains a list environment (quote, quotation, theorem, definition, enumerate, itemize...), the line after the list may have some extra indentation. If this is the case, add \parshape=0 to the end of the list environment.

\section{Introduction}
The strict control of cell division and apoptosis is  critical  for tissue development and maintenance~\cite{Barres1992}. 
Dysfunctional cell birth and death control mechanisms lead to several physiological diseases including cancers~\cite{Weinberg13}. 
Together with genetic cues controlling birth-death processes, mechanical behavior of a collection of cells is thought to be of fundamental importance in 
biological processes such as embryogenesis, wound healing, stem cell dynamics,  morphogenesis,
tumorigenesis and metastasis~\cite{Ingber2003,Guilak2003,Kumar2009,Stetler1993,tambe2011collective,marchetti2013hydrodynamics}. 
Due to the interplay between birth-death processes and cell-cell interactions, we expect that collective 
motion of cells ought to exhibit unusual non-equilibrium dynamics, whose understanding might hold the key to describing tumor invasion and related phenomena. 
Interestingly, characteristics of glass-like behavior such as diminished motion (jamming) of a given cell in a dense environment created by neighboring cells (caging effect), dynamic 
heterogeneity, and possible viscoelastic response have been reported in confluent tissues~\cite{angelini,sadati2013collective}. %{\bf Differentiation {\bf 86}, 121-125 (2013)}  
Using imaging techniques that track cell motions, it has been shown that in both two (kidney cells on a 
flat thick polyacrylamide gel~\cite{angelini,angelini2010cell}) and three dimensions (explants from zebrafish embedded in agarose~\cite{schotz})  
the mean displacement exhibits sub-diffusive behavior, reminiscent of dynamics in supercooled liquids at intermediate time scales. This behavior, which 
can be rationalized by noting that the core of a growing collection of cells is likely to be in a jammed state, is expected on time scales less than the cell division time. 

A theory to capture the essence of tumor invasion  must consider the interplay of the cell mechanics, adhesive interaction 
between cells, and the dynamics  associated with cell division and apoptosis, over a wide range of time scales. 
In an attempt to capture collective dynamics in cells a number of models based on cellular 
automaton~\cite{gonzalez2002metapopulation,durrett1994stochastic}, vertex and Voronoi models~\cite{nagai2001dynamic,bi,farhadifar2007influence,li2014coherent,fletcher2014vertex},
subcellular element model~\cite{newman2005modeling}, cell dynamics based on Potts model~\cite{graner1992simulation,szabo2010collective}, and phase field description for collective migration~\cite{camley2014polarity,camley2017physical} have been proposed. 
Previous works have investigated a number of two-dimensional (2D)  models in various contexts~\cite{hilgenfeldt2008physical,angelini,bi} 
including probing the dynamics in a homeostatic state where cell birth-death processes are balanced~\cite{ranft,matoz}. 
Existing three-dimensional (3D) models focus solely on tumor growth kinetics, spatial 
growth patterns~\cite{drasdo,schaller} or on cell migration at low cellular density on time scales shorter than the cell division time~\cite{wu2014three,gal2012intracellular,dieterich2008anomalous}.  
A recent interesting study~\cite{ranft} shows that cell dynamics in a confluent tissue 
is always fluidized by cell birth and death processes, on time scales  
comparable to cell division time.  A more recent  two-dimensional  model~\cite{matoz} investigates glass-to-liquid transition in confluent tissues using simulations. 
However, both these instructive models~\cite{ranft,matoz} focus on the steady state regime where the number of tumor 
cells is kept constant by balancing the birth and death rates. Consequently, they  do not address  the non-equilibrium dynamics of the evolving tumor in the  early stages, which is of great interest in cancer biology~\cite{valencia,pickl,wolf}. %In addition to the models cited above, which are based on the physics governing interactions between cells, several vertex models have been used successfully to probe dynamics in confluent tissues~\cite{Manning} and other cellular processes (see~\cite{Fletcher14BJ} for a review){\bf THERE ARE OTHERS MAKE SURE THEY GET CITED INCLUDING THE FIRST ONE IN PHIL. MAG.} 

Here,  we use a minimal physical 3D model that combines both cell mechanical characteristics, cell-cell adhesive interactions and variations in cell 
birth rates to probe the  non-steady state tumor evolution. Such a model, which has the distinct advantage that it can be generalized to include mutational effects naturally, was first introduced by Drasdo and H{\"o}hme~\cite{drasdo}. One of our primary goals is to understand quantitatively the complex invasion dynamics of tumor into a collagen matrix, and provide a mechanism for the observation of super-diffusive behavior.  We use  free boundary conditions for tumor evolution to 
study  dynamical fingerprints  of invasion in order to quantitatively compare the results to experimental observations. 
%Our focus is to use simulations and theory to study the challenging problem of 
 %  tumor growth dynamics at the early stages, a regime 
%of great interest~\cite{valencia,pickl,wolf}, using a physically realistic model. 
We  model the proliferation behavior of tumor cells, and investigate the effect of pressure
dependent growth inhibition. Good agreement  between our  results and  {\it in vitro}  experiments on three-dimensional 
growth of multicellular tumor spheroids lends credence to the model. 
On time scales less than the cell division time, the dynamics of cell movement within the tumor exhibits
glassy behavior, reflected in a sub-diffusive behavior of the mean square displacement, $\Delta(t)$. However, at times exceeding cell division time we find super-diffusive behavior with $\Delta(t) \sim t^{\alpha}$ ($\alpha = 1.26\pm 0.05$). The duration for which sub-diffusion persists decreases as the cell growth rate increases,  in sharp contrast to the dynamics in confluent tissues. 
Detailed analyses of the individual cell trajectories reveal complex  heterogeneous 
spatial and time dependent cell migration patterns, thus 
providing insights into how cells are poised for invasion into regions surrounding the tumor. 
%Our work  paves a path to understanding the physical factors behind the epithelial-mesenchymal transition (EMT) 
%and mesenchymal-epithelial transition (MET) in cancers~\cite{yang,Thiery2009}. 
We find that activity due to cell division coupled with cell mechanical interactions plays a critical role 
in the non-equilibrium dynamics and the physical structure of the polarized tumor invasion process.  The 
dynamical properties of cells in our model share considerable similarities to those found in non-living soft materials such as soap foams and toothpaste~\cite{Hwang2016}. %abd removed gopal, durian reference 
In all these cases the transition from a glass-like behavior to super diffusion occurs as a result of cell growth and death (or creation and destruction of particles), resulting in non-conservation of number density without the possibility of reaching homeostasis.  
In other words, the non-equilibrium dynamics arising due to forces that result from key biological events (cell birth and death) that we have investigated here are qualitatively 
different from dynamics in systems which do not take into account such forces.
%abd need to be moved or modified

\section{Multicellular Tumor Growth Model}
\label{secmo}
We simulated the spatiotemporal dynamics of a multicellular tumor using a three dimensional (3D)
agent-based model in which the cells in the tumor are represented as interacting objects. 
In this model, the cells grow stochastically as a function of time and divide upon reaching a critical size. 
The cell-to-cell interaction is characterized by direct elastic and adhesive forces. 
We also consider cell-to-cell and cell-to-matrix damping as a way of accounting for the effects of friction experienced
by a moving cell due to other cells, and by the extracellular matrix (ECM) (or collagen matrix), respectively.

%\textbf{Details of the model:}
Each cell is modeled as a deformable sphere with a time dependent radius. 
Several physical properties such as the radius, elastic modulus, membrane receptor and ligand 
concentration, adhesive interaction, characterize each cell. 
Following previous studies~\cite{schaller, pathma, drasdo}, we use the Hertzian contact mechanics
to model the magnitude of the elastic force between two spheres of radii $R_{i}$ and $R_{j}$(Fig.~\ref{cellcellinter}),
\begin{equation}
\label{rep}
F_{ij}^{el} = \frac{h_{ij}^{3/2}(t)}{\frac{3}{4}(\frac{1-\nu_{i}^2}{E_i} + \frac{1-\nu_{j}^2}{E_j})\sqrt{\frac{1}{R_{i}(t)}+ \frac{1}{R_{j}(t)}}},
\end{equation}
where $E_{i}$ and $\nu_{i}$, respectively, are the elastic modulus and
Poisson ratio of cell $i$. The overlap between the spheres, if they interpenetrate without deformation, is $h_{ij}$, which 
is defined as $\mathrm{max}[0, R_i + R_j - |\vec{r}_i - \vec{r}_j|]$ with $|\vec{r}_i - \vec{r}_j|$ being the center-to-center 
distance between the two spheres (see Fig.~\ref{cellcellinter}). 
The repulsive force in Eq.~(\ref{rep}) is valid for small virtual overlaps such that $h_{ij} << \mathrm{min}[R_i,R_j]$, and is
likely to underestimate the actual repulsion between the cells~\cite{schaller}. Nevertheless, the model incorporates 
measurable mechanical properties of the  cell, such as $E_i$ and $\nu_i$, and hence we use this form for the repulsive force.

Cell adhesion, mediated by receptors on the cell membrane,  
is the process by which cells interact and attach to one another. 
For simplicity, we assume that the receptor and ligand molecules are evenly 
distributed on the cell surface. Consequently, the magnitude of the adhesive force, $F_{ij}^{ad}$, 
between two cells $i$ and $j$ is expected to scale as a function of their contact area, 
$A_{ij}$~\cite{palsson}. We estimate $F_{ij}^{ad}$ using~\cite{schaller},
\begin{equation}
\label{ad}
F_{ij}^{ad} = A_{ij}f^{ad}\frac{1}{2}(c_{i}^{rec}c_{j}^{lig} + c_{j}^{rec}c_{i}^{lig}),
\end{equation}
where the $c_{i}^{rec}$ ($c_{i}^{lig}$) is the receptor (ligand) concentration 
(assumed to be normalized with respect to the maximum receptor or ligand concentration so that  
$0 \leq c_{i}^{rec},  c_{i}^{lig} \leq 1$). The coupling constant $f^{ad}$ allows us to 
%\leq c_{i}^{(rec/lig)/max} 
rescale the adhesion force to account for the variabilities in the maximum densities of the receptor and ligand concentrations.  
We calculate the contact surface area, $A_{ij}$, using the Hertz model prediction,  $A_{ij} = \pi h_{ij}R_{i}R_{j}/(R_{i}+R_{j})$.
The Hertz contact surface area is smaller than the proper spherical contact surface area. However, 
in dense tumors many spheres overlap, and thus the underestimation of the cell surface overlap may be advantageous 
in order to obtain a realistic value of the adhesion forces~\cite{schaller}. 

Repulsive and adhesive forces considered in Eqs.(\ref{rep}) and (\ref{ad}) 
%act on the center of the spheres 
act along the unit vector $\vec{n}_{ij}$ pointing from the center of cell $j$ to the center of cell $i$ (Fig.~\ref{cellcellinter}). 
The force exerted by cell $i$ on cell $j$, ${\bf F}_{ij}$, is shown in Fig.~\ref{forcejkr}.
The total force on the $i^{th}$ cell is given by the sum over its nearest neighbors ($NN(i)$), 
\begin{equation}
\vec{F}_{i} = \Sigma_{j \epsilon NN(i)}(F_{ij}^{el}-F_{ij}^{ad})\vec{n}_{ij}. 
\end{equation}
We  developed a distance sorting algorithm to efficiently provide a list of nearest neighbors in contact with the $i^{th}$ cell
for use in the simulations. 
For any given cell, $i$, an array containing the distances from cell $i$ to all the other cells 
is initially created. We then calculated, $R_i + R_j - |\vec{r}_i - \vec{r}_j|$ and sorted the cells $j$
satisfying  $R_i + R_j - |\vec{r}_i - \vec{r}_j|~>~0$ - a necessary condition for any cell $j$ to be in contact with cell $i$. 
%The force experienced by each cell is calculated only 

\section{Simulations}
\textbf{Equations of Motion:} %Once the force is calculated, 
The spatial dynamics of the cell is computed based 
on the equation of motion~\cite{schaller, dallon, palsson1} for a cell of mass $m_{i}$,
\begin{equation}
\label{eom}
m_{i}\ddot{r}_{i}^{\alpha'} = F_{i}^{\alpha'}(t) - \Sigma_{\beta'} \gamma_{i}^{\alpha' \beta'} \dot{r}_{i}^{\beta'}(t) - \Sigma_{\beta'}\Sigma_{j} \gamma_{ij}^{\alpha' \beta'} [\dot{r}_{i}^{\beta'}(t) - \dot{r}_{j}^{\beta'}(t)],
\end{equation}
where the Greek indices $[\alpha', \beta'] = [x,y,z]$ are for coordinates, and the Latin indices $[i, j] = [1,2,...,N]$ are the cell indices. 
In Eq.~(\ref{eom}), $\gamma_{i}^{\alpha' \beta'}$ is the cell-to-medium friction
coefficient, and $\gamma_{ij}^{\alpha' \beta'}$ is the cell-to-cell friction coefficient. The adhesive and repulsive forces are included in the term $F_{i}^{\alpha'}$. 
The cell-to-ECM friction coefficient is assumed to be given by the Stokes relation,
\begin{equation}
\gamma_{i}^{\alpha' \beta', visc} = 6\pi \eta R_{i} \delta^{\alpha' \beta'},
\end{equation}
based on the friction of a sphere in a medium of viscosity $\eta$.
Here, $\delta^{\alpha' \beta'}$ is the Kronecker delta.

Because the Reynolds number for cells in a tissue is  small~\cite{dallon}, overdamped 
approximation is appropriate implying that the neglect of the inertial term $m_{i}\ddot{r}_{i}^{\alpha'}  \approx 0$ is justified~\cite{schaller} (see Appendix~\ref{secdet} for further discussion). 
Since additional adhesive forces are also present, cell movement is
further damped~\cite{drasdo1}.
We simplify the equations of motion (Eq.~(\ref{eom})), by replacing the intercellular drag term 
with a modified friction term, given that the movement of the bound cells is restricted. The 
modified friction term will contribute to the diagonal part of the damping matrix with 
$\gamma_{i}^{\alpha' \beta'} = \gamma_{i}^{\alpha' \beta', visc} + \gamma_{i}^{\alpha' \beta', ad} $
where, 
\begin{eqnarray}
\gamma_{i}^{\alpha' \beta', ad} =&& \gamma^{max}\Sigma_{j \epsilon NN(i)} (A_{ij}\frac{1}{2}(1+\frac{\vec{F}_{i} \cdot \vec{n}_{ij}}{|\vec{F}_{i}|})\times \\ \nonumber &&\frac{1}{2}(c_{i}^{rec}c_{j}^{lig} + c_{j}^{rec}c_{i}^{lig}))\delta^{\alpha' \beta'} \, .
\end{eqnarray}
Notice that the added friction coefficient $\gamma_{i}^{\alpha' \beta', ad} $ is proportional to the cell-to-cell contact surface, 
implying that a cell in contact with multiple cells would move less. 
The non-isotropic nature of the adhesive friction is evident from the factor $(1+\frac{\vec{F}_{i} \cdot \vec{n}_{ij}}{|\vec{F}_{i}|})$
where the maximum contribution occurs when the net force $\vec{F}_{i}$ is parallel to a given unit vector, $\vec{n}_{ij}$, among the nearest neighbors. 
With these approximations, the equations of motion (Eq.~(\ref{eom})) are now diagonal, 
\begin{equation}
\label{eqforce}
\dot{\vec{r}}_{i} = \frac{\vec{F}_{i}}{\gamma_i}.
\end{equation}

\textbf{Cell Cycle:}
In our model, cells can be either in the dormant ($D$) or in the growth ($G$) phase.
We track the sum of the normal pressure that a particular cell $i$ experiences due to contact with its neighbors, using, 
\begin{equation}
\label{pressure}
p_{i} =  \Sigma_{j \epsilon NN(i)} \frac{|\vec{F}_{ij} \cdot  \vec{n}_{ij}|}{A_{ij}}.
\end{equation}
If the local pressure, $p_{i}$, exceeds a critical limit ($p_c$) the cell stops growing and enters the dormant phase (see the left panel in Fig.~\ref{celldorgro}). 
%The cell is considered to be in the dormant state. 
For growing cells, their volume increases at a constant rate $r_V$. 
The cell radius is updated from a Gaussian distribution with the mean rate $\dot{R} = (4\pi R^2)^{-1} r_V$.  
Over the cell cycle time $\tau$, 
\begin{equation}
r_V = \frac{2\pi (R_{m})^3}{3\tau},
\end{equation}
where $R_{m}$ is the mitotic radius. 
A cell divides once it grows to the fixed mitotic radius. 
To ensure volume conservation, upon cell division, we use 
$R_d = R_{m}2^{-1/3}$ as the radius of the daughter cells  (see the right panel in Fig.~\ref{celldorgro}). 
The two resulting cells are placed at a center-to-center distance 
$d = 2R_{m}(1-2^{-1/3})$. The direction of the new cell location 
is chosen randomly from a uniform distribution on the unit sphere. 
One source of stochasticity in the cell movement in our model is due to random choice for  
the mitotic direction. Together with stochasticity in the cell cycle duration, we obtain 
fairly isotropic tumor spheroids.

%\section{Data Analyses}
%\label{secdata}

%\hl{\textbf{Influence of cell adhesion strength on self-intermediate scattering function $F_{s}(q,t)$:}}
%Most of our results are obtained using a fixed value for the cell adhesion strength $f^{ad} = 10^{-4}\mathrm{\mu N/\mu m^2}$ (see Table I). 
%However, during tumor progression cell adhesion strength could vary~\cite{Polyak}. 
%To check the influence of this variation,  we calculated the self-intermediate scattering functions $F_{s}(q,t)$ (see the main text for definition)
%for different values of $f^{ad}$ (see Fig.~\ref{figs4}). 
%Compared to Fig.~2(b) in the Main text, the time scale associated with the first relaxation process corresponding 
%to cell movement within the cage is slower as $f^{ad}$ increases. 
%Cells are trapped by their neighbors in this time regime making it more difficult for cells to move within the cage
%as $f^{ad}$ increases, leading to an increase in the relaxation time. The relaxation time dependence 
%on the values of $f^{ad}$ is weaker for the second process corresponding 
%to transitions out of the cage. 
%This finding is in agreement with our 
%theoretical prediction that in the long time regime, the dynamics of cell migration is determined by birth and death processes, 
%and is impervious to the details of interactions between the cells. As a consequence there is a 
%universal power law dependence of the mean square displacement (Eq.~\ref{msdsi}). 

\textbf{Tumor invasion distance:}
The invasion or spreading distance %(see Fig.~\ref{figs6}) 
of the growing tumor, $\Delta r(t)$, is determined by measuring the 
average distance from the tumor center of mass ($\vec{R}_{CM}~= ~(1/N)\Sigma_{i} \vec{r}_i$) to the cells at the tumor periphery,
\begin{equation}
\label{deltart}
\Delta r(t) = \frac{1}{N_{b}} \sum_{i}^{N_b} |\vec{r}_{i} - \vec{R}_{CM}|,
\end{equation}
where the sum $i$ is over $N_b$, the number of cells at the tumor periphery. 
In order to find the cells at the tumor periphery ($N_b$), we denote the collection of 
all cells $N$ as a set of vertices $\{1, 2, 3 ......... N\}$ in $\mathbb{R}^{3}$ where the vertices 
represent the center point of each cell. We generate a 3D structure of tetrahedrons using these vertices, 
where each tetrahedron 
is comprised of $4$ vertices. 
Let $T$ be the total number of tetrahedrons. Since each tetrahedron has $4$ faces, the total number of faces is $4T$. 
Any face that is not on the boundary of the 3D structure is shared by 2 tetrahedrons but the boundary faces are not shared. 
Thus, our aim is to find the set of unshared boundary faces out of the total number, $4T$, of faces. 
Once the boundary faces are obtained, we know the list of vertices, and hence the positions of cells at the tumor periphery, allowing us 
to compute $\Delta r(t)$ in Eq.~(\ref{deltart}). 

\section{Results}

\textbf{Calibration of the model parameters:}
%There are many stages in 
%cancer progression. In the early stages, 
%tumor is relatively small and is localized in 
%the organ where it is initiated. Because we are 
%interested in studying the initial stages of tumor growth when the 
%cell numbers are small.
%, for the purposes of calibrating 
%the various parameters in our model (see Table I), 
We compare the normalized 
volume of the growing tumor to 
experimental data~\cite{montel}, as a way of assessing if the parameters (Table I) used in our model
are reasonable. 
The tumor volume, $V(t)$, normalized 
by the initial volume of the spheroid ($V_{0}$), was 
tracked experimentally using colon carcinoma cells~\cite{montel} through experimental methods that are very different from the way 
we simulated tumor growth.
%at various applied pressures. 
The tumor growth was measured by imposing stress~\cite{montel}, known to 
%The role of mechanical stress in 
inhibit cancer growth~\cite{helm}.  These effects are included 
in our model, which allows us to make quantitative comparisons between our simulations and experiments. %abd changed to quantitative
The tumor volume is obtained in the simulation from, $R_g$, the 
radius of gyration,
\begin{equation}
R_{g}^{2} = \frac{1}{N} \sum_{i}^{N} (\vec{r}_{i} - \vec{R}_{CM})^2,
\end{equation}
where $N$ is the total number of cells. 
The volume $V(t)$ is given by $(4\pi /3)R_{g}^{3}(t)$.
Our simulation of the growth of the 
spheroid tumor volume in the early stages is in good agreement with experimental data
(see Fig.~\ref{figs2}).  Thus, our model captures quantitative aspects of tumor growth. 
%Good agreement with experiment is obtained. 
%\floatsetup[figure]{style=plain,subcapbesideposition=top}

In the experiment~\cite{montel}, $V(t)/V_{0}$ was measured for external pressure 
ranging from $0~-~20$kPa. In Fig.~(\ref{figs2}), we compared our simulation results 
with the $500$Pa result from experiments~\cite{montel}. This is rationalized as follows. 
Unlike in experiments, the pressure is internally generated as the tumor grows (Fig.~\ref{pressd}) 
with a distribution that changes with time. The mean value of the pressure (see dashed lines in Fig.~\ref{pressd}) 
at the longest time is $\approx 100$Pa. Thus, it is most appropriate to compare our results 
obtained using $p_{c}=100$Pa with experiments in which external pressure is set at $500$Pa.

\textbf{Predicted pressure-dependent growth dynamics is consistent with experiments.}
Visual representation of the tumor growth process generated in simulations is vividly illustrated in the movie (see Movies 1 and 2) in the Appendix~\ref{secmov}. 
Snapshots of the evolving collection of cells 
% 100s, $10^5$s, $3\cdot10^5$s, $5.5\cdot10^5$s 
at different times are presented in Figs.~\ref{figx}a-d. 
As the tumor evolves,  the cells aggregate into 
a spheroidal shape due to cell division plane being isotropically 
distributed (Fig.~\ref{figx}d and the movie in the Appendix~\ref{secmov}). In spheroidal cell aggregates, it is known that  pressure 
inhibits cell proliferation~\cite{Quail2013,Jain2014,montel}. 
We expect the pressure (see Eq.~\ref{pressure}) experienced by the cells  in the interior of the tumor 
to be elevated due to crowding effects, 
%This inhibits cell division and therefore lead to slower dynamics in the interior. 
causing the cells to enter  a dormant state if the pressure  from the neighbors  
reaches a preset threshold value, $p_{c}$. 
Tumor growth behavior is strongly dependent on the value of $p_{c}$ (see Fig.~\ref{figx}e). 
At  $p_{c} = 10^{-3} $ MPa, 
the total number ($N(t)$) of tumor cells during growth is well approximated 
as an exponential  $N(t) \propto {\mathrm exp(const\times t)}$. %as showed by the exponential fitting of solid red line to the simulation results (purple squares). 
As $p_{c}$  is lowered, implying  growth is inhibited at smaller pressures, increase in the tumor size is described by a power law, $N(t) \propto t^{\beta}$, at long timescales,
%quickly as we reduce the $p_{c}$ although it 
while $N(t)$ retains exponential growth at early stages (see the inset of Fig.~\ref{figx}e).  Our simulations also show that $\beta$ is $p_{c}$-dependent, increasing as $p_{c}$ increases. 
Power-law growth in 3D tumor spheroid size  has been observed in many tumor cell lines with $\beta$ varying from one to three~\cite{conger,Mandonnet2003,Simeoni2004,Hart1998,grimes}.   
The overall growth of the  tumor slows down as the value of pressure experienced by cells increases, which is also consistent with recent experimental results~\cite{Jain2014}. The known results in Fig.~\ref{figx}e, are in near quantitative agreement with several experiments, thus validating the model.

\textbf{Cell motility within the tumor spheroid.} Using direct imaging techniques it has become possible to monitor the overall invasion of the tumor as well as the movement of individual cells within the spheroid~\cite{Pampaloni2007,Alessandri2013}. In order to compare our  results to experiments we calculated the mean square displacement, $\Delta(t) = \langle[{\bf r}(t)-{\bf r}(0)]^2\rangle$, of individual cells.
By tracking the movement of all the initial cells within the tumor, we calculated $\Delta(t)$  by averaging over hundreds of trajectories. 
The growth rate-dependent $\Delta(t)$  (displayed in Fig.~\ref{msdexp}a on a log-log scale) shows that there is a rapid increase in   $\Delta(t)$ at early times ($t \leq 0.01 \tau_{min} $, where $\tau_{min}$ is the benchmark value of the cell cycle time given in Table I)  because the cells move unencumbered, driven by repulsive interaction with other cells. At intermediate timescales (0.01$\tau_{min} < t < \tau$ with $\tau$ being the average cell cycle time),  $\Delta(t)$  exhibits  sub-diffusive behavior ($\Delta(t) \sim t^{s}$ with $s < 1$). The signatures of the plateaus in $\Delta(t)$ (together with other characteristics discussed later) in this time regime indicate that cells are caged by the neighbors (see the left inset in Fig.~\ref{msdexp}a),  
and consequently undergo only small displacements.  Such a behavior is reminiscent of  a supercooled liquid undergoing a glass transition, as  illustrated in colloidal 
particles using direct imaging as their densities approach the glass transition~\cite{Weeks2000,Kegel2000}.
As $\tau$ increases, the plateau persists for longer times because of a decrease in the outward stress, which slows the growth of the tumor. 
When $t$ exceeds $\tau$ the average cell doubling time, the $\Delta(t)$ exhibits 
super-diffusive motion $\Delta(t) \approx t^{\alpha}$ ($\alpha > 1$). In order to determine $\alpha$ we performed multiple simulations and calculated $\Delta(t)$ 
for each of them by generating a large number of trajectories. All the independent simulations show that $\Delta(t)$ has the characteristic plateau at intermediate 
times followed by super-diffusion at long times in Fig.~\ref{msdexp}b. From each of these simulations we determine $\alpha$, whose distribution is shown in 
Fig.~\ref{msdexp}c. In all cases, we find that $\alpha$ is greater than unity. The estimate from the distribution in Fig.~\ref{msdexp}c is $\alpha = 1.26 \pm 0.05$ 
where $0.05$ is the standard deviation.

On long time scales,  cells can escape the cage created by their neighbors, as illustrated in the middle inset of Fig.~\ref{msdexp}a. 
Our observation of super-diffusion in $\Delta(t)$  at long times agrees well with the experimental result ($\alpha  \approx  1.40 \pm 0.04$) 
obtained for fibrosarcoma cells in a growing tumor spheroid~\cite{valencia}.  
%The onset of super-diffusive motion is a consequence of collective motion, which does not occur in **.
The onset of super-diffusive behavior in $\Delta(t)$ shifts to earlier times as we decrease the average cell cycle time (see Fig.~\ref{msdexp}a), implying 
that cell division is the mechanism resulting in super-diffusion (see below for further discussion of this crucial finding). 

We provide another rationale for robustness of the long time super diffusive behavior. This comes from examining the time-dependent changes in  the invasion distance, $\Delta r(t)$ in Eq.\ref{deltart}. 
%In Fig. 14 of the previous version of manuscript, we noted that $\Delta r \propto t^{\xi}$ with $\xi$ being $0.63$. 
The finding that the invasion distance does not increase as a function of time with exponent $\Delta r \propto t^{0.5}$
but rather at a higher exponent at the long time regime necessarily implies that 
cells do not execute random walk motion (see Ref.~\cite{valencia}).  
%We have now extended the time of the simulations to show more clearly an exponent of 
%$\approx 0.63$ in Fig.~\ref{deltar}. 
The dependence of $\Delta r(t)$ on $t$ in Fig.~\ref{figs6} shows that for $t~\le~\tau_{min}$, the invasion radius is roughly constant. As cells divide the tumor 
invasion distance, $\Delta r(t)$, increases as $t^{\xi}$ with $\xi~\sim~0.63$ (implies $\alpha \approx 1.26$) for $t~>~\tau_{min}$, a value that is not inconsistent with experiments~\cite{valencia}. 

\textbf{Theoretical predictions.} In order to understand the role of  cell growth and apoptosis in the observed sluggish dynamics at intermediate times and 
 super-diffusive   behavior at long times, we developed a theory to study the 
dynamics of a colony of cells in a dissipative environment (Appendix~\ref{secth}).     
The interactions between cells contain both attractive (adhesive) and excluded 
volume terms.
%The cells interact via adhesion and excluded volume repulsion. Each cell 
%experiences a thermal white noise and moves under the force generated 
%from the potential due to its neighbors. 
Starting from the Langevin equation describing  the dynamics of the $i^{th}$ cell, and incorporating the 
birth reaction,  $X \xrightarrow[]{k_{a}} X+X$  with the rate constant $k_{a}$ (= 1/$\tau$) for each cell, and the apoptosis reaction
$X+X\xrightarrow[]{k_{b}} X$  with the rate $k_{b}$, an  equation  for the 
 time dependence of the density $\rho({\bf k},t)$ (Eq.~\ref{rho12} in the Appendix~\ref{secth}) can be derived.  
 The cell division and apoptosis processes drive the system far from equilibrium,  thus violating the Fluctuation Dissipation Theorem (FDT). As a consequence, we cannot use standard methods used to calculate response 
 and correlation functions from which the $t$-dependence of $\Delta (t)$ can be deduced. To overcome this difficulty we used the Parisi-Wu stochastic quantization method~\cite{Parisi1981} in which the evolution of $\rho({\bf k},\omega)$ ($\omega$ is the frequency) is described in a fictitious time in which FDT is preserved. From 
 the analysis of the resulting equation (Appendix~\ref{secth} contains the sketch of the calculations) the scaling of $\Delta (t)$ may be obtained as,
%\begin{eqnarray}
%\label{phirt}
%%\begin{flalign*}
%\frac{\partial \phi({\bf r},t)}{\partial t}&=& {\bf \nabla }\cdot \left(\phi({\bf r},t)\int d{\bf r'} \phi({\bf r'},t){\bf \nabla}U({\bf r-\bf{r'}})\right)+k_{b} \phi(\frac{k_a}{k_b}-\phi)\\ \nonumber
%&+&{\bf \nabla} \cdot \left(\eta({\bf r},t) \phi^{1/2}({\bf r},t)\right)+\sqrt{k_a \phi+k_b \phi^2} f_\phi \, ,
%%\frac{\partial \phi({\bf r},t)}{\partial t} &=& {\bf \nabla }\cdot (\phi({\bf r},t)\int d{\bf r'} \phi({\bf r'},t){\bf \nabla}V({\bf r-\bf{r'}})) 
%%+k_{C} \phi(\frac{k_{A}}{k_{C}}-\phi) \\ \nonumber
%%&+&{\bf \nabla} \cdot (\eta({\bf r},t) \phi^{1/2}(({\bf r},t)))+\sqrt{k_{A} \phi+k_{C} \phi^2} f_\phi,
%%\end{flalign*}
%\end{eqnarray}
%where $\langle f_\phi({\bf r},t)f_\phi({\bf r'},t')\rangle=\delta({\bf r}-{\bf r}')\delta(t-t')$. 
%, which we neglect for the purposes of this calculation. 
%By performing an analysis of Eq.~(\ref{phirt}), we obtain,
\begin{equation}
 \Delta(t) =  \langle[{\bf r}(t)-{\bf r}(0)]^2\rangle~\sim t^{2/z} \ .
\end{equation}
In the intermediate time regime, 
 $z=5/2$, implying $\Delta(t) \approx t^{4/5}$. The predicted
sub-diffusive behavior of $\Delta(t)$  is qualitatively consistent with  simulation results.  
It is likely that the differences in the  scaling exponent between simulations ($2/z \approx 0.33$) and theoretical predictions ($2/z \approx 0.80$) 
in this non-universal time regime  may be due to the differences in the cell-to-cell interactions used in the two models. 
 
In the long time limit, the cell birth-death process (the fourth term in  Eq.~\ref{rho12}) dominates 
 the interactions between cells. As a result, we expect that the exponent $2/z$ should be universal, independent  of the forces governing cell motility. Our theory predicts that  $z = 3/2$, which shows that 
$\Delta(t) \approx t^{4/3}$,  in excellent agreement with the simulations  (Fig.~\ref{msdexp}a) and experiments~\cite{valencia}. It is clear from our theory that  the interaction-independent biologically important birth-death 
processes drive the {\it observed fluidization} during tumor (or tissue) development, 
resulting in super-diffusive cell motion at long times. The underlying mechanism for obtaining super-diffusive behavior is  that cells must move 
persistently in a given direction for  a long time leading to polarized tumor growth,   ultimately  resulting in invasion driven predominantly by birth. We provide additional numerical evidence for this assertion below.

\textbf{Dependence of relaxation times on cell cycle time.} We first characterized the structural evolution as the tumor evolves. In order to assess the spatial variations in the positions of the cells as the tumor grows, we calculated 
the pair correlation function using,
\begin{equation}
 g(r) =\frac{V}{4\pi r^2 N^{2}} \sum_{i}^{N}\sum_{j\neq i}^{N}\delta(r-|\vec{r}_{i}-\vec{r}_{j}|). 
\end{equation}
The pair correlation functions (Fig.~\ref{figs3}), at different cell cycle times  
($\tau$), show that 
at longer cell cycle times, the cells 
are packed more closely.  
There is a transition from a liquid-like to a glass-like structure as $\tau$ is increased,  
as indicated by the peaks in $g(r)$.

To further quantify the fluidization transition driven by cell birth-death processes, 
we calculated the isotropic self-intermediate scattering function 
$F_{s}({q},t) = \langle e^{i {\bf q}\cdot ({\bf r}(t) -{\bf r}(0))} \rangle$ at $|{\bf q}| = 2\pi/r_{0}$, where $r_{0}$
is the position of the first maximum in the pair correlation function (see Fig.~\ref{fig:gr}).  
The average is taken over all the initial cells, which are  alive during the entire simulation 
time and the angles of ${\bf q}$.  We note that  $F_{s}({q},t)$ exhibits a two-step relaxation process (Fig.~\ref{figcell}a) characterized by two time scales. 
The initial relaxation time, corresponding  
to the motion of cells in a cage formed by neighboring cells, depends only weakly on the  cell cycle time. 
The second relaxation time ($\tau_{\alpha}$), extracted by fitting $F_{s}(q,t)$ to an exponential function ($F_{s}(q,t) \approx a_{0} e^{-t/\tau_{\alpha}}$,  see colored solid lines in Fig.~\ref{figcell}a), 
depends strongly on the average  cell cycle time.  As in the relaxation of supercooled liquids, $\tau_{\alpha}$ is associated with  the collective motion of cells leaving the cage~\cite{kirkpatrick1989scaling,Berthier2011}. %{\bf abdul new ref is Kirkpatrick, Thirumalai, and Wolynes PRA 1989}
As the average  cell cycle time is reduced,
$\tau_{\alpha}$ decreases (see Fig.~\ref{figcell}b), 
and the $F_{s}({q},t)$  begins to approach a single 
relaxation regime, as expected for a normal fluid. 
%The first relaxation time (observed by fitting  $F_{s}(q,t)$ by a single 
%exponential function in cyan solid line in Fig.~\ref{figx2}b) is not influenced by changing the growth rate. 
The second relaxation process in $F_{s}(q,t)$  (Fig.~\ref{figcell}a) can be collapsed onto one master
curve by  rescaling time by $\tau_{\alpha}$, resulting in the independence of $F_{s}({q},t)$ on  the cell cycle time (Fig.~\ref{figcell}c). We surmise that the cage relaxation  is driven by 
the same mechanism (the cell birth-death processes) that gives rise to super-diffusive behavior in $\Delta(t)$.

\textbf{Diffusion of tracer cells}: Elsewhere~\cite{matoz} using a two-dimensional model $\Delta(t)$ was computed for cells 
as well as tracer cells. In their study, using periodic boundary conditions, the choice of birth and death rates is such 
that in the long time limit homeostasis is always reached where birth and death of cells is balanced. 
It was found that $\Delta(t)$ for live cells (those that can be born and die) show a plateau at intermediate times followed by normal diffusion ($\Delta(t) \sim t$) 
at long times. In contrast, $\Delta_{tr}(t)$, the mean squared displacement computed for tracer cells (ones which have all the characteristics of cells except 
they are alive throughout the simulations and do not grow or divide) shows no caging effects but grows linearly with time~\cite{matoz}, suggestive of normal diffusion. 
In light of this dramatically different behavior reported in~\cite{matoz}, we performed simulations using our model by  including $100$ randomly placed tracer cells. The interactions between the
tracer cells with each other and the cells that undergo birth and death are identical. 
The calculated dependence of  $\Delta_{tr}(t)$ for tracer cells, as a function of $t$ (Fig.~\ref{figtr}a) 
is qualitatively similar to that for cells (compare Fig.~\ref{msdexp}a and Fig.~\ref{figtr}a). In particular at varying values of $\tau$,  $\Delta_{tr}(t)$ exhibits 
a plateau followed by super-diffusive behavior, $\Delta_{tr}(t) \sim t^{\alpha_{tr}}$ %(\alpha_{tr} varying from 1.45 to 1.73)$ 
at long times. However, we find that $\alpha_{tr}(>1.4)$ depends on $\tau$ in contrast to the universal exponent 
for cell dynamics. Similarly, $F_{s}(q,t)$ for tracer cells also displays two-step relaxation for the three values of 
$\tau$ investigated, as shown in Fig.~\ref{figtr}b. Interestingly, the values of the first relaxation times are longer than for the 
corresponding dynamics associated with the cells. The results in Fig.~\ref{figtr}b show that the dynamics 
of tracer cells is qualitatively similar to that calculated for the actual cells (see also Fig.~\ref{figcell}a for comparison).

\textbf{Heterogeneity during tumor growth.} The effect of glass or liquid-like state of tumor growth is  illustrated by following the
 trajectories of individual cells in the growing tumor. Figs.~\ref{figx3}a and \ref{figx3}b highlight the trajectory of  cells  during a time of 
$\approx 3$ days for the average cell division time of $15\tau_{min}$ and $0.25\tau_{min}$, respectively.          
In the glass-like phase (intermediate times corresponding to $t/\tau < 1$ for $\tau=15\tau_{min}$),  the displacements
are small, exhibiting caging behavior %especially at the longer cell cycle time 
(Fig.~\ref{figx3}a), resulting in the localization of the cells near 
their starting positions. On the other hand, 
cells move long distances and show 
signatures of persistent directed motion at the shorter cell cycle time in the long time superdiffusive regime corresponding to $t/\tau > 1$ (see Fig.~\ref{figx3}b). These observations suggest that  the anisotropic growth of cells, manifested largely in the evolution 
of cells at the periphery of the tumor, depends on the cell growth rate, a factor that determines tumor virulency. 
%While Figs.~\ref{figx3}a - \ref{figx3}b, show cell trajectories 
%with the observation times until the intermediate glassy regime and superdiffusive regime respectively, in Fig.~\ref{tracknorm} the observation time is scaled proportional to the cell cycle time. 
%The clear distinction between the motion of cells in the intermediate glassy  
%and the long time superdiffusive  regimes no longer stands out even as slightly pronounced movement of cells at shorter cell cycle time is palpable.

To  quantify spatial heterogeneity, 
we divided the tumor into two regions - interior
and periphery. 
We note that such a division is not applicable in a system with 
periodic boundary conditions~\cite{matoz}. 
After obtaining the invasion distance $\Delta r(t=t_{E})$ 
(see Eq.~(\ref{deltart})), 
we calculated the distance from the center of mass 
for the colony of all the initial cells that are alive at $t_{E}$.
Let us call this vector \textit{$\vec{r}_{I}$}. 
We grouped \textit{$\vec{r}_{I}$} into two distinct 
categories: interior region if  
\textit{$|\vec{r}_{I}(i)| < 0.4\Delta r(t=t_{E})$} and 
boundary region if  $|\vec{r}_{I}(i)| > \Delta r(t=t_{E}) - 2\langle R_i \rangle $. 
The average radius 
of all the cells in the tumor is denoted by $\langle R_i \rangle$. We chose $t_{E}$ = 350,000$s$ = 6.48$\tau$. With this choice of $t_E$ we obtain good statistics allowing us to  glean both the sub-diffusive and super diffusive behavior from the time dependence of $\Delta(t)$ (see Fig. \ref{figrep5}). Once the initial 
cells are classified 
in this manner, we obtain their entire trajectory 
history and calculated their $\Delta (t)$. 
%Having sorted the initial cells into the two regions, we plot  
In Fig. \ref{figrep5}, a plot of $\Delta (t)$ for the
interior and boundary cells is shown. 
%We see that 
The dynamics associated with interior 
cells is sub-diffusive through their entire lifetime while 
the boundary cells show sub-diffusive motion 
at intermediate times and super-diffusive motion at long times. 
Interestingly, the cells at the boundary
also show the intermediate glassy regime, which is {\it a priori} hard to predict.

Because the nature of cell movement determines cancer progression and 
metastasis~\cite{friedl}, it is critical to understand how  various factors affecting collective cell migration
emerge from individual cell movements (Figs.~\ref{figx3} \& \ref{figx3part2}).  Insights into cell migrations may be obtained by using analogies to 
spatial heterogenous dynamics  in supercooled liquids~\cite{Thirumalai1993,Barrat1990}. 
In simple fluids, the distribution of particle displacement is Gaussian while 
in supercooled liquids the displacements of a subset of particles  
 deviate from the Gaussian distribution~\cite{Thirumalai1993}. 
 In Fig.~\ref{figx3part2}a, the van Hove function 
of  cell displacement (or the probability distribution of  step size)  is shown. 
The single time step 
distance covered by a cell is defined as $|\delta r_{i}(\delta t)| = |r_{i}(t+\delta t) - r_{i}(t)|$. By normalizing $\delta t$ by the average 
cell cycle time i.e. $\delta r_{i} (\delta t/ \tau = 0.0074)$, 
%for all initial cells throughout the simulation
we obtain a long-tailed
$\delta r$ probability distribution ($P(\delta r)$).  
The distribution 
$P(\delta r)$, has a broad, power law tail cut off at large values 
of $\delta r$, that depends on the cell cycle time. 
%shows evidence of moving to smaller distances. 
As we approach the glass-like phase for longer average  cell cycle time, 
$P(\delta r)$ is suppressed by an order 
of magnitude over a wide range of $\delta r$. 
Interestingly, we do not observe an abrupt change in the behavior of 
$P(\delta r)$ as the average cell cycle time is changed. 
%The peak in distribution at smaller 
%distances is slightly increased as the cell growth rate decreases. 
The transition between glass-like and liquid-like regimes 
occurs continuously. 
To further analyze the displacement distribution, we fit the van-Hove function 
for squared displacements ($P(\delta r^2)$)
at normal cell division time ($ \tau_{min}$) to 
both exponential and power law.
The distribution is 
considerably broader than the Gaussian distribution (see Fig.~\ref{figs1} and Table II), providing one indication of heterogeneity~\cite{Kirkpatrick2015}. 
%The non-Gaussian parameter\cite{Vorselaars2007} $\Gamma \equiv \langle \delta r(\delta t)^4\rangle/(3\langle \delta r(\delta t)^2\rangle^2) -1$ is about 200 which is far greater than 0 expected for a Gaussian distribution. 
%\textbf{van-Hove Function:}
%Probability distribution, $P(\delta r)$, of multi time step cell displacement 
%was shown in Fig.~\ref{figx3b} for different values of the cell cycle time. 
%To analyze the displacement distribution, \textcolor{blue}{we fit the van-Hove function 
%at normal cell division time ($ \tau_{min}$) to 
%both exponential and power law.} 
%%, which is the expected behavior for isotropic diffusive motion. 
% For the movement of cancer cells, the distribution 
%$P(\delta r^2)$ is decidedly non-Gaussian, as evidenced by the power law fat-tail in $P(\delta r^2)$ (Fig.~\ref{figs1}). 
%{\bf The probability of large displacements, above $1.3 \mu m$, is much larger 
%than what would be expected from a Gaussian distribution. WILL NEED TO BE CHANGED with figure P($\delta r^2$)}
%The presence of fat-tail contributes to super-diffusive behavior (see below). 

Cell-to-cell phenotypic heterogeneity is considered to be one of the greatest challenges 
in cancer therapeutics~\cite{brooks,marusyk}. 
Within the context of our model, spatio-temporal  heterogeneity in dynamics can be observed in tissues by analyzing the movement of individual cells. 
While the simulated time-dependent variations in the average mean-squared displacement 
is smooth, the movement of 
the individual cell is not (see Fig~\ref{figx3part2}b). Cells move slowly 
and periodically undergo rapid `jumps' or hops similar to the phenomenon in supercooled liquids~\cite{Thirumalai1993,Barrat1990}. 
The squared displacement  of individual cells (Fig~\ref{figx3part2}b) vividly shows the heterogeneous behavior of different cells.

\textbf{Polarized tumor growth.} From our simulations, we constructed a spatial map of the velocities 
of the individual cells in the tumor. Using these maps, we characterized the 
spatial heterogeneity in the dynamics in order to elucidate regions of 
coordinated activity in the movement of cells. Fig.~\ref{figx4}a shows a snapshot of 
the spatial map of the single cell velocities.  The velocity map, which spans more than eight orders of magnitude, reveals that there are cell-to-cell variations in the dynamics. More importantly, it also reveals 
the existence of spatial correlations between cell dynamics. 
In the tumor cross-section (Fig.~\ref{figx4}b), faster moving cells are concentrated at the 
outer periphery of the tumor. By calculating the average magnitude of cell velocity as
a function of radius, we show in Fig.~\ref{figx4}c that 
faster moving cells are located at the outer periphery of the tumor quantitatively. 
We calculate the average velocity of cells at different radii of the tumor using,
\begin{equation}
\langle v(r) \rangle = \frac{\Sigma_{i} v_{i} \delta(r-(|\vec{R}_{CM} - \vec{r}_{i}|))}{\Sigma_{i} \delta(r-(|\vec{R}_{CM} - \vec{r}_{i}|))}. 
\end{equation}
Arrows indicating the velocity direction 
show that cells in the periphery tend to move farther away from the center
of the tumor as opposed to cells closer to the center of the tumor whose direction of motion 
is essentially isotropic. 
This prediction agrees well with the experiments~\cite{valencia}, which showed that cells at the periphery of the tumor spheroid move persistently
along the radial direction, resulting in polarized tumor growth.

Mean angle $\theta$ between cell velocity and the position vector with respect 
to the center of the tumor plotted in Fig.~\ref{figx4}d further illustrates  that cell movement becomes persistently directed 
outward for cells closer to the outer layer of the tumor.  
%\textbf{Average Velocity and Cell Velocity Polarization:}
%where $\vec{R}_{CM}~= ~(1/N)\Sigma_{i} \vec{r}_i$ is the tumor center of mass.  
To calculate the radius-dependent average polarization in cell velocity, we first define a vector pointing 
from the center of mass of the tumor to the cell position $\vec{c}_i = \vec{r}_i - \vec{R}_{CM}$ (see the green arrow in the inset of Fig.~\ref{figx4}d). 
The angle $\theta$ (see the inset of Fig.~\ref{figx4}d) between each cell velocity (orange arrow) and the the vector (green arrow) from the center of mass to the tumor 
periphery can be calculated from 
$\cos(\theta)_i = \vec{c}_i \cdot \vec{v}_i/(|\vec{c}_i||\vec{v}_i|)$. 
The average of this angle as a function of radius is calculated using, 
\begin{equation}
\langle \cos(\theta[r]) \rangle = \frac{\Sigma_{i} \cos(\theta[r])_{i} \delta(r-(|\vec{R}_{CM} - \vec{r}_{i}|))}{\Sigma_{i} \delta(r-(|\vec{R}_{CM} - \vec{r}_{i}|))}. 
\end{equation}
The results are presented in  Fig.~\ref{figx4}d in the main text. 
The distribution of the $\theta$ angle at different distances ($r$)  
(see Fig.~\ref{figx4a}a) also illustrates that  cell movement is isotropic close to the 
tumor center, while they move outward in a directed fashion (see the peak of the histogram in blue) at the periphery of the tumor.
To  quantify the heterogeneity in cell velocity, we plot the 
probability distribution of the velocity magnitude (normalized by the mean velocity - $\langle v \rangle$), $P(|v|/\langle v \rangle)$, (Fig.~\ref{figx4a}b)
 accessible in experiments using direct imaging or particle image velocimetry methods~\cite{Pampaloni2007,Alessandri2013}. 
There is a marked change in the velocity distribution 
as a function of  cell cycle time. At the longer cell cycle time ($\tau=15\tau_{min}$), 
corresponding to the glass-like phase, $P(|v|/\langle v \rangle)$ distribution %is sharply peaked 
is clustered around smaller values of $|v|/\langle v \rangle$ 
while quickly decaying to zero for higher velocities. 
For the shorter  cell cycle time, 
%at both $1\times \tau_{min}, 0.25\times \tau_{min}$, 
the velocity distribution is considerably broader. 
The broader velocity distribution indicates the presence of more invasive cells
within the tumors characterized by high proliferation capacity.

\textbf{Consistency with experiments.} %Even though three dimensional agent based models have been utilized to study tissues properties previously, quantitative understanding of the physical principles underlying complex spatiotemporal dynamical properties of cells within $in~vitro$ tissue spheroids has been lacking. 
We show here that the minimal model captures the three critical aspects of %{\color{red} collective cell invasion
%Our model captures all the three essential features of the growing spheroidal tumors discovered 
%as reported in} 
a recent single-cell resolution experiment probing the invasion of cancer cells into a collagen matrix~\cite{valencia}:
$(i)$ %on average, directed but not random cell motion is observed for the individual cells within the spheroids as indicated by the power-law relation  for the 
Ensemble-averaged mean square displacement of individual cells exhibit a power-law behavior at long times ($\Delta(t) \sim t^{\alpha}$ with $\alpha \approx 1.40 \pm 0.04$ from experiments compared with 
simulation results in Fig.~\ref{msdexp}(a) with $\alpha \approx 1.26 \pm 0.05$)
indicating that, on an average, directed rather than random cell motion is observed. $(ii)$
%Our model shows perfect agreement with this experimental observation (see Fig.~\ref{figx2}(a)). 
%Secondly, 
Cells exhibit a distinct topological motility profiles.  At the spheroid periphery 
cell movement is persistently along the radial direction while stochastic movement is observed for cells closer to the center. Such spatial topological heterogeneity is 
well-described as arising in our model from pressure dependent inhibition (see Figs.~\ref{figx3}, \ref{figx3part2} \&  \ref{figx4}). 
$(iii)$ The highly invasive spheroid boundary (deviating from what would be expected due to an isotropic random walk) as experimentally observed 
is qualitatively consistent with simulation results (see Fig.~\ref{figs6}). 

A salient feature of the dynamics of living cells is that birth and death processes break number conservation, having consequences 
on their collective behavior~\cite{toner2012birth}. To account for these processes leading to the super-diffusive behavior at long times, we establish a field  theory based on stochastic quantization that account for the physical interactions of the cells as well as birth and death processes. Simulations and theory suggest a mechanism of the plausible universality in the onset of super-diffusive behavior in tumor growth and unrelated  systems. Remarkably, the theory predicts the dynamics of invasion at all times that are in good agreement with recent experiments~\cite{valencia}.

%change1 ends here
%We investigated three-dimensional tumor evolution using a  spatial model with 
%unbalanced cell growth and apoptosis. The system size grows 
%exponentially at early times, followed by a power-law growth 
%behavior at longer times. The change in the growth law occurs because pressure acting on cells due to their 
%neighbors inhibits cell growth.  
\textbf{Onset of super-diffusion depends on cell division time.} In previous studies~\cite{ranft,matoz}, fluidization of tissues due to 
cell division and apoptosis was observed at the homeostatic state.  Our work shows that  a 
glass-to-fluid transition is driven by cell division at  
non-steady states and under free boundary conditions, relevant during early stages of cancer invasion. 
%which is reminiscent of the metastasis of cancer. 
The transition 
from glass to fluid-like behavior is determined by the average cell division 
time. Super-diffusion of cells in the mean-squared displacement due to highly polarized tumor growth is observed on a time scale 
corresponding to the cell division time 
with a universal scaling exponent $\alpha = 1.26 \pm 0.05$. 

\textbf{Comparison with previous studies:}
The startlingly contrasting results that we find for the dynamics of tracer cells 
and live cells compared to the results reported elsewhere~\cite{matoz} could arise for the following reasons:
(i) In our simulations, we use  free boundary conditions and the tumor 
grows continuously (with birth rate being always higher than the death 
rate). The effect of a free boundary  is particularly 
pronounced at the periphery of the tumor where the cells undergo rapid division. 
%We focus on the evolution of
% tumor growth at early stages which is far from equilibrium state. 
(ii) Imposing a birth rate that
depends on the local cell density (Ref.~\cite{ranft}) or on the number of nearest neighbors
as in Ref.~\cite{matoz} eventually results in a homeostatic state where birth and death are balanced. 
%(as noted in Ref. [25] and by this referee). 
The super-diffusive behavior, observed in our study would  not be
present, when there is a possibility of reaching a homeostatic dense liquid-like state. In our model, this situation could be mimicked by arbitrarily increasing the cell cycle time. 
For example,  when cell cycle times are very long $\tau >\sim 10~\tau_{min}$, 
 the super-diffusive MSD exponent  in the long time regime begins to deviate from $\approx~1.3$ to lower values (see Fig.~\ref{msdexp}(a)). 
%- see Fig.~\ref{msdvaried} and explanation below. 
(iii) Even in our model, simple diffusion at long times is obtained if the death rate is modified. In Fig.~\ref{msdvaried}, 
normal diffusion (red symbols) is 
 observed at long times when the death rate is modified to
 \begin{equation}  
 k_{b}=[1/(\tau_{min})]N(t)/(N(0)+N(t)),
 \label{deathrate}
 \end{equation}
  where $N(t)$ and $N(0)$ are the tumor 
 size at time $t$ and $t=0$ respectively. At $t=0$, cells have a higher birth rate, $k_{a}=1/(\tau_{min})$, 
 compared with the death rate ($k_{b}/k_{a} =0.5$). As the tumor grows, the death rate becomes 
 higher and a homeostatic state is reached once  the birth and death 
 rates are balanced, giving rise to normal diffusion as found in Ref.~\cite{ranft,matoz}. 
Therefore, the super diffusive behavior can only be found
if birth and death processes are not balanced, the regime which is the focus of our study. Most importantly, if there is a mechanism for reaching homeostasis by balancing birth and 
death rates or making the cell division time arbitrarily long we predict that normal diffusion would result, as shown here using Eq.\ref{deathrate} and previously found elsewhere~\cite{ranft,matoz}.

%,  in perfect agreement with a recent experimental observation ($\alpha \approx 1.4 \pm 0.04$)~\cite{valencia}.  
%We explained the universality of the exponent using a novel theory that takes into account birth 
%and death processes.
%Similar super-diffusive 
%motion is also found in other soft materials such as soap foams
%(with $\alpha \approx 1.37$~\cite{Hwang2016}). Our theoretical model 
%(using the Doi-Peliti formalism) 
%shows that the super-diffusive behavior can be induced by birth-death processes 
%in the long time limit resulting in a general scaling exponent $\alpha = 1.33$ 
%irrespective of the interaction between agents in the system. 
%It provides a perfect 
%explanation for the super-diffusion phenomena observed in the system studied here 
%and also other similar systems\cite{Hwang2016}.

{\textbf{Lack of time translational invariance:}
The field theory shows that the dynamics is not time translationally invariant, 
which is supported by simulation of $\Delta (t)$. 
Fig.~\ref{msdrep7} shows two different methods used to calculate the 
$\Delta (t)$: (i) the definition of MSD, $\Delta(t)=\langle \frac{1}{N(t)}\Sigma_{i}[r_{i}(t)-r_{i}(0)]^2 \rangle$, 
which always utilizes the original
position of cells (at the initial simulation time) ($r(0)$). Here, $N(t)$ is the number of initial cells and 
$\langle ... \rangle$ is the average over multiple simulation runs; (ii) 
$\Delta(t)=\langle \frac{1}{N_t}\Sigma_{t_{1}}^{(t_{S}-t)}[r(t_{i}+t)-r(t_{i})]^2 \rangle$, a time shift average (varying $t_{i}$), 
used routinely in simulations of periodic systems %with time translational invariance gives good statistics. 
with $N_t$ being the number of possible time intervals for a given $t$. Here, $\langle ... \rangle$ 
denotes average over initial cells. In generating the results in Fig.~\ref{msdrep7} we chose $t_{S}$ = 500,000$s$.

These two methods for computing $\Delta (t)$ produced different results, as shown in 
Fig.~\ref{msdrep7}. The inequivalence of the two methods in obtaining $\Delta (t)$ 
shows that the system with free boundary conditions violates time translational invariance. 
The intermediate regime with sub-diffusive behavior (red circles) 
using method (i) disappears when the second method (ii) is used to compute the $\Delta (t)$ 
(green squares). 
%This clearly demonstrates that our system is not time-translation invariant, and 
It is also the reason why we focused on 
initial cells for $\Delta (t)$ calculations. 
In calculating $\Delta (t)$ using the method (ii) commonly employed 
in simulations of periodic systems, large amount of statistics 
is extracted for the super-diffusive regime (as it has larger time range on the order of $\approx 10^5 \mathrm{s}$). 
Therefore, when we average $\Delta (t)$ over the various time intervals, the beginning time regime  
(which is comparatively short $\approx 10^4\mathrm{s}$) is suppressed. 
It should be noted, however, that the time shift averaged method of computing $\Delta (t)$ also clearly 
shows evidence for super diffusive behavior over three decades in time (Fig.~\ref{msdrep7}).

\section*{Conclusions}
Heterogeneity is a hallmark of cancer~\cite{Almendro2013}. It is difficult to 
capture this characteristic of cancers in well-mixed models that exclude 
spatial information. 
%By implementing an agent based off-lattice model, 
%however, heterogeneity in dynamics is clearly demonstrated. 
An important signature of cell dynamic heterogeneity - 
large variations in the squared 
displacement of cells in the tumor - is observed in our simulations. We find a broad velocity 
distribution among tumor cells driven by cell growth rate. 
The formation of spatial niches, with tumor periphery and center as being topologically distinct is
characterized by differences in proliferative and cell signaling activities. Such a distinct behavior, 
alluded to as the driving factor behind intratumor heterogeneity (ITH)~\cite{hoefflin2016spatial,lee2017melanoma}, is not well understood. 
Our results predict that the pressure dependent inhibition of cell growth to be the critical factor behind the development 
of distinct topological niches, implying that the dynamics of cells is dependent on the microenvironment~\cite{hoefflin2016spatial,lee2017melanoma}.
Cells closer to the center of the tumor spheroid are surrounded by many other cells 
causing them to be predominantly in the dormant state and can move in 
random directions, while cells closer to the periphery can divide and move in a directed manner by pushing against the extracellular matrix, thus promoting tumor growth and invasiveness. 
We provide experimentally testable hypotheses on the signatures of heterogeneity -  
the onset of ITH could occur at very early stages of tumor growth (at the level of around 10,000 cells).

Although the context of our work being rooted in understanding tumor growth, we expect our model 
to be relevant to the study of soft glassy materials. 
The motion of cells in our model is surprisingly consistent with the complex motion of 
bubbles in a foam, also shown to be super-diffusive \cite{Hwang2016} 
with the MSD exponent of $\alpha \approx 1.37 \pm 0.03$, consistent with both our theoretical predictions ($\alpha \approx 1.33$) 
and simulation results ($\alpha =  1.26 \pm 0.05$).  The bubbles are characterized by birth and death processes and pressure dependent growth, which we predict to be the 
driving factors behind super-diffusive behaviors observed in these diverse systems.  
Emergence of underlying similarities in the motion of constituent particles between living systems, such as cells, and soft glassy materials, such as foams, 
suggest that many of the shared, but, 
as of yet unexplained dynamic behavior may 
emerge from a common underlying theme - an imbalance in the birth and death processes and pressure dependent growth inhibition.

\begin{acknowledgments} We acknowledge Anne D. Bowen at the Visualization Laboratory (Vislab), Texas Advanced Computing Center for help with figure and movie visualizations. We are grateful to  Mauro Mugnai, Naoto Hori and Upayan Baul for discussions and comments on the manuscript. This work is supported by the National Science Foundation (PHY 17-08128 and 16-32756). Additional support was provided by the Collie-Welch Chair (F-0019). A. N. M and X. L. contributed equally to this work.
\end{acknowledgments}

\appendix

\section{Simulation Test}
\label{secdet}

{\bf Effects of random forces:} The neglect of random forces, which should be taken into account to satisfy the FDT, 
might seem like a drastic simplification. There are, however, two considerations. First, the tumor growth model 
involves birth and apoptosis. Hence, it behaves like an active system. Indeed, the theory outlined in Section~\ref{secth} shows that 
under these conditions FDT is not satisfied, 
forcing us to adopt the stochastic quantization methods to compute response and correlation %missing word
functions (Fig.~(\ref{fig:rg5})). Second, from practical considerations we note that the 
%It is possible to include random forces within the model. 
%However, the corresponding 
cellular diffusion constant is $10^{-4}$ $\mathrm{\mu m^2/s}$ or smaller~\cite{schaller}, resulting in only small displacements for a large fraction of cells. 
%The expansion of the tumor spheroid front driven by proliferating cells will generally engulf cells 
%that have separated due to random motion under physiological conditions. 

In order to verify that the contributions to the dynamics arising from the random noise is small, we modified Eq.~(\ref{eqforce}) 
to include the random forces,
\begin{equation}
\label{eqforcenoise}
\dot{\vec{r}}_{i} = \frac{\vec{F}_{i}}{\gamma_i} + \sqrt{2k_{B}T/\gamma_{i}^{visc}}\zeta_{i}(t),
\end{equation}
where $k_B$ is the Boltzmann constant, $T$ the temperature and $\zeta$ is
white noise with zero mean and variance, $\langle\zeta_{i}(t)\zeta_{i}(t')\rangle = \delta(t-t')$. 
The corresponding diffusion constant, $k_{B}T/\gamma^{visc}$, is small. 
%No effect on the MSD exponent 
%is observed due to random noise (see Fig.~\ref{msdnoise}) as expected. 
%The growth rate of the cell at each time step is also stochastic as explained below.   
%a normal distribution with average . 
Thus, inclusion of random force has no consequence on the dynamics of tumor evolution. 
The results for $\Delta(t)$ as a function of $t$ obtained using Eqs.~(\ref{eqforce}) and (\ref{eqforcenoise})
are identical (Fig.~\ref{msdnoise}). 

\section{MOVIES}
\label{secmov}
In order to visualize the dynamic growth of the tumor we generated movies from the simulations. 
They demonstrate vividly the polarized growth of the tumor, which we have quantified using various measures in the main text. 
%Simulation movies can be found here \url{https://utexas.box.com/s/voitlbt4zv7l8ygc1g0n4ya5ob98octf}

{\bf Supplementary Movie 1. 3D growth of tumor}

This movie shows the three dimensional growth of the tumor over $\approx$ 8 days.  Each frame is at $1000$ seconds. The cell cycle time $\tau=\tau_{min}$.  
Colormap indicates the life time of the cells. 
Newborn cells are shown in blue and older cells that have lived longer are in red (color bar in video shows cell lifetime in seconds). Cell division and death events are explicitly depicted. 

{\bf Supplementary Movie 2. Cross section view through the growing tumor spheroid}

Illustration of an alternate view of the growing tumor shown in Supplementary Movie 1. Cells with longer lifetimes are
mostly localized near the center of the tumor with some of them moving to the periphery. Newly born cells are mostly located  
in the periphery and division events are amplified in the periphery compared to the center of the tumor. Color bar shows cell lifetime.  

{\bf Supplementary Movie 3. Moving clip through a tumor showing velocity heterogeneity}

This video visualizes the velocity heterogeneity within the tumor. 
Colormap indicates the speed of cells (shown in log scale) and the direction of velocity is indicated by an arrow. The video begins 
with a snapshot of the tumor after $\approx$ 3 days of growth at $\tau=0.25\tau_{min}$. A clip moving 
through the tumor shows the velocity distribution of cells over different slices. 
It is clear  that cells move slowly closer to the center 
while faster moving cells are mostly in the periphery. Direction of the velocity is more randomly oriented in the tumor center 
but is mostly polarized outward as  the periphery is approached. 
\par \bigskip

\section{Theory}
\label{secth}
The behavior of the Mean Square Displacement ($\Delta(t)$), especially the time dependence of $\Delta(t)$ at intermediate 
and long times, can be theoretically obtained for the tumor
growth model mimicking the one used in the simulations. 
We consider the dynamics of a colony of cells in a dissipative environment with negligible inertial effects. 
The interaction between cells is governed by  adhesion and excluded volume repulsion.
%Each cell experiences a thermal white noise, and moves under the force generated from the potential due to the neighbors. 
The equation of motion for a single cell $i$ is,
\begin{equation}
\label{eqmo}
\frac{\partial {\bf r}_i}{\partial t}=-\sum_{j=1}^{N}{\bf \nabla} U({\bf r}_i(t)-{\bf r}_j(t))+\eta_i(t),
\end{equation}
where $U$ contains the following form of repulsive interactions with range $\lambda$, and favorable attractive interactions between cells with range $\sigma$,
\begin{eqnarray}\label{HamiltVH}
&&{U}({\bf r}(i)-{\bf r}(j))=\\ \nonumber &&\frac{v}{(2\pi \lambda^2)^{3/2}}
e^{-\frac{({\bf r}(i)-{\bf r}(j))^2}{2\lambda^2}}-
\frac{\kappa}{(2\pi \sigma^2)^{3/2}} e^{-\frac{({\bf r}(i)-{\bf r}(j))^2}{2\sigma^2}}.
%\label{HamiltV}
\end{eqnarray}
$v$ and $\kappa$ above are the strengths of the repulsive and attractive interactions, respectively. 
The $\kappa$ parameter in Eq.~\ref{HamiltVH} mimics adhesion between cells. 
The noise ($\eta_{i}$ in Eq.~(\ref{eqmo})) is uncorrelated in time. 

The simplified form for $U$, which captures minimally the interactions between cells but differs from the more elaborate model used in the simulations, allows us to obtain analytical results
for $\Delta(t)$ as a function of $t$. %in harmony with simulations. 
In terms of the density field of a cell, $\phi_i({\bf r},t)=\delta[{\bf r}-{\bf r}_i(t)]$,
a closed form Langevin equation for the density, $\phi({\bf r},t)=\sum_i \phi_i$ can be obtained using the 
approach introduced by Dean~\cite{Dean}. %The time evolution of $\phi({\bf r},t)$ is given by Eq.~(1)  in the main text. 
%$\phi({\bf r},t)=\sum_i \delta[\bf r-{\bf r}_i(t)]$
%\begin{equation}
%\frac{\partial \phi({\bf r},t)}{\partial t}={\bf \nabla} \cdot \left(\eta({\bf r},t) \phi^{1/2}({\bf r},t)\right)+ {\bf \nabla }\cdot \left(\phi({\bf r},t)\int d{\bf r'} \phi({\bf r'},t){\bf \nabla}U({\bf r-\bf{r'}})\right),
% \end{equation}
%where $\eta$, the Gaussian white noise, satisfies $<\eta({\bf r},t)\eta ({\bf r'},t')>=\delta({\bf r-r'})\delta(t-t')$.
In order to study tumor cell dynamics, we extend the model phenomenologically %in order to 
to describe both cell division and death, and introduce a noise term that breaks the cell number conservation.
These crucial features needed to describe tumor growth can be investigated using the Doi-Peliti (DP) 
formalism~\cite{Doi1976,peliti}, introduced in the context of reaction-diffusion processes. A related approach was used recently by Gelimson and Golestanian~\cite{gelimson2015collective} to describe collective dynamics in a dividing colony of chemotactic cells.

We use a scheme to study the interplay between stochastic growth and apoptotic process, 
and use it to derive a Langevin type equation for logistic growth. 
%The DP formalism allows us to derive an expression for the density dependence of the noise strength describing number fluctuations. 
The birth reaction, $X \xrightarrow[]{k_{a}} X+X$, occurs with the rate constant $k_a$ for each cell, and the 
backward reaction (apoptosis)  $X+X\xrightarrow[]{k_{b}} X$ occurs with rate $k_b$. 
By incorporating birth and apoptosis, and assuming that 
%we obtain the following 
%Langevin equation, for the time-dependent changes in the density, $\phi({\bf r},t)$, 
%\begin{eqnarray}
%\label{phi10}
%\frac{\partial \phi({\bf r},t)}{\partial t}=&& {\bf \nabla }\cdot \left(\phi({\bf r},t)\int d{\bf r'} \phi({\bf r'},t){\bf \nabla}U({\bf r-\bf{r'}})\right)+k_{b} \phi(\frac{k_a}{k_b}-\phi)\\ \nonumber && 
%+{\bf \nabla} \cdot \left(\eta({\bf r},t) \phi^{1/2}({\bf r},t)\right)+\sqrt{k_a \phi+k_b \phi^2} f_\phi \, ,
%\end{eqnarray}
%with $f_\phi$ $<f_\phi({\bf r},t)f_\phi({\bf r'},t')>=\delta({\bf r}-{\bf r}')\delta(t-t')$.
%%First term in the r.h.s of above equation represents the growth rule, where $\frac{\alpha}{\beta}$ is the carrying capacity, and $\beta$ is the effective growth rate per unit concentration. 
%The third term, on the right hand side (RHS) of Eq.~(\ref{phi10}), arising due to Brownian noise, 
%has negligible influence in producing the scaling behavior for the MSD, and hence may be discarded. 
%
%To simplify the multiplicative noise term (last term in Eq.~(\ref{phi10})), we assume that 
the density fluctuates around a constant value, 
%$C_0$,
%allowing us to write 
$\phi({\bf r},t)=C_0+\rho({\bf r},t)$, we obtain the approximate 
%If we expand $\phi({\bf r},t)$ in powers of $\frac{\rho}{C_0}$
%to the lowest order in non-linearity, the 
equation for the density fluctuation, which
%becomes,
%\begin{eqnarray}\label{rho}
%\frac{\partial \rho({\bf r},t)}{\partial t}=&&{\bf \nabla}\cdot \left(C_0 \int d{\bf r'} \rho({\bf r'},t) \nabla U({\bf r-r'})\right)+{\bf \nabla}\cdot \left(\rho({\bf r'},t)  \int d{\bf r'} C_0 \nabla U({\bf r-r'})\right) \\ \nonumber
%&&+{\bf \nabla }\cdot \left(\rho({\bf r},t)\int d{\bf r'} \rho({\bf r'},t){\bf \nabla}U({\bf r-\bf{r'}})\right)+(a-2b C_0)\rho({\bf r'},t) -b \rho^2+\sqrt{a C_0+b C_0^2} f_\phi .
%\end{eqnarray}
in Fourier space reads,
%\begin{widetext}
\begin{eqnarray}\label{LF}
\label{rho12}
\frac{\partial \rho({\bf k},t)}{\partial t}=&&-C_0 k^2 U({\bf k}) \rho({\bf k},t)+(k_{a}-2k_{b}C_0)\rho({\bf k},t)\\ \nonumber && +\int d{\bf q} (-{\bf q}\cdot {\bf k})U({\bf q})\rho({\bf q},t)\rho({\bf k}-{\bf q},t) 
-k_{b} \int d{\bf q} \rho({\bf q},t)\rho({\bf k}-{\bf q})+\sqrt{k_{a}C_0+k_{b}C_0^2} f_\phi . 
\end{eqnarray}
%\end{widetext}
%Without non-linear terms one can infer from Eq. (\ref{LF}) that the uniform density phase is 
%stable if $(C_0 k^2 U({\bf k}) -(a-2b C_0))>0$ or $(a-2b C_0)<0$ in the hydrodynamic limit. 
%An instability,  corresponding to a phase transition,  occurs at $(C_0 k^2 U({\bf k}) -(a-2b C_0))=0$.  
We derived Eq.~(\ref{LF}) by expanding the density to lowest order in $\frac{\rho}{C_0}$ non-linearity. 
The noise $f_\phi$ satisfies $\langle f_\phi({\bf r},t)f_\phi({\bf r'},t') \rangle =\delta({\bf r}-{\bf r}')\delta(t-t')$. 
In the hydrodynamic, $k\rightarrow 0$ and $t\rightarrow \infty$ limit, the first and third terms 
in the RHS of Eq.~(\ref{rho12}) vanish, and hence the scaling behavior of $\Delta(t)$ at long times 
is determined solely by the death-birth terms. 
%implying that mechanical interactions, embodied in $U(\bf{k})$, do not play any role in determining 
%the scaling behavior. This is also 
%reflected in the simulation results in the long time and long distance limit. However, in the finite time 
%and space limit, all the terms contribute to the scaling behavior. We assume the strength of 
%the interactions is such that mechanical interaction dominates over the death-birth (first and third terms in Eq.~(\ref{rho12})) process   
%and will be important in determining the scaling behavior in the finite time limit.  
%To get the scaling behavior, we use the change of scale ${\bf r}\rightarrow s {\bf r}$, $\rho \rightarrow s^{\chi} \rho $ and $t\rightarrow s^z t$ 
%where $\chi $ is the cell density fluctuation exponent and $z$ is the dynamic exponent.

The scaling of $\Delta(t)$ can be obtained by treating the non-linear terms in Eq.~(\ref{rho12}) perturbatively 
using the Parisi-Wu stochastic quantization scheme~\cite{Parisi1981,HSSPRE2006,HSSPLA2006}, 
which is needed because Fluctuation Dissipation Theorem (FDT) is not satisfied in Eq.~(\ref{rho12}) due to 
%violation of number conservation. 
cell birth and death processes. 
In order to outline the essence of the theory, let us consider the probability distribution corresponding to the noise term given by
\begin{equation}
\label{pfphi}
P(f_\phi)({\bf k},\omega) \propto \text{exp}\left[-\int \frac{d^D{\bf k}}{(2\pi)^D}\frac{d\omega}{2\pi}\frac{1}{2}f_{\phi}({\bf k},\omega)f_{\phi}(-{\bf k},-\omega)\right ].
\end{equation} 
By re-expressing $P(f_{\phi}({\bf k},\omega))$ in terms of $P(\rho({\bf k},\omega))$, a Langevin equation of motion 
in the fictitious time, $\tau_f$, may be derived in which FDT is satisfied. Consequently, 
in the $\tau_f\rightarrow \infty$ limit, the distribution function $P(\rho({\bf k},\omega))~\propto~\text{exp}(-S({\bf k},\omega))$, where an expression for the effective action 
$S({\bf k},\omega)$ is derivable from Eqs.~(\ref{LF}) and (\ref{pfphi}). 
Using this formalism, the %correlation function needed to obtain $\Delta(t)$ may be computed 
Green's function can be obtained using perturbation theory
by solving the Dyson equation, 
\begin{equation}
\label{green1}
[G]^{-1}=[G^{(0)}]^{-1}+\Sigma({\bf k},\omega, \omega_{\tau_f}), \, 
\end{equation}
where $\omega_{\tau_f}$ is the frequency related to $\tau_f$ and $G_0^{-1}=i\omega_{\tau_f}+\frac{1}{2(k_a C_0+k_b C_0^2)}[  \omega^2 +\{C_0 k^2 U({\bf k})-(k_a-2k_bC_0)\}^2]$.
Diagrammatic representation of self energy term $\Sigma({\bf k},\omega, \omega_{\tau_f})$ is shown in 
Fig.~\ref{fig:rg5} to one loop order. We obtain, $\Sigma({\bf k},\omega, \omega_{\tau_f})\sim  \int \frac{d^D {\bf k_{1}}}{(2\pi)^D} \frac{d\omega_{1}}{2\pi} \frac{d\omega_{\tau_{f1}}}{2\pi} V V G_{0}C_{0}$, 
where the vertex term is of the form: $V=\frac{1}{2(k_a C_0+k_b C_0^2)}[\{i \omega+C_0 k^2 U({\bf k})-(k_a-2 k_b C_o )\} \{(-{\bf k_1} \cdot {\bf k}) U({\bf k_1})-k_b\}+
	 \{i \omega_1+C_0 k_1^2 U({\bf k_1})-(k_a-2k_b C_o )\} \{(-{\bf k_1}\cdot {\bf k}) U(-{\bf k})-k_b\}+\{i \omega_1+C_0 k_1^2 U({\bf k_1})-(k_a-2 k_b C_o )\}  \{(-{\bf k_1} \cdot ({\bf k}-{\bf k_1})) U({\bf k}-{\bf k_1})-k_b\}]$, 
the correlation $C_{0}=G_{0}G_{0}^{*}$, and $D$ is the spatial dimension.
After computing the self energy to second order in non-linearity, Eq. (\ref{green1}) can be written as,
\begin{equation}\label{green2}
[G]^{-1}({\bf k},\omega,\omega_{\tau_f})=-i\omega_{\tau_f}+\frac{1}{2(D_0)}[  \omega^2 ]+\frac{1}{2(\bar{D})}[\nu_{eff}^2 k^4] \, ,
\end{equation}
where $D_0=k_{a}C_0+k_{b} C_0^2$. The above equation allows us to determine an effective coefficient $\bar{D}$ from $G^{-1}({\bf k},0,0)$,
%and $\bar{D}=\bar{k}_a C_0+\bar{k}_b C_0^2$, and
\begin{equation}\label{green3}
\frac{1}{2(\bar{D})}[ \nu_{eff}^2 k^4]=\frac{1}{2(D_0)}(\nu k^2 )^2+\Sigma({\bf k},\omega, \omega_{\tau_f}), 
\end{equation}
with $\nu=C_0  U({\bf k})$. In obtaining Eq.~(\ref{green3}), needed for calculating the scaling of $\Delta(t)$ in the intermediate time, 
the strength of the interactions are such that $C_0 k^2 U({\bf k})$ 
dominates over $(k_{a}-2k_{b}C_0)$. 
%We neglect the term $(k_{a}-2k_{b}C_0)$ in the Green's function equation in the finite time regime. %ask
Expanding $\nu_{eff}$ about $\nu$ and $\bar{D}$ around $D_0$, and noting that the renormalization of $\nu $ dominates, we write 
using $\Delta \nu = \nu_{eff} - \nu$,
\begin{equation}\label{scale}
%\nu_{eff} l^2 \simeq \nu l^2 +\frac{1}{2\nu l^2} \Sigma({\bf l},\Omega, \Omega_{\tau_f}),~~\text{or}~ ~
\Delta\nu k^2=\frac{1}{2\nu k^2} \Sigma({\bf k},\omega, \omega_{\tau_f}). 
\end{equation}
In the spirit of self-consistent mode coupling theory, we now replace $\nu$ by $\Delta \nu$ in 
the self energy term $\Sigma({\bf k},\omega, \omega_{\tau_f})$, use $G$  as given by 
Eq.  (\ref{green2}), and the correlation function $C~=~GG^{*}$, as follows from the FDT. 
According to scale transformation, $\omega \sim k^z$, $\omega_{\tau_f} \sim k^{2z}$, $G \sim k^{-2z}$, 
$C \sim k^{-4z}$ and the vertex factor $V \sim k^{z+2}$. The self energy term in Fig.~(\ref{fig:rg5}) can be written as 
$\Sigma({\bf k},\omega, \omega_{\tau_f})\sim  \int \frac{d^D {\bf k}'}{(2\pi)^D} \frac{d\omega'}{2\pi} \frac{d\omega'_{\tau_f}}{2\pi} V V GC$ (Fig.~(\ref{fig:rg5}) 
provides a diagrammatic representation of the theory). 
By carrying out the momentum count of $\Sigma({\bf k},\omega, \omega_{\tau_f})$, and 
noting that $\nu k^2\sim k^z$, we find $\Sigma(k,\omega, \omega_{\tau_f})\sim k^{D-z+4}$. 
Using Eq. (\ref{scale}), we obtain $k^{z+2}\sim k^{D-z+4}$, leading to $z=1+\frac{D}{2}$.

The scaling of $\Delta(t)$ at intermediate and long times may be gleaned using the relation $C~=~(1/\omega_{\tau_f})$Im~$G$. Assuming dynamic scaling holds, the single cell mean-square displacement should behave as,
\begin{equation}
\Delta(t)=<[{\bf r}(t)-{\bf r}(0)]^2>\sim t^{2/z}=t^\alpha.
\label{msdsi}
\end{equation}
In 3D, $\alpha=\frac{4}{5}= 0.8$, implying $\Delta(t)$ should display sub-diffusive behavior. 
%Deviation from the simulation 
%result is due to differences in the potential used in the theory. 
The theoretical prediction is in accord with the behavior of $\Delta(t)$ in the caging regime. 
In the long time limit, the non-linearity due to death-birth dominates over mechanical interactions ($\propto~U({\bf k})$). A similar procedure, as mentioned above, produces the dynamic exponent $z=D/2$. In this regime, $\alpha =1.33$,
implying super-diffusive motion, a prediction that is also in agreement with our simulations and experimental results~\cite{valencia}.
Thus, the theory explains the simulation results, and by extension the experimental data, nearly quantitatively. 
\newpage

%\begin{widetext}
\textbf{Table I:}
The parameters used in the simulation. \par \bigskip

\begin{tabular}{ |p{7cm}||p{4cm}|p{5cm}|p{3cm}|  }
 \hline
 \bf{Parameters} & \bf{Values} & \bf{References} \\
 \hline
 Timestep ($\Delta t$)& 1$\mathrm{s}$ - 10$\mathrm{s}$  & This paper \\
 \hline
Critical Radius for Division ($R_{m}$) &  5 $\mathrm{\mu m}$ & ~\cite{schaller}\\
 \hline
 Extracellular Matrix (ECM) Viscosity ($\eta$) & 0.005 $\mathrm{kg/ (\mu m~s)}$   & ~\cite{galle}  \\
 \hline
 Benchmark Cell Cycle Time ($\tau_{min}$)  & 54000 $\mathrm{s}$  & ~\cite{freyer, casciari, landry}\\
 \hline
 Adhesive Coefficient ($f^{ad})$&  $10^{-4} \mathrm{\mu N/\mu m^{2}}$  & ~\cite{schaller} \\
 \hline
Mean Cell Elastic Modulus ($E_{i}) $ & $10^{-3} \mathrm{MPa}$  & ~\cite{galle}    \\
 \hline
Mean Cell Poisson Ratio ($\nu_{i}$) & 0.5 & ~\cite{schaller}  \\
 \hline
 Death Rate ($b$) & $10^{-6} \mathrm{s^{-1}}$ & This paper \\
 \hline
Mean Receptor Concentration ($c^{rec}$) & 1.0 (Normalized) & ~\cite{schaller} \\
\hline
Mean Ligand Concentration ($c^{lig}$) & 1.0 (Normalized) & ~\cite{schaller}  \\
\hline
Adhesive Friction $\gamma^{max}$ &  $10^{-4} \mathrm{kg/ (\mu m^{2}~s)}$  &  This paper\\
\hline
Threshold Pressue ($p_c$) & $10^{-4} \mathrm{MPa}$  & ~\cite{schaller,montel}    \\
\hline
%\label{table:1}
%\caption{Table 1}
\end{tabular}
%\end{widetext}
\par \bigskip
\clearpage

%\begin{widetext}
\textbf{Table II:}
The goodness of fit in the inset of Fig.~\ref{figs1}. R-square value (middle column) and RMSE, the root mean squared error (last column) for power law and exponential fits are provided. 
The digits in parenthesis in the last column refer to powers of 10. \par \bigskip
%\begin{table}
%\centering
\begin{tabular}{|p{3cm}||p{4cm}|p{4cm}|} 
 \hline
\bf{ Fit Type} & \bf{$R^2$} & \bf{RMSE} \\ [0.5ex] 
 \hline\hline
 Power & 0.93 & $2.4(-05)$ \\ 
 \hline
 Exponential & 0.57 & $5.7(-05)$ \\ 
 \hline
\end{tabular}
%\caption{The R-square and root mean square error (RMSE) of fits.}
\label{table:2}
%\end{table}
%\end{tabular}
%\end{widetext}

%\showacknow % Display the acknowledgments section
%\pnasbreak
%\subsection*{References}
%\bibliographystyle{h-physrev}
%\bibliographystyle{apsrev4-1}
%\bibliographystyle{abbrv}
%\bibliographystyle{unsrt}
\newpage
\bibliographystyle{unsrt}

\bibliography{PRX_nocolor}
\newpage
{FIG.~1:\textbf{(a)} Illustration of two interpenetrating cells $i$ and $j$ with radii $R_{i}$ and $R_{j}$, respectively.  The distance between the centers of the two cells is $|{\bold {r}}_{i} -{\bold {r}}_{j}|$, and their overlap is $h_{ij}$.
\textbf{(b)} Force on cell $i$ due to $j$, ${\bf F}_{ij}$, for $R_{i}=R_{j}=4~\mu m$ using mean values of elastic modulus, poisson ratio, receptor and ligand concentration (see Table I). ${\bf F}_{ij}$ is plotted 
as a function of distance between the centers of the two cells. 
Inset shows the region where ${\bf F}_{ij}$ is attractive. When $|{\bold {r}}_{i} -{\bold {r}}_{j}|~\ge~R_{i}+R_{j}=8~\mu m$ the cells are no longer in contact, and hence, ${\bf F}_{ij}=0$.}

\bigskip

{FIG.~2: Cell dormancy (left panel) and cell division (right panel). If the local pressure $p_i$ that the $i^{th}$  cell experiences 
(due to contacts with the neighboring cells) exceeds the critical pressure $p_{c}$, it enters 
the dormant state ($D$).  Otherwise, the cells grow (G) until they reach the mitotic radius, $R_{m}$. 
At that stage, the mother cell divides into two identical daughter cells with the same radius $R_{d}$. 
We assume that the total volume upon cell division is conserved. A cell that is dormant at a given time can transit from that state at subsequent times.}

\bigskip

{FIG~3: \textbf{(a)} Normalized volume, $V(t)/V_0$, of a tumor spheroid as a function of time. The result of simulation 
(red) agrees nearly quantitatively with experimental data obtained for the tumor spheroid growth 
at an applied pressure of $500$Pa~\cite{montel}. We used a critical pressure  
$p_{c}=100$Pa and cell division time of $\tau=\tau_{min}$ in these simulations. The reason for comparing the results from the $p_c=100$Pa simulations with the growth dynamics 
obtained in the colon carcinoma cells with an external pressure of $500$Pa is explained in the text. 
\textbf{(b)} Distribution of pressure as a function of total growth time with cell division time $\tau=\tau_{min}$ = 0.625 days. The mean values 
are indicated by the dashed lines.}

\bigskip
{FIG~4: \textbf{ Dynamics of Tumor growth.} \textbf{(a-d)} show instantaneous snapshots of 
the tumor during growth at different times. 
Each cell is represented by a sphere.  
There are approximately 2,000 cells in {\bf (d)}. The color of each cell is to aid visualization. In figures \textbf{(a-d)} the cell sizes are rescaled for illustration purposes only. 
Note that even at $t = 0$ ({\bf a}) the sizes of the cells are different because they are drawn from a Gaussian distribution.  {\bf (e)}, The total number of
tumor cells, $N(t)$, as a function of time at different values of the threshold pressure 
$p_{c}$, which increases from bottom to top
($10^{-4}$, $2\times 10^{-4}$, $3\times10^{-4}$, $5\times10^{-4}$, $10^{-3}$ MPa). 
The dashed red line is an exponential function while other lines show  
power-law behavior $N(t) \approx t^{\beta}$ where $\beta$ ranges from 2.2, 2.5, 2.8 to 3.1 (from bottom to top).  The inset in {\bf (e)} 
shows $N(t)$ with $p_{c} = 10^{-4}$ on a log-log scale with both exponential and power law fits. 
The dash-dot curve in the inset is an exponential function while the power-law trend
is illustrated by the solid line. The average cell cycle time  $\tau = \tau_{min}$ = 54,000s and other parameter values are taken from Table I.}

\bigskip
{FIG~5: \textbf{Super-diffusive behavior in $\Delta(t)$ at long times.}  
{\bf (a)} The mean-squared displacement ($\Delta(t)$) of cells. From top to bottom, the curves correspond to 
increasing average cell cycle time ($\tau$ is varied from  $\tau_{min}$ to 10$\tau_{min}$ where $\tau_{min} = 54,000$ seconds). Time taken for  reaching the 
super-diffusive regime increases by increasing $\tau$. The blue and red lines have a slopes of 
0.33, and 1.3, respectively. The sub-diffusive (super-diffusive) behavior corresponds to dynamics in the intermediate (long) times.  The left and middle inset
 illustrate the ``cage effect" and ``cage-jump" motion, respectively. %The right inset shows a zoom-in for the dashed-line rectangle at long timescales. 
 The unit for y-axis is  ($\mu m^2$).
{\bf (b)} Fit of the mean-square displacement, $\Delta (t)$ to $t^{\alpha}$ for several 
simulation runs. Each $\Delta (t)$ plot is averaged over $\approx$ 300 cell trajectories. The six $\Delta(t)$ plots consistently show $\alpha > 1.20$ over $\approx$ 1800 cell trajectories.  
The plots are separated for clarity. {\bf (c)} Distribution of $\alpha$ values from multiple independent simulations.  Mean value of $\alpha$ is $1.26$ with a standard deviation of  $\pm 0.05$.}

\bigskip

{FIG~6: The invasion distance, $\Delta r(t)$ (Eq.~\ref{deltart}), as a function of time. The exponent of the invasion distance, as indicated by the dashed line, 
determined using $\Delta r \propto t^{\xi}$ is $\xi \approx 0.63$. Note that this value and the one extracted from experiments~\cite{valencia} are in reasonable agreement. 
The inset shows $\Delta r(t)$ for $t > \tau_{min}$ for average cell cycle time $\tau=1\tau_{min}$.}

\bigskip
{FIG~7: Pair correlation function at four different cell division times 
($\tau$): $0.2\tau_{min}$ (red), $0.5\tau_{min}$ (green), $\tau_{min}$ (blue), and $10\tau_{min}$ (cyan). 
%10,800, 27,000, 54,000 & 540,000s
The cells are packed more closely at longer cell cycle times as reflected by the sharper peak for the cyan line compared 
to the others. The distance, $r$, at which $g(r)$ approaches zero is considerably smaller for $\tau = 10\tau_{min}$. The distance at which 
the first peak appears is $\approx$ 2$R_m$ $\approx~10\mu m$ (Table I), which implies that despite being soft the cells in the interior are densely packed, as in a body centered cubic lattice. }

\bigskip
{FIG~8: \textbf{Self intermediate scattering function at different cell cycle times.}  
{\bf (a)} The self-intermediate scattering function, $F_{s}(q,t)$, shows that relaxation occurs in two steps.  From left to right, the second relaxation for $F_{s}(q,t)$ 
 slows down as $\tau$ increases (from  0.5$\tau_{min}$ to 5$\tau_{min}$ ).  
The solid lines are exponential fits. The upper inset shows a zoom-in of the dashed-line rectangle at long timescales. 
%{\bf (c)} Time dependence of $\Delta (t)$ of tracer cells at different cell division times, $\tau$. 
%\textcolor{blue}{{\bf (d)} The self-intermediate scattering function for tracers. In {\bf c}, MSD fit at intermediate time 
%and long times are shown with exponents in the inset. Biexponential fits to the decay of tracer 
%$F_{s}(q,t)$ are shown by solid lines in {\bf (d)}. }
{\bf (b)} The second relaxation time $\tau_{\alpha}$ 
obtained from {\bf (a)} as a function of cell division time (rescaled by $\tau_{min}$). 
The red solid line is a linear fit ($\tau_{\alpha} \propto 0.69\tau$). {\bf (c)} The rescaled self-intermediate 
scattering function $F_{s}(q,t)/a_{0}$ as a function of the rescaled time t/$\tau_{\alpha}$.}

\bigskip
{FIG~9: \textbf{Dynamics of the tracers at different cell cycle times.}  
{\bf (a)}Time dependence of $\Delta (t)$ of tracer cells at different cell division times, $\tau$. 
Fits to $\Delta (t)$ at intermediate time 
and long times are shown in the inset.
{\bf (b)} The self-intermediate scattering function, $F_{s}(q,t)$, for tracers. Biexponential fits to the decay of tracer 
$F_{s}(q,t)$ are shown by solid lines.}

\bigskip
{FIG~10: \textbf{Trajectories displaying spatial heterogeneity.}  
{\bf (a)} Trajectories (randomly chosen from the whole tumor) for slowly growing cells are shown. Cell cycle time is 15$\tau_{min}$.  Dynamic arrest due to caging in the glass-like phase is vividly illustrated.
{\bf (b)} Trajectories for rapidly growing cells with cell cycle time $\tau = 0.25\tau_{min}$.  Displacements of the cells are shown over 3 days representing the initial stages of tumor growth in {\bf (a)} and {\bf (b)}. Two representative trajectories (time dependence of the $x-z$ and $x-y$ coordinates) for the labelled cells are shown on the right of {\bf (a)} and {\bf (b)}. Length in {\bf (a)} and {\bf (b)} is measured in units of $\mu m$. The two colored spheres in {\bf a} and {\bf b} show the approximate extent of the tumor. }

%\bigskip
%{FIG~11: {\bf (a)} Trajectories (randomly chosen from the whole tumor) for slowly growing cells are shown. Cell cycle time is 15$\tau_{min}$.  %Dynamic arrest due to caging in the glass-like phase is vividly illustrated.
%{\bf (b)} Trajectories for rapidly growing cells with cell cycle time $\tau = 0.25\tau_{min}$.  Displacements of the cells are shown over time $t$ normalized by average cell cycle time $\tau$ in {\bf a} and {\bf b}. 
%Two representative trajectories (time dependence of the $x-z$ and $x-y$ coordinates) for the labelled cells in the cuboid are shown on the right of {\bf a} and {\bf b}. 
%Length in {\bf a} and {\bf b} is measured in units of $\mu m$. The two colored spheres in {\bf (a)} and {\bf (b)} show the approximate extent of the tumor.}

\bigskip
{FIG~11:
Plot of the mean squared displacement, $\Delta (t)$, for interior, boundary and the total initial cells. 
The interior cells exhibit sub-diffusive behavior through their entire lifetime. 
The cells at the boundary show both  sub-diffusive motion at intermediate times followed by 
super-diffusive behavior at long-times. The interior $\Delta (t)$ is multiplied by $0.1$ and the boundary $\Delta (t)$ by $10$ 
%and total $\Delta (t)$ unchanged 
for clarity. Cartoon depicting the way we have divided the tumor into interior and the boundary regions is also shown.}

\bigskip
{FIG~12: \textbf{Quantifying spatial heterogeneity in tumor cell growth.}  
{\bf (a)} Probability distribution of distance $\delta r$ (in unit of $\mu m$), moved by cells over
$\delta t= 100s, 400s, 6000s$ respectively for varying average cell cycle time $\tau=0.25 \tau_{min}$, $1 \tau_{min}$ and $15 \tau_{min}$. 
$\delta t$ is normalized by $\tau$ to $0.0074$. {\bf (b)} Time resolved squared displacements, $\Delta(t)$ (in unit of $\mu m^2$), of individual cells in a model for growing tumor ($\tau = \tau_{min}$). The average, shown as a dashed line for $\approx$ 800 such individual trajectories, is not meaningful because of dynamic heterogeneity.}

\bigskip
{FIG~13: The probability distribution $P(\delta r(t_{i}^2))$ of cell displacements ($\delta r(t_{i})^{2} = |r(t_{i}+\delta t)-r(t_{i}) |^{2}$ in units of $\mu m^2$) at $\delta t=100 s$ is shown. 
Cell trajectories until $t=5\tau$ are analyzed at $\tau=1\tau_{min}$. 
The histogram was constructed 
by varying $t_i$ in $\delta r(t_{i})^{2}$ for over $400$ cell trajectories and after obtaining $\approx10^{6}$ data points for $\delta r(t_{i})^{2}$. 
For comparison, the inset shows fits to both exponential (dashed line) and power law (solid line). 
$P(\delta r^2) \sim A \times (\delta r^2)^{B}$ fit the trend best where $A$ is a constant and $B$ is 
$\sim 0.9$.
The striking non-Gaussian behavior, with fat power law tails,
is one indication of heterogeneity. Goodness of fit can be assessed using the parameters listed in Table~II.}

\bigskip
{FIG~14: \textbf{Heterogeneity in the tumor cell dynamics.}  
{\bf (a)} Instantaneous snapshot of a collection $N \approx1.3\times 10^4$ cells at $\approx$ 3 days with $\tau = 0.25\tau_{min}$. 
Colors indicate the different velocities of the individual cells (in $\mu m/s$).  
{\bf (b)} Cross section through the clump of cells shown in Fig.~\ref{figx4}a. Arrows denote the direction of velocity. {\bf (c)} 
Average speed of the cells as a function of the tumor radius at different $\tau$. 
Observation time is at $18.5\tau$, $14.8\tau$ and $11.1\tau$ for $\tau=0.25 \tau_{min}$, $0.5 \tau_{min}$ and $1 \tau_{min}$ respectively. 
{\bf (d)} Mean angle $\theta$ (see the inset figure) between cell velocity and a line through the center of the tumor to the periphery as function of the tumor radius at different $\tau$. 
Observation time is the same as in {\bf c}.}

\bigskip
{FIG~15: \textbf{Heterogeneity in the tumor cell dynamics(continued).}  
{\bf (a)} Distribution of the angle ($\theta$) at different distances ($r$) from tumor center at $\approx$ 3 days ($\tau = 0.25\tau_{min}$). 
{\bf (b)} Probability distribution of the cell speed normalized by mean cell velocity, $\langle v \rangle$, at two different cell cycle times at the long time regime ($t = 5 \tau$). }

\bigskip
{FIG~16: The mean-squared displacement ($\Delta (t)$) of cells when  
 cell death rate depends on time. The red circles shows the results obtained by averaging
 over initial cell position, $\Delta(t)=<[r(t)-r(0)]^2>$. The green squares 
 show the results calculated using time shift average, $\Delta(t)=<[r(t_{i}+t)-r(t_{i})]^2>$. 
 The solid lines show power-law fits to the simulation data. Normal diffusion results because the use of Eq.\ref{deathrate} leads to homeostasis at long times.}

 \bigskip
{FIG~17: 
 Mean squared displacement, $\Delta (t)$, of cells as a function of time. The red circles show the results obtained by averaging 
over the initial cell positions, $\Delta(t)= \langle [r(t)-r(0)]^2 \rangle$. The green 
squares are the results under time shift average, 
$\Delta(t)=\langle [r(t_{i}+t)-r(t_{i})]^2 \rangle$. The solid lines show power-law fitting of the simulation data. 
The long time super-diffusive behavior is evident in both the plots.}
 
 \bigskip
{FIG~18: Mean square displacement, $\Delta(t)$, with (blue) and without (red) random noise. The slope obtained from the long time limit are both 1.3 (dashed green). 
The two curves are almost identical, thus justifying the neglect of the random noise (second term in Eq.~(\ref{eqforcenoise})) in the simulations.}

\bigskip
{FIG~19: %The $\otimes$ is the noise correlations and the solid lines are the Green functions. 
	The diagrams correspond to perturbation expansions 
	of the theory (Eq.~(\ref{rho12})) in which the dynamical equations for the density field is expressed in fictitious time. 
	Self-energy term ($\Sigma$) is obtained by contracting the two density $\rho$ fields. 
	The first diagram is the two loop contribution generated from the first order term (contains two $\rho$ fields) in the 
	time-dependent equation for the density fields. The second diagram, with one loop contribution from the second order term (contains three $\rho$ fields), resulting in the correction to $\omega^2 +\{C_0 k^2 U({\bf k})-(k_a-2 k_bC_0)\}^2$ 
	does not have any new momentum dependance. Hence, only the first term is significant in producing the scaling results.}

\clearpage
\newpage
\floatsetup[figure]{style=plain,subcapbesideposition=top}
\begin{figure}
%\begin{turn}{-90}
%\begin{subfigure}[b]{.24\linewidth}
%\subfigure{\textbf{a} \includegraphics[width=0.9\linewidth] {cellcellinter.eps} } %oa
\sidesubfloat[]{\includegraphics[width=0.7\linewidth] {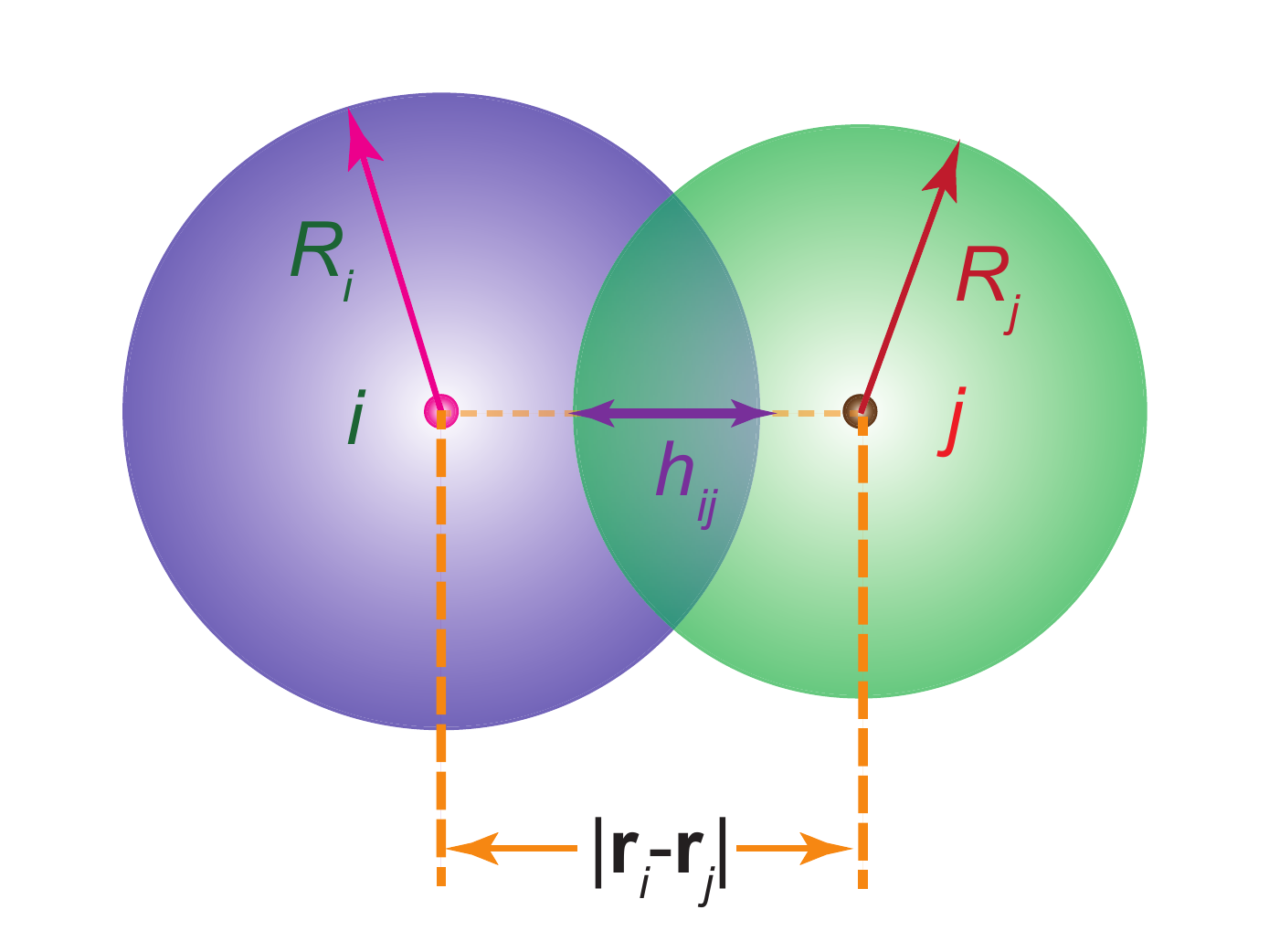}\label{cellcellinter}}
	\par\bigskip
%\caption{} 
% \includegraphics[width=0.9\linewidth] {cellcellinter.eps} 
% \label{cellcellinter}} 
% \end{subfigure}
%\end{turn}
%\par\bigskip
%\subfigure{ \textbf{b} \includegraphics[width=0.9\linewidth] {forcejkr.eps}}
\sidesubfloat[]{\includegraphics[width=0.70\linewidth] {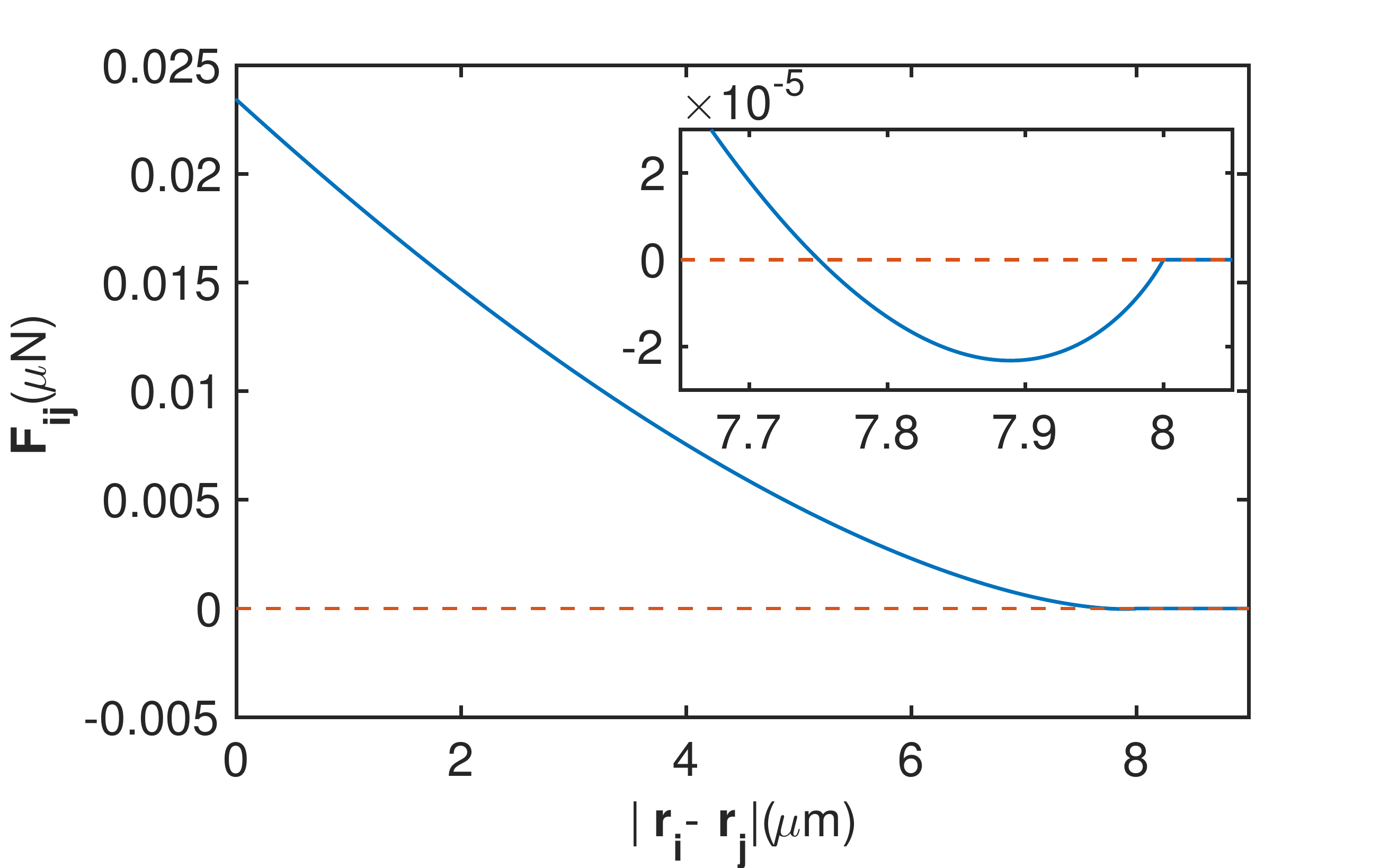}\label{forcejkr}} 
%\caption{ \textbf{b}}
%\begin{subfigure}[b]{.24\linewidth}
% \includegraphics[width=0.9\linewidth] {forcejkr.eps} 
% \label{forcejkr}}
%  \end{subfigure}
%\caption{(Continued on the following page.)}
%\caption{ }
%\end{figure}
%\begin{figure}
%\caption{(Continued on the following page.)}
%\end{figure}
%\begin{figure}
\caption{}
%\caption{\textbf{(a)} Illustration of two interpenetrating cells $i$ and $j$ with radii $R_{i}$ and $R_{j}$, respectively.  The distance between the centers of the two cells is $|{\bold {r}}_{i} -{\bold {r}}_{j}|$, and their overlap is $h_{ij}$.
%\textbf{(b)} Force on cell $i$ due to $j$, ${\bf F}_{ij}$, for $R_{i}=R_{j}=4~\mu m$ using mean values of elastic modulus, poisson ratio, receptor and ligand concentration (see Table I). ${\bf F}_{ij}$ is plotted 
%as a function of distance between the centers of the two cells. 
%Inset shows the region where ${\bf F}_{ij}$ is attractive. When $|{\bold {r}}_{i} -{\bold {r}}_{j}|~\ge~R_{i}+R_{j}=8~\mu m$ the cells are no longer in contact, and hence, ${\bf F}_{ij}=0$. }
%\label{cellcellinter}
% \label{forcejkr}
\end{figure}
\bigskip

\clearpage
\newpage
 \begin{figure}
%\begin{turn}{-90}
\includegraphics[width=1.0\linewidth] {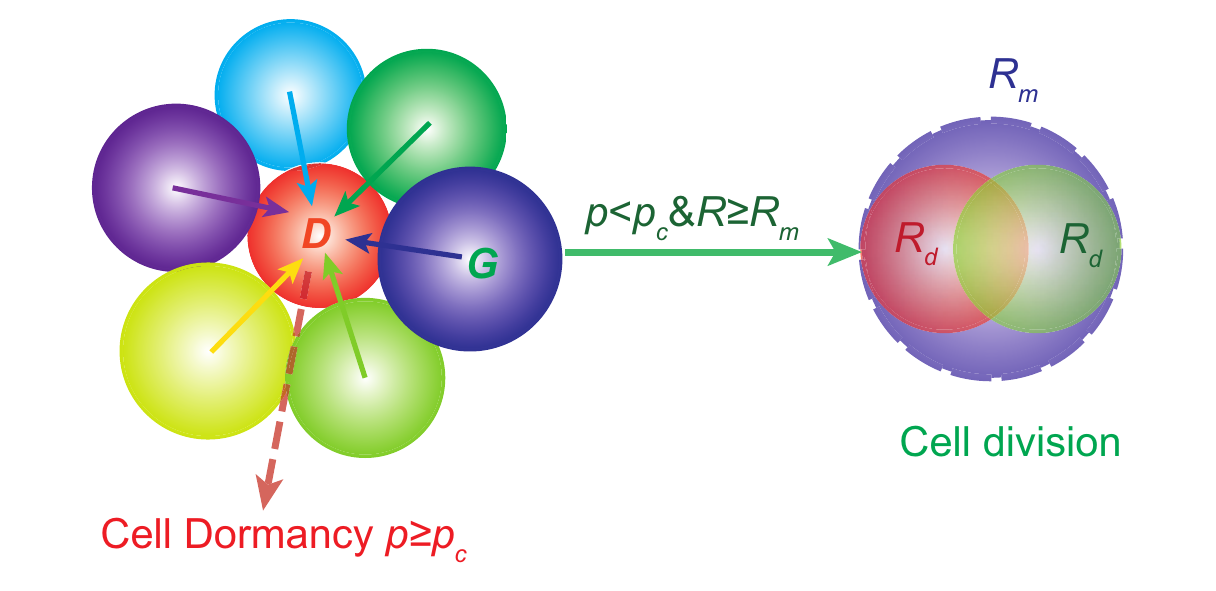} %oa
%\end{turn}
\caption{}
\label{celldorgro}
\end{figure}

\clearpage
\newpage
\floatsetup[figure]{style=plain,subcapbesideposition=top}
\begin{figure}%[h!]
%\begin{turn}{-90}
%\centering
	%\begin{subfigure} [t]{0.50\textwidth}
	\sidesubfloat[]{\includegraphics[width=0.70\linewidth] {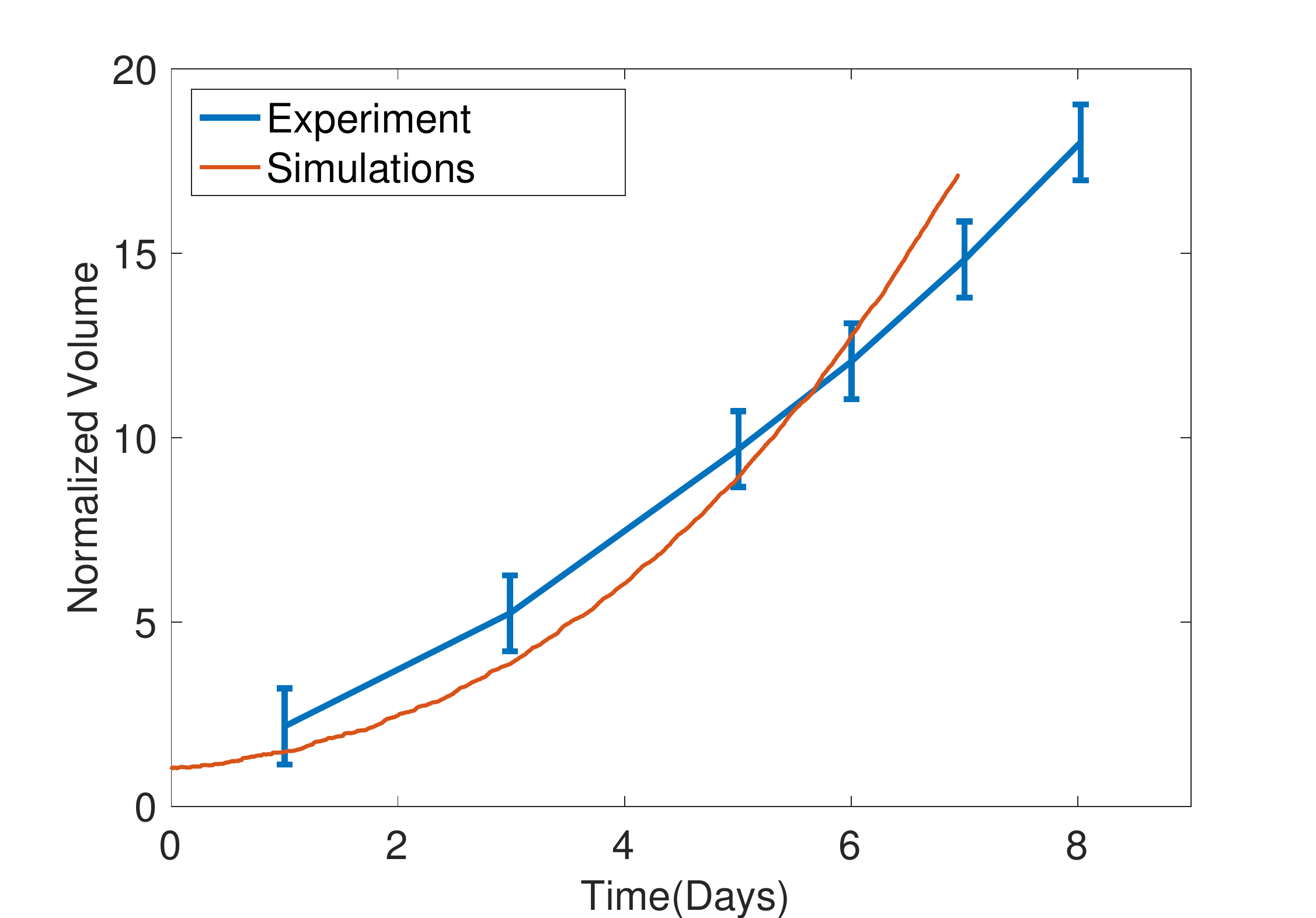}\label{figs2}}
	\par\bigskip
	\sidesubfloat[]{\includegraphics[width=0.70\linewidth] {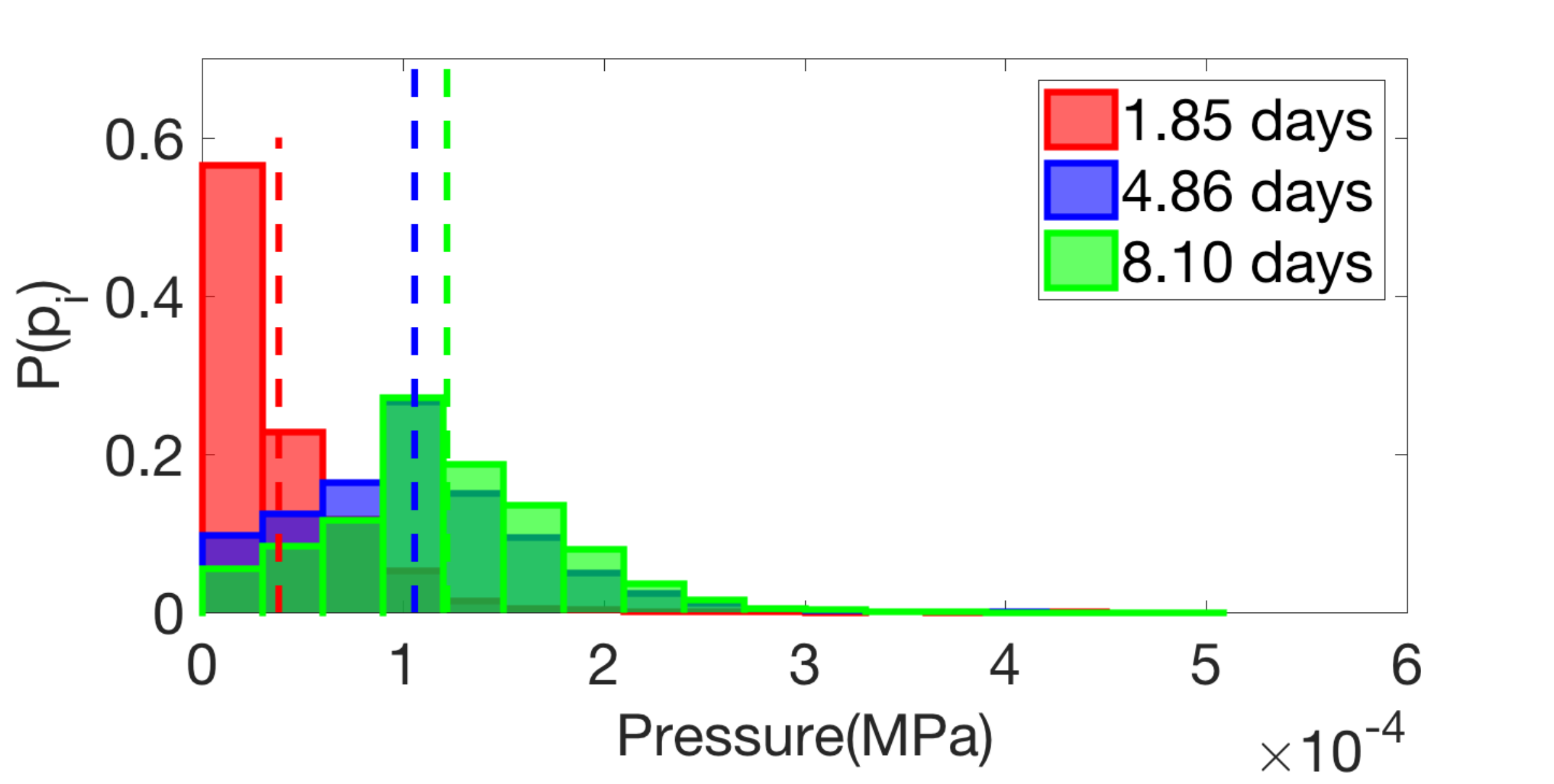}\label{pressd}} 
%	\subfigure{\textbf{(a)}\includegraphics[width=1.0\linewidth] {norm_vol1.eps}}%oa	
%	\subfigure{\textbf{(b)} \includegraphics[width=1.0\linewidth] {pressd.eps}}
%	 \label{pressd}
\caption{}
\end{figure}

\newpage
\begin{figure}[h]
\centerline{\includegraphics[width=0.9\linewidth]{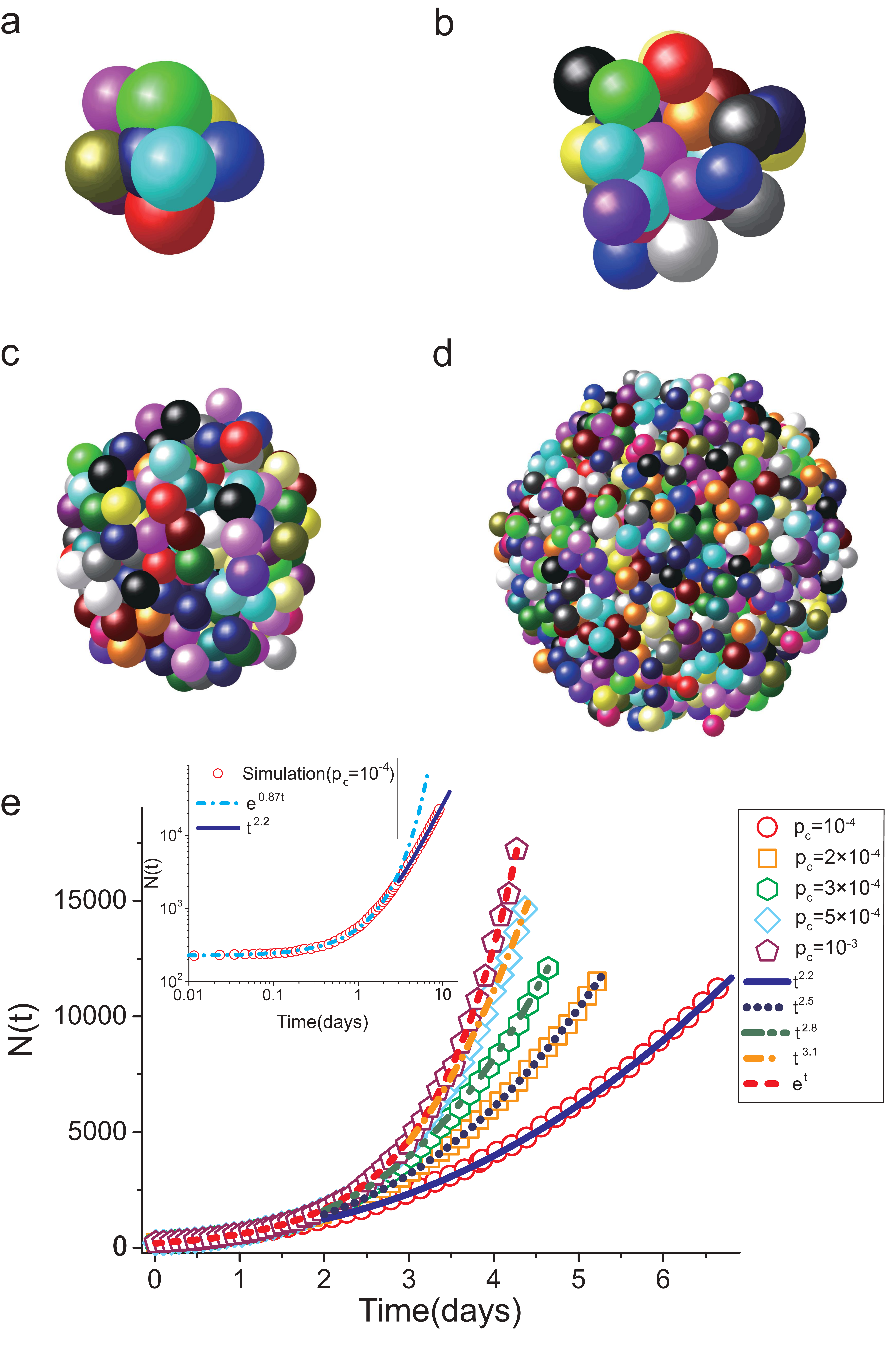}}
\caption{}
\label{figx}
\end{figure} 

\clearpage
\newpage
\floatsetup[figure]{style=plain,subcapbesideposition=top}
\begin{figure}[t!]
\centering
\sidesubfloat[]{\includegraphics[width=0.55\textwidth] {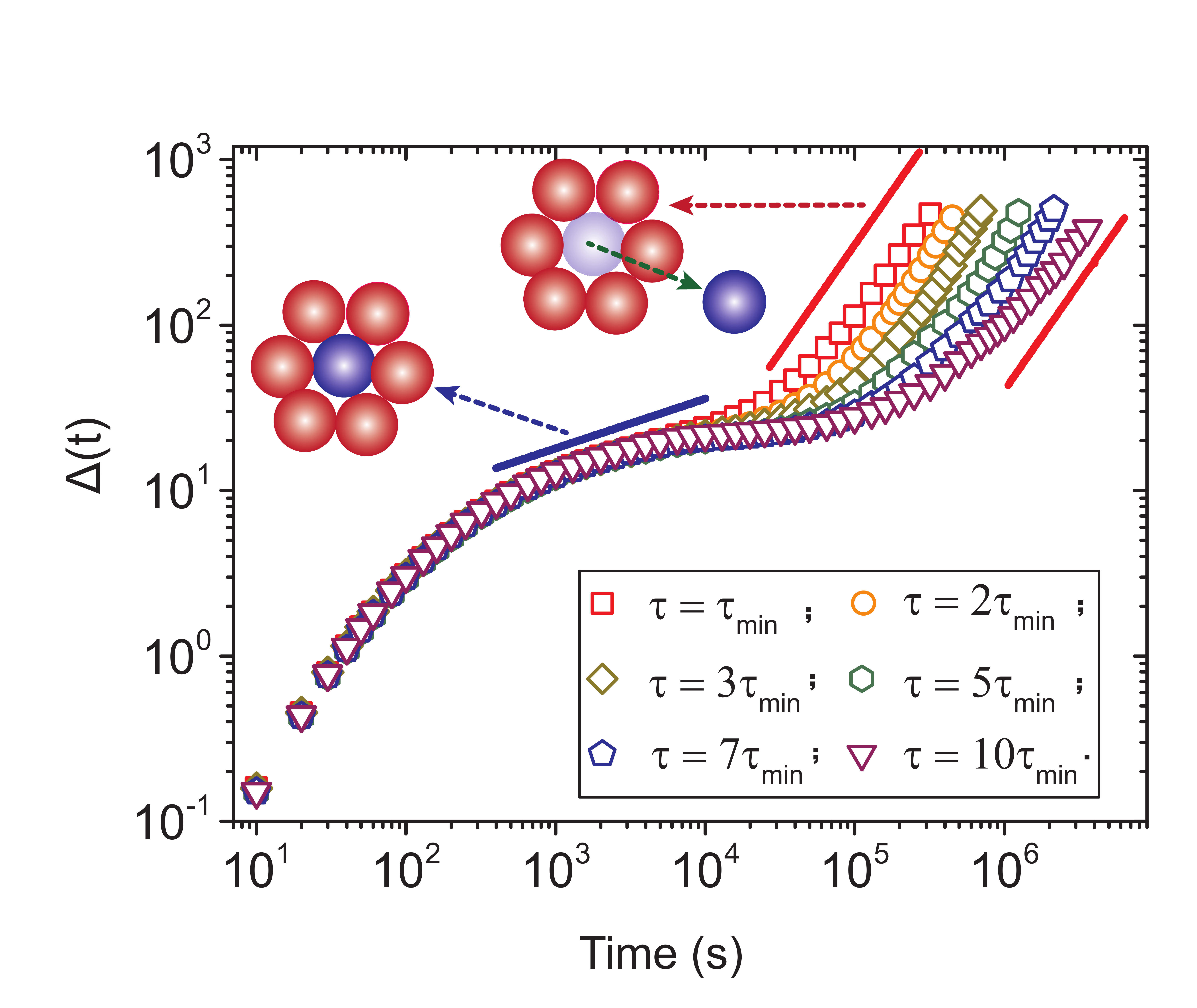} \label{figx2a}} \par %\bigskip
%\subfloat[]{\includegraphics[width=1.0\textwidth] {fig3-11-20.pdf} \label{figx3a}} \par 
\sidesubfloat[]{\includegraphics[width=0.55\textwidth] {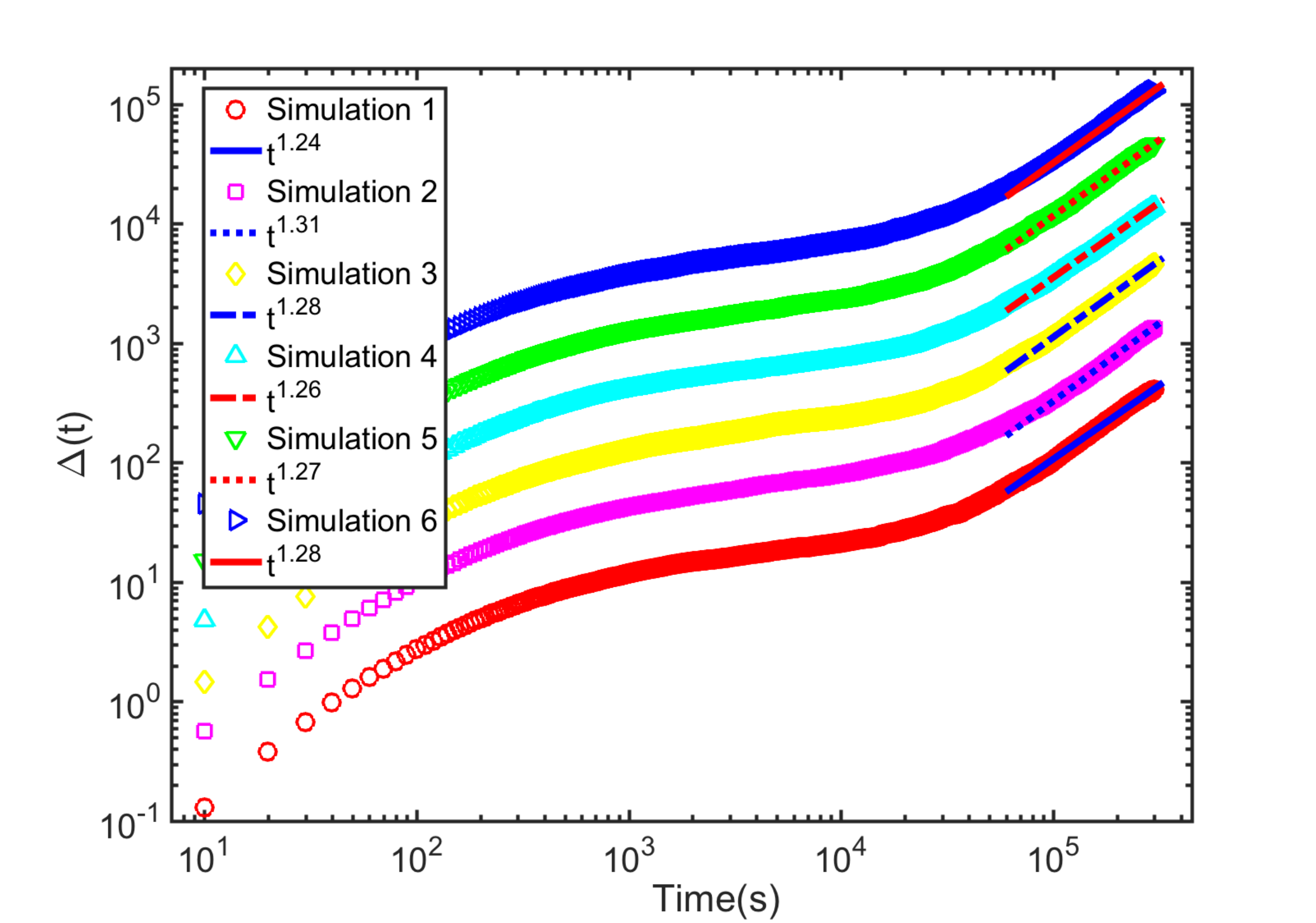} \label{fitexp}} \par %\bigskip
\sidesubfloat[]{\includegraphics[width=0.55\textwidth] {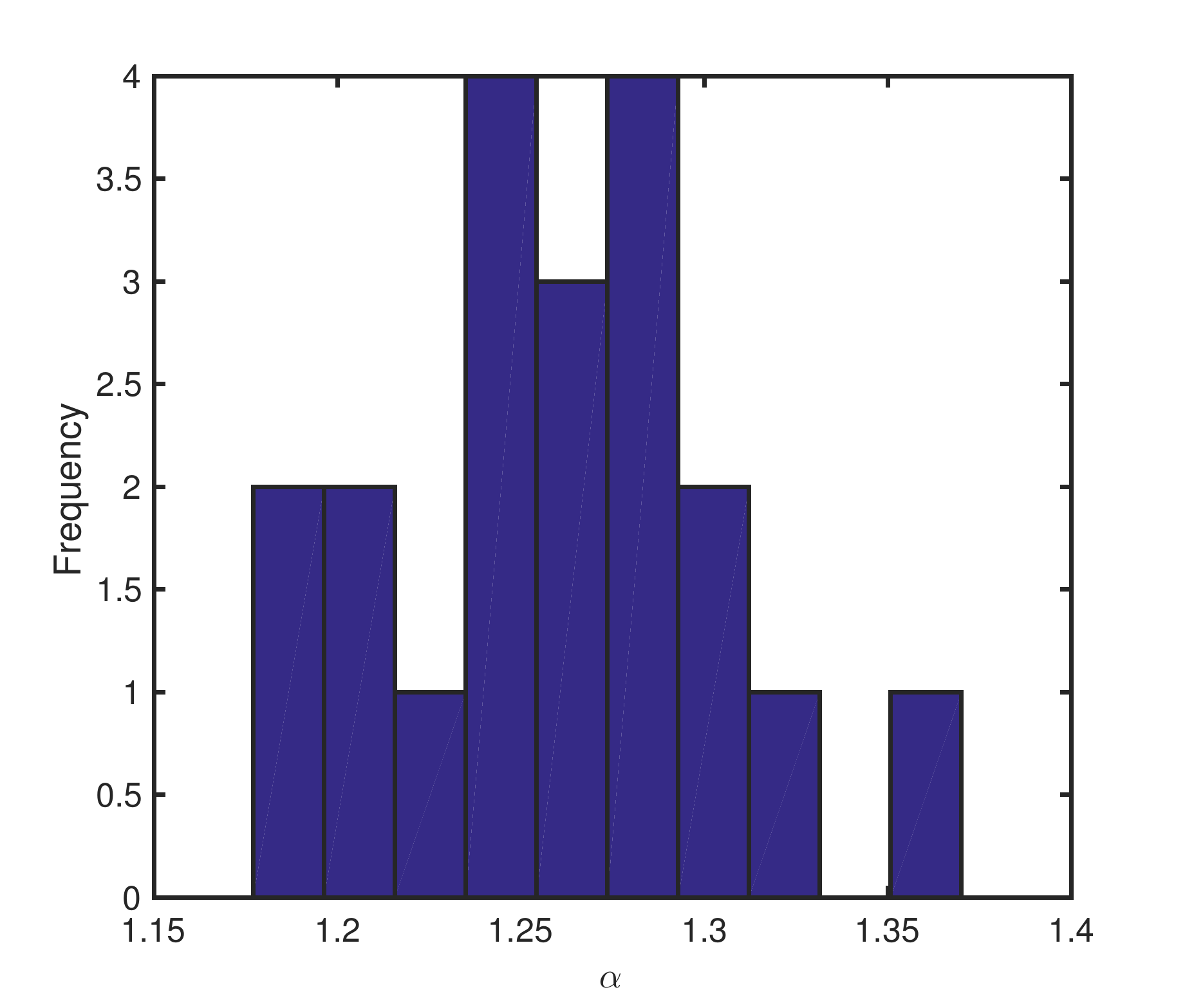} \label{fitexp1}}
%\centerline{\includegraphics[width=0.65\linewidth]{fig3.pdf}}
\caption{}
\label{msdexp}
\end{figure}
%\begin{figure}
%\contcaption
%\end{figure}

\clearpage
\newpage
\begin{figure}
%\begin{turn}{-90}
\includegraphics[width=1.0\linewidth] {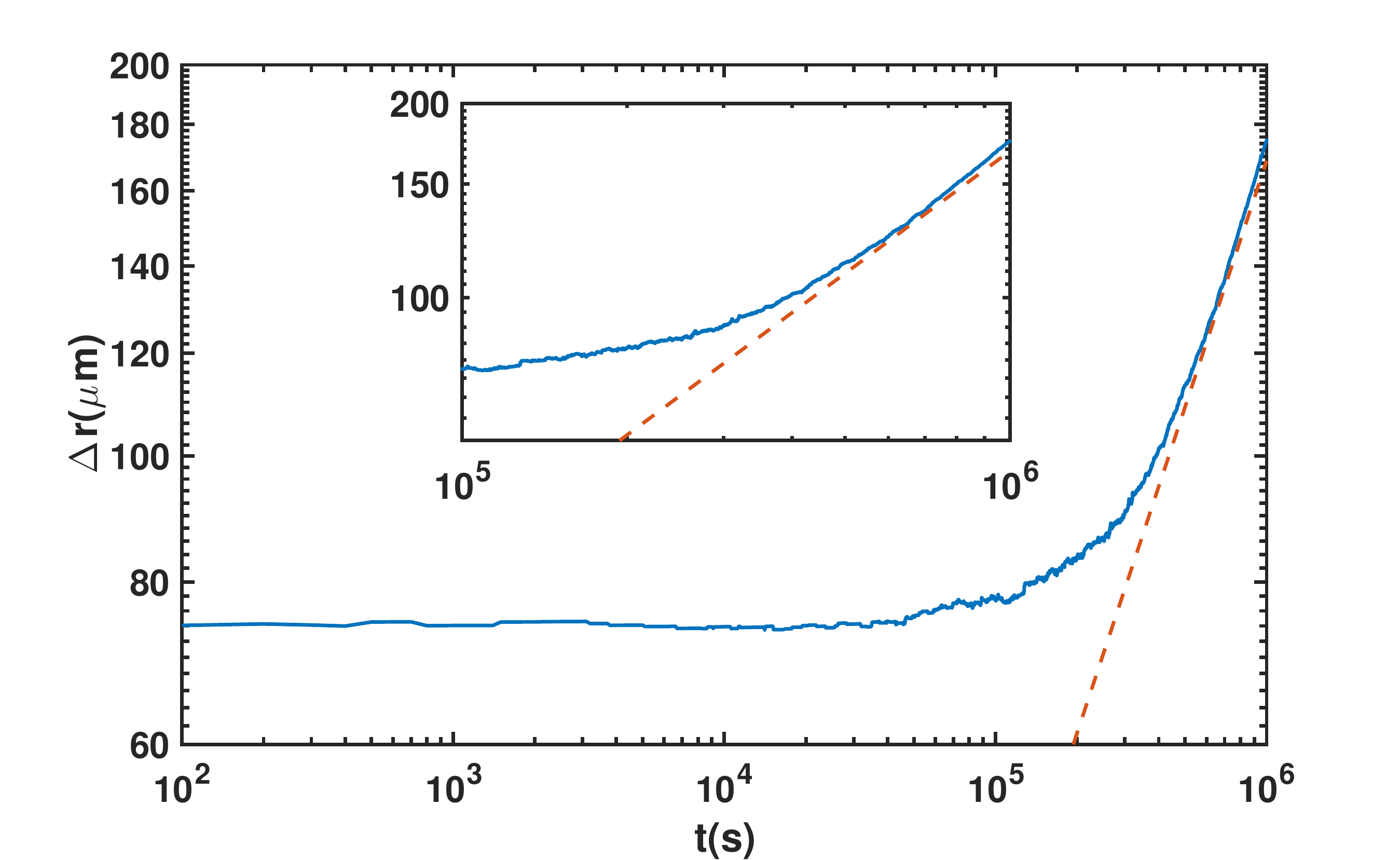} %oa
%\end{turn}
\caption{}
%~=~54,000s$ (Table I). }
\label{figs6}
\end{figure}

\clearpage
\newpage
\begin{figure}
\includegraphics[width=1.0\linewidth,clip=true] {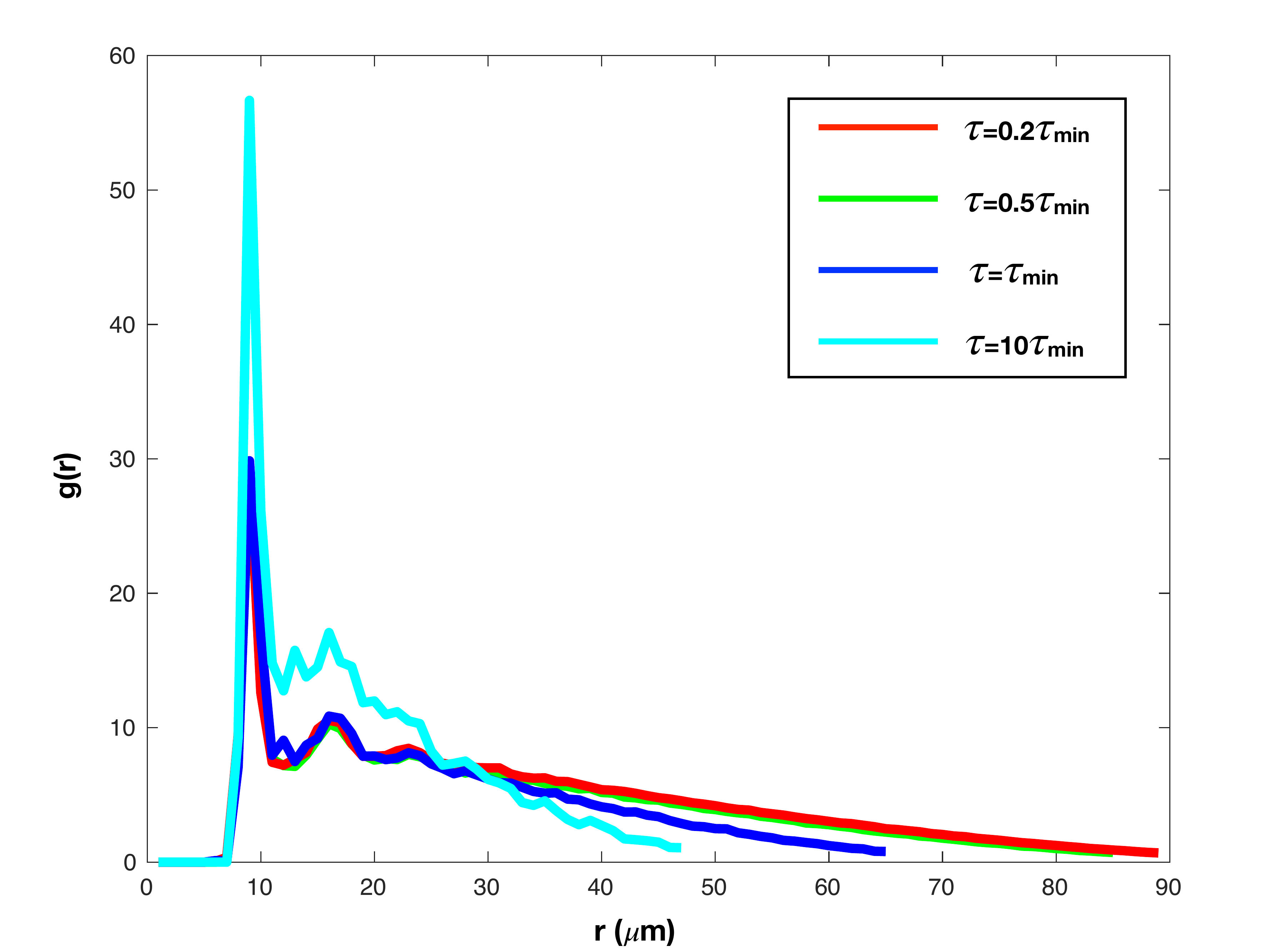} %oa
\caption{}
\label{figs3}
\label{fig:gr}
\end{figure}

\clearpage
\newpage
\begin{figure}
\subfloat{\includegraphics[width=0.60\textwidth] {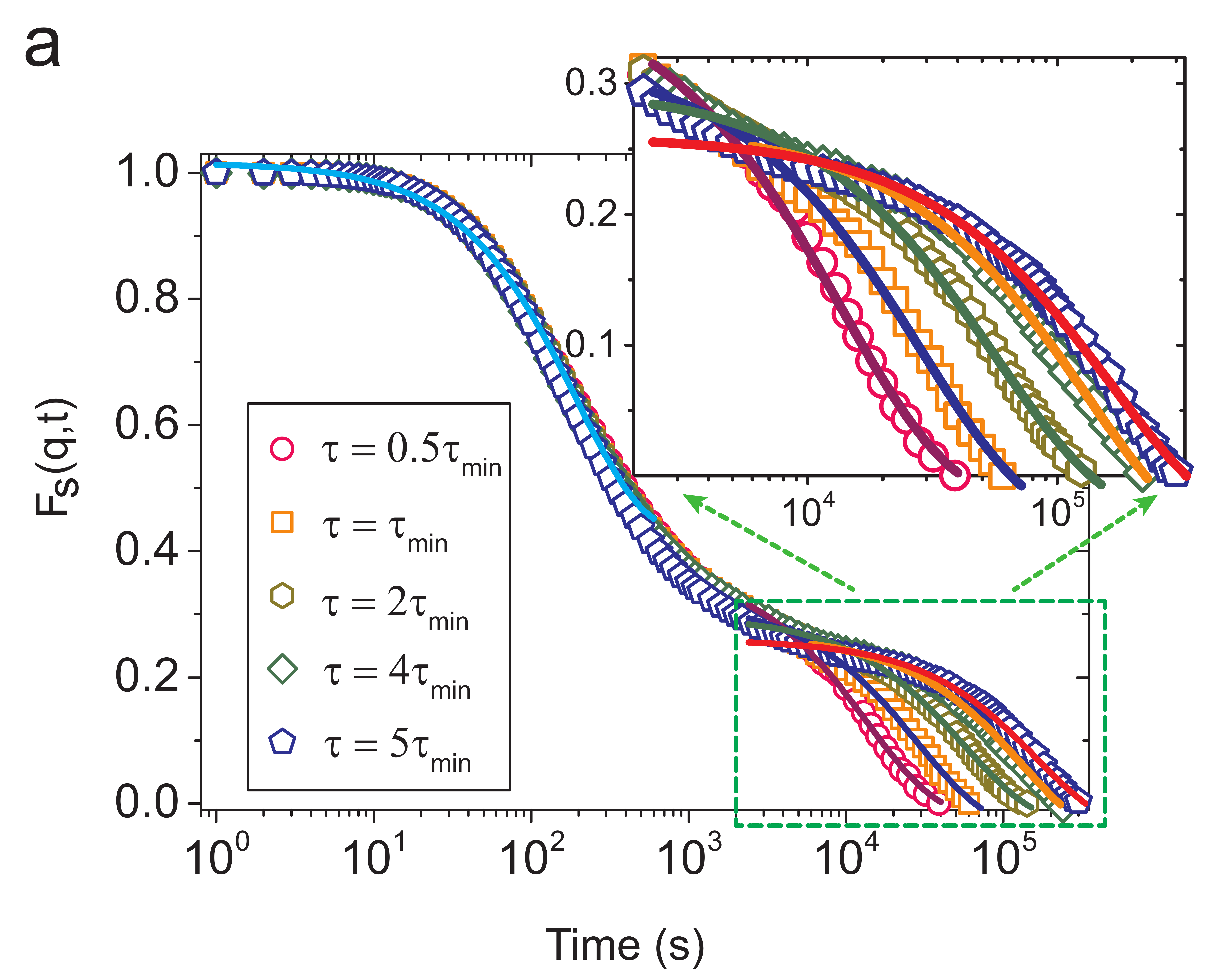} \label{figcella}} \par
\subfloat{\includegraphics[width=0.60\textwidth] {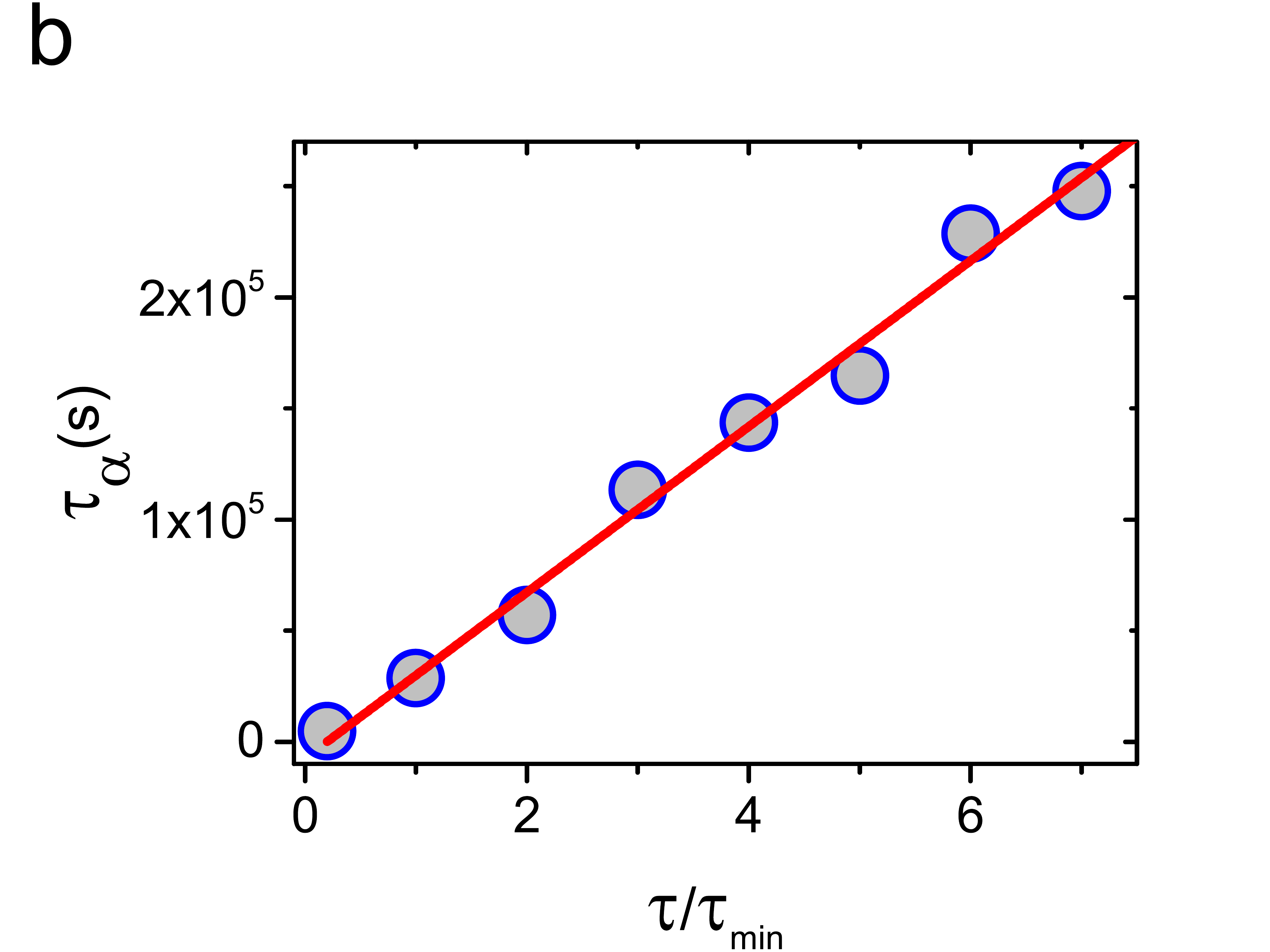} \label{figcellb}} \par 
\subfloat{\includegraphics[width=0.60\textwidth] {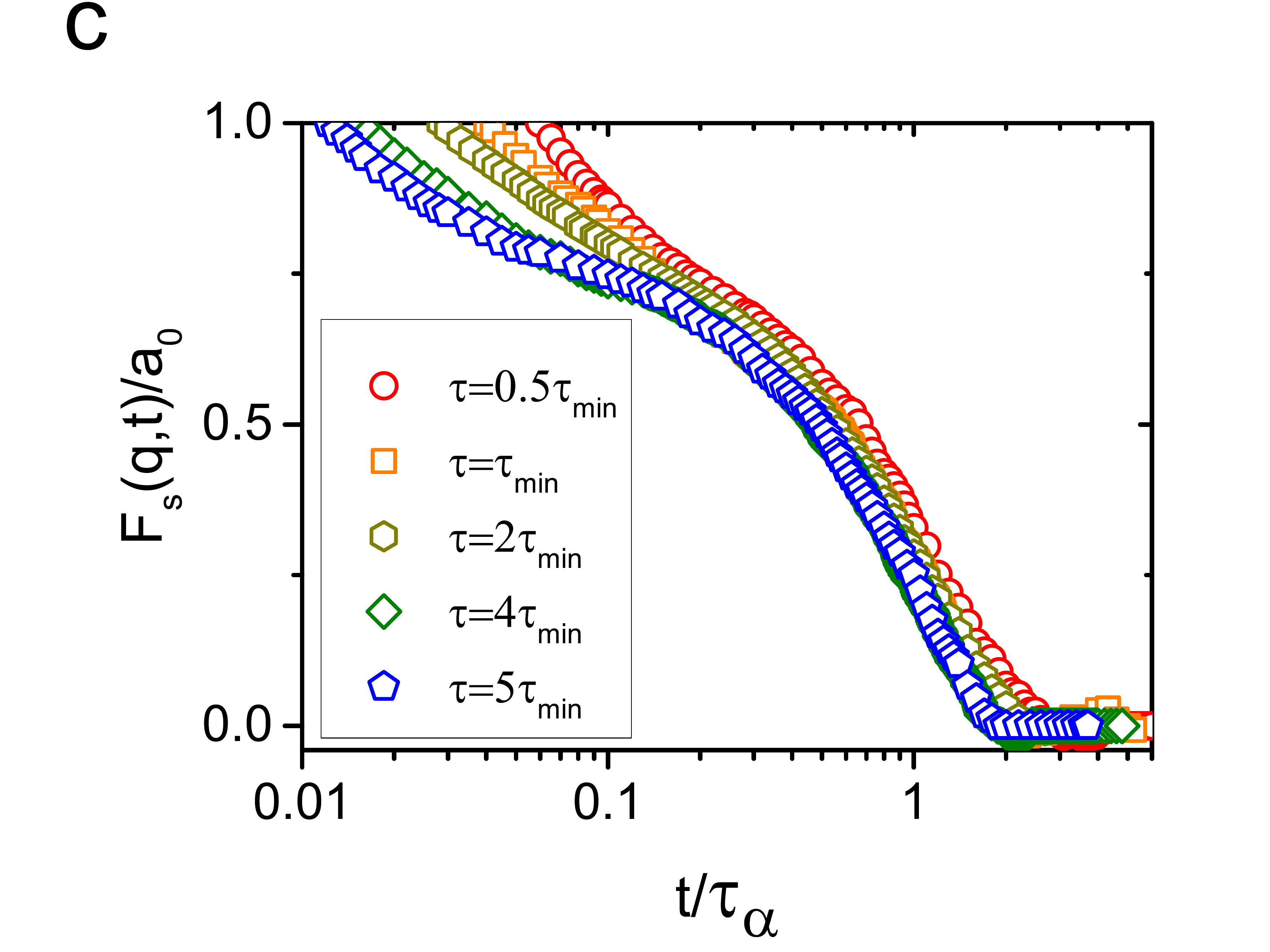} \label{figcellc}}
\caption{}
\label{figcell}
\end{figure}
%\begin{figure}
%\contcaption{}
%\end{figure}
%\hrule\bigskip

\clearpage
\newpage
\begin{figure}
\subfloat{\includegraphics[width=0.8\textwidth] {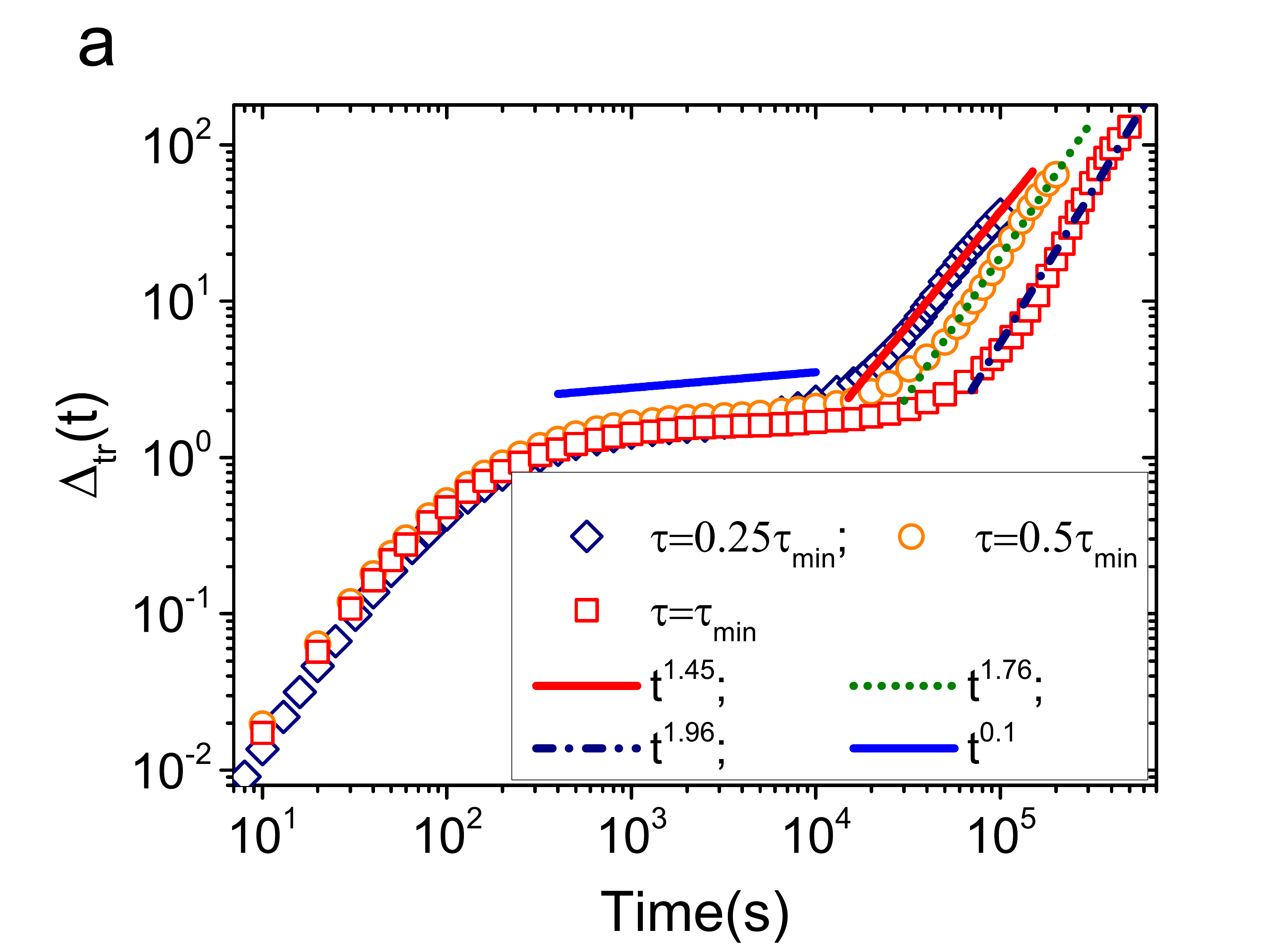} \label{figtra}}\par
\subfloat{\includegraphics[width=0.8\textwidth] {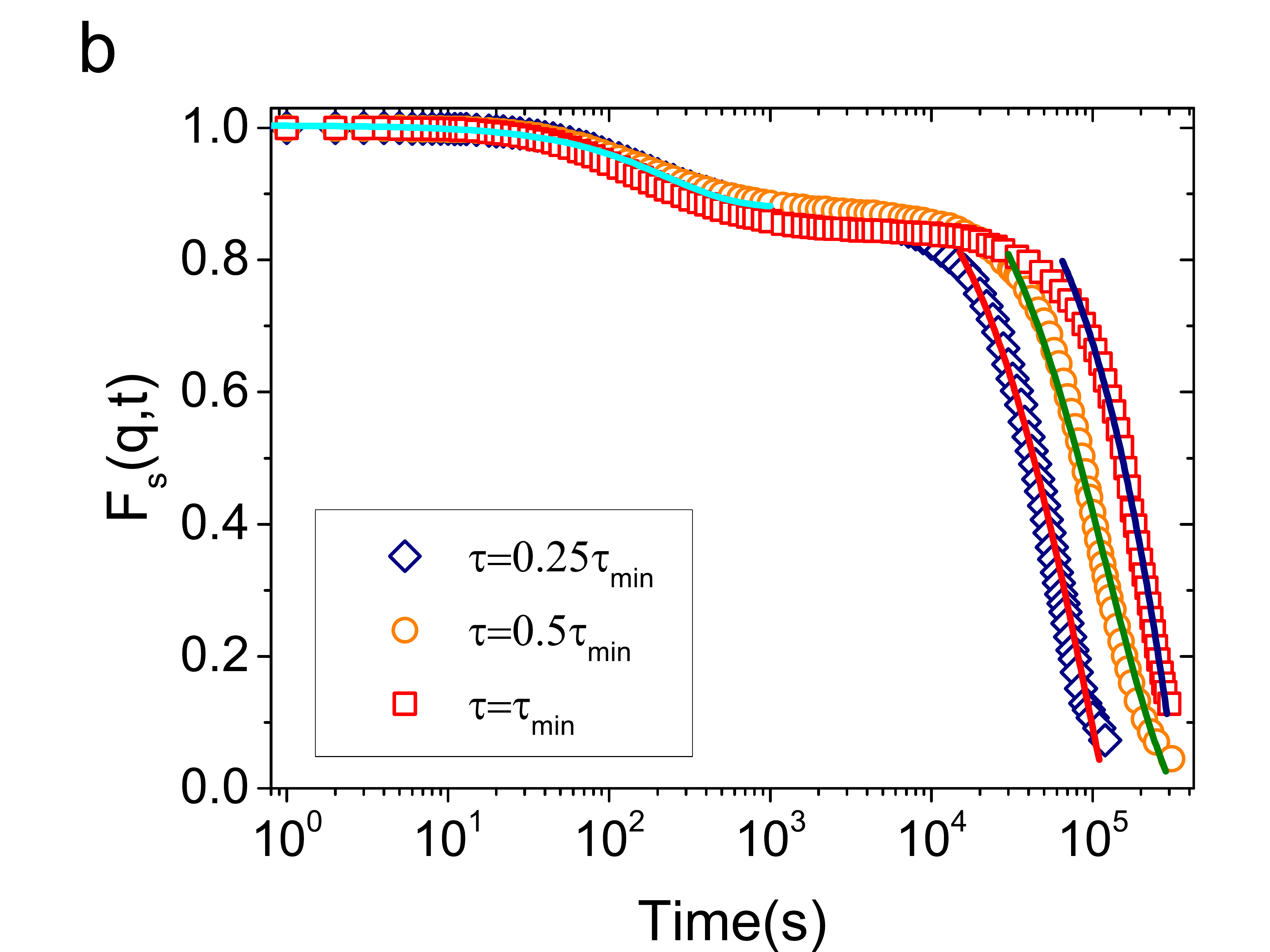} \label{figtrb}}
\caption{}
\label{figtr}
\end{figure}
%\begin{figure}
%\contcaption{\textcolor{blue}{\textbf{Dynamics of the tracers at different cell cycle times.}  
%{\bf (a)}Time dependence of $\Delta (t)$ of tracer cells at different cell division times, $\tau$. 
%MSD fits at intermediate time 
%and long times are shown in the inset.
%{\bf (b)} The self-intermediate scattering function for tracers.  Biexponential fits to the decay of tracer 
%$F_{s}(q,t)$ are shown by solid lines. }}
%\end{figure}

\clearpage
\newpage
\begin{figure}
\centerline{\includegraphics[width=1.3\linewidth]{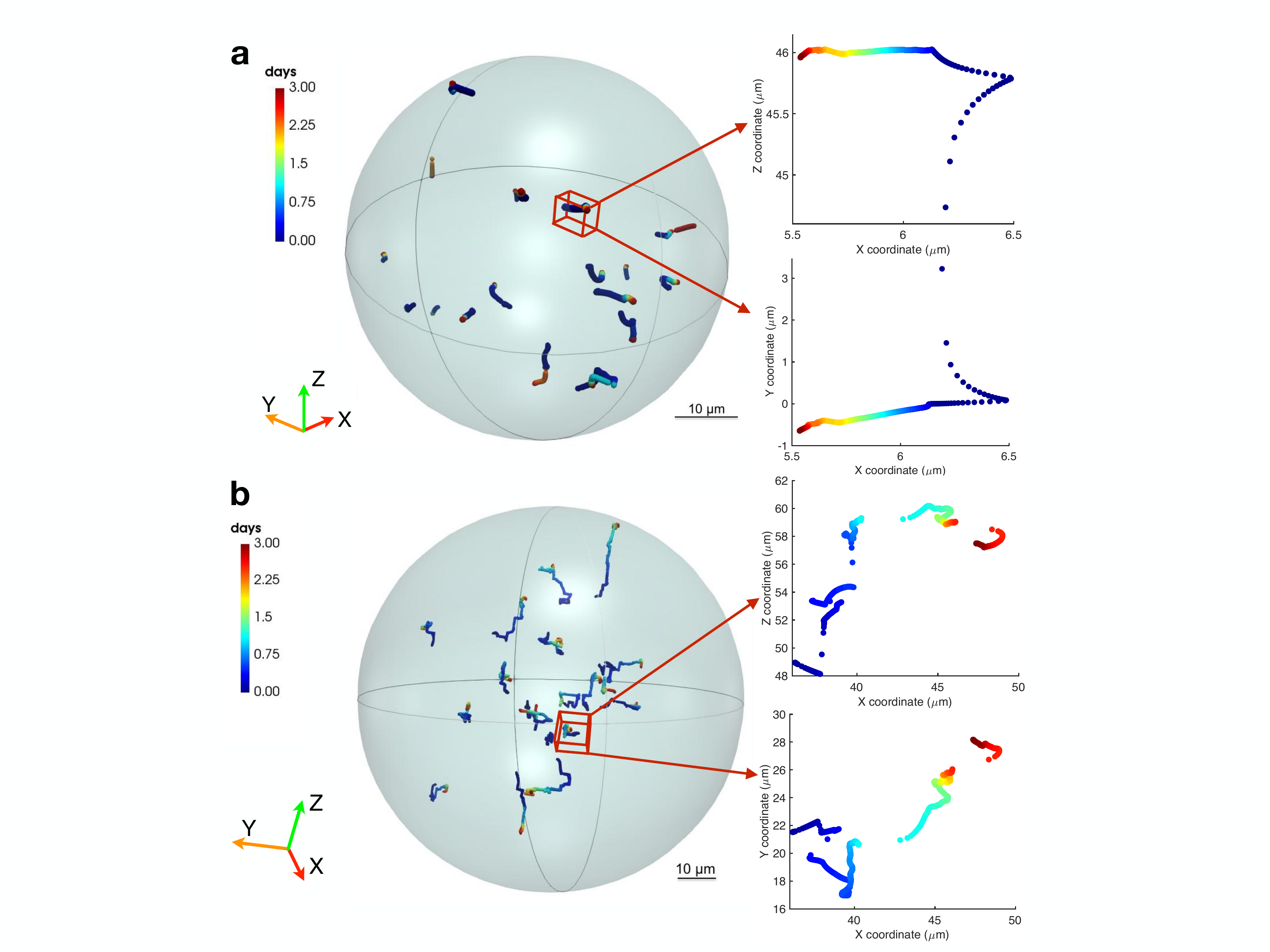}}
\caption{}
\label{figx3}
\end{figure}
%\begin{figure}
%\contcaption{}
%\end{figure}

\begin{figure}
\includegraphics[clip,width=0.65\textwidth]{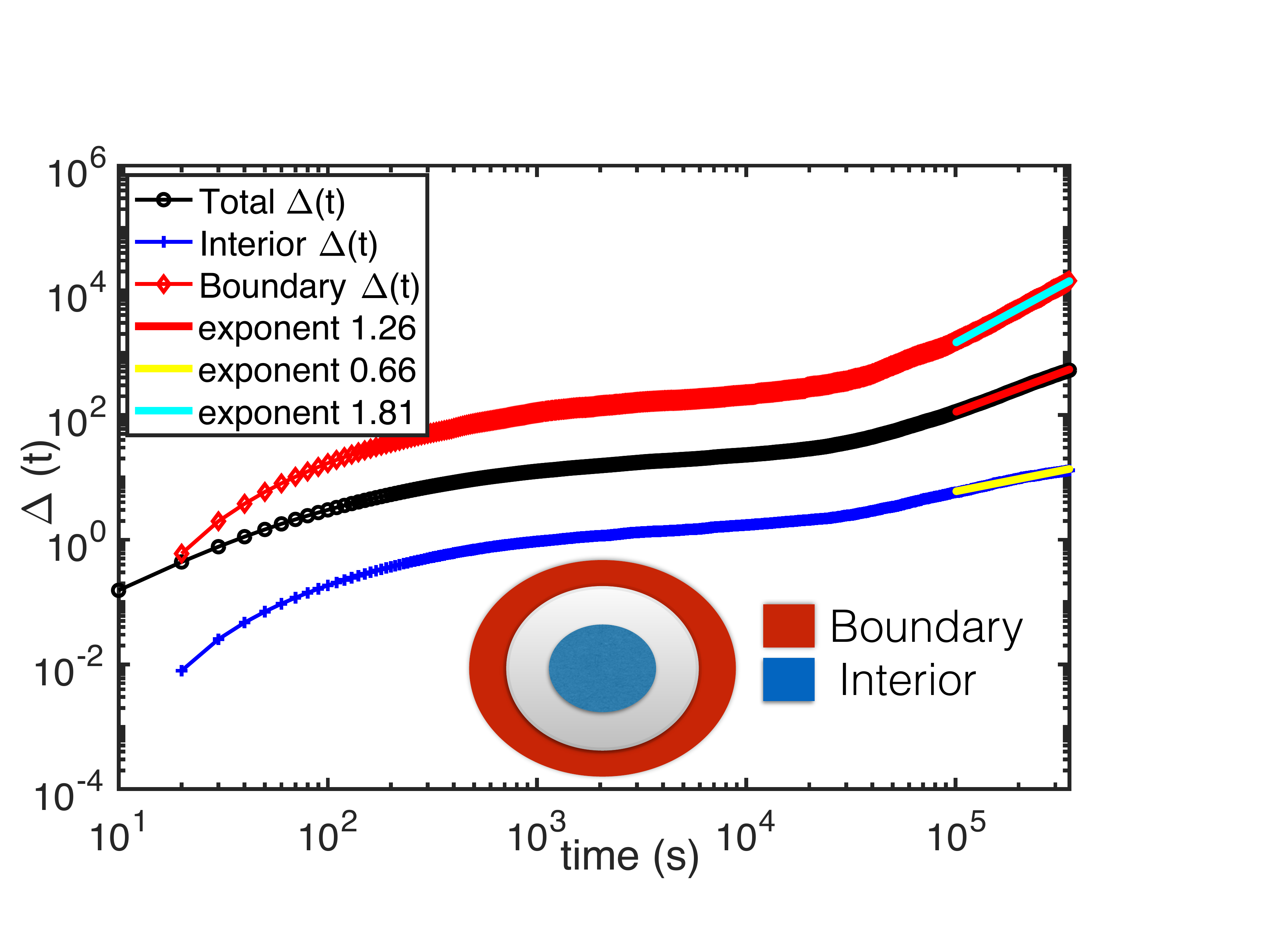}
\caption{}
\label{figrep5}
\end{figure}

%\clearpage
%\newpage
% \begin{figure}
%\includegraphics[width=1.1\textwidth,left] {figsuppl.pdf}
%\caption{}
%%\includegraphics[clip,width=0.5\textwidth]{msdcompvarieddeath.pdf}
% \label{tracknorm}
%\end{figure}
%\begin{figure}
%\contcaption{}
%\end{figure}

\clearpage
\newpage
\floatsetup[figure]{style=plain,subcapbesideposition=top}
\begin{figure}
%\subfloat[]{\includegraphics[width=1.0\textwidth] {fig3-11-20.pdf} \label{figx3a}} \par \bigskip
\sidesubfloat[]{\includegraphics[width=1.0\textwidth] {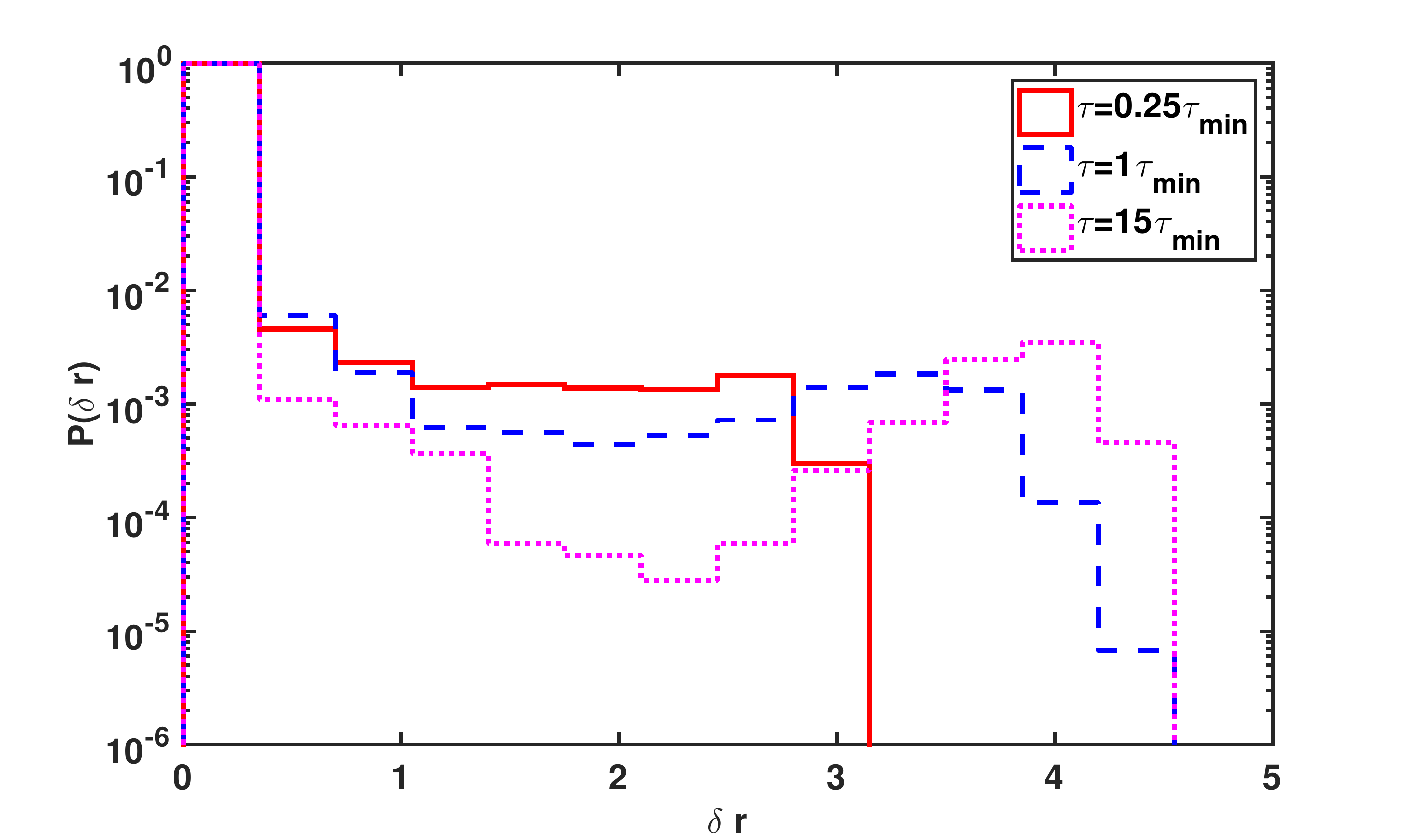} \label{figx3b}} \par \bigskip
\sidesubfloat[]{\includegraphics[width=1.0\textwidth] {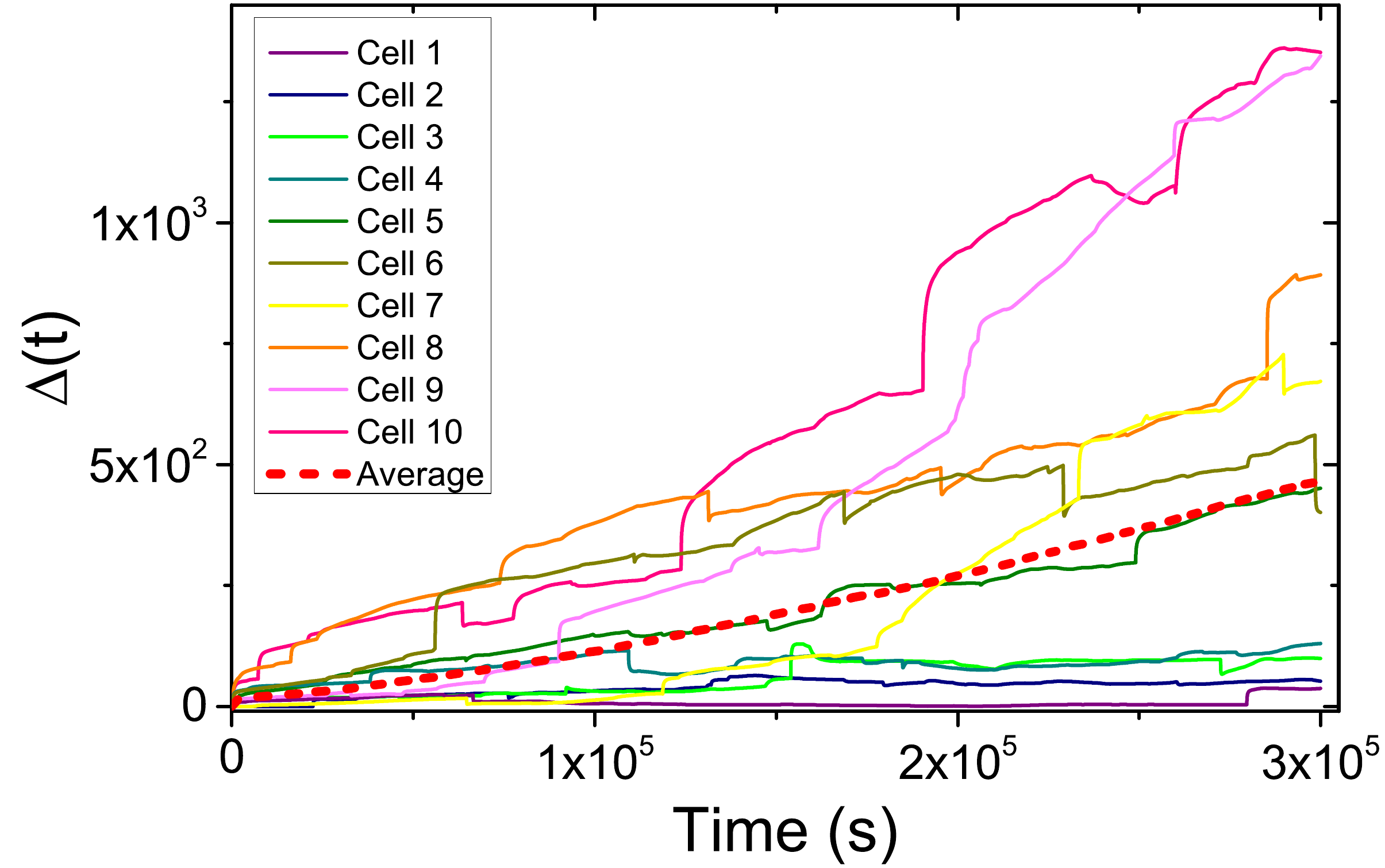} \label{figx3c}}
%\centerline{\includegraphics[width=0.65\linewidth]{fig3.pdf}}
\caption{}
\label{figx3part2}
\end{figure}
%\begin{figure}
%\contcaption{}
%\end{figure}

\clearpage
\newpage
 \begin{figure}
%\begin{turn}{-90}
\includegraphics[width=1.0\linewidth] {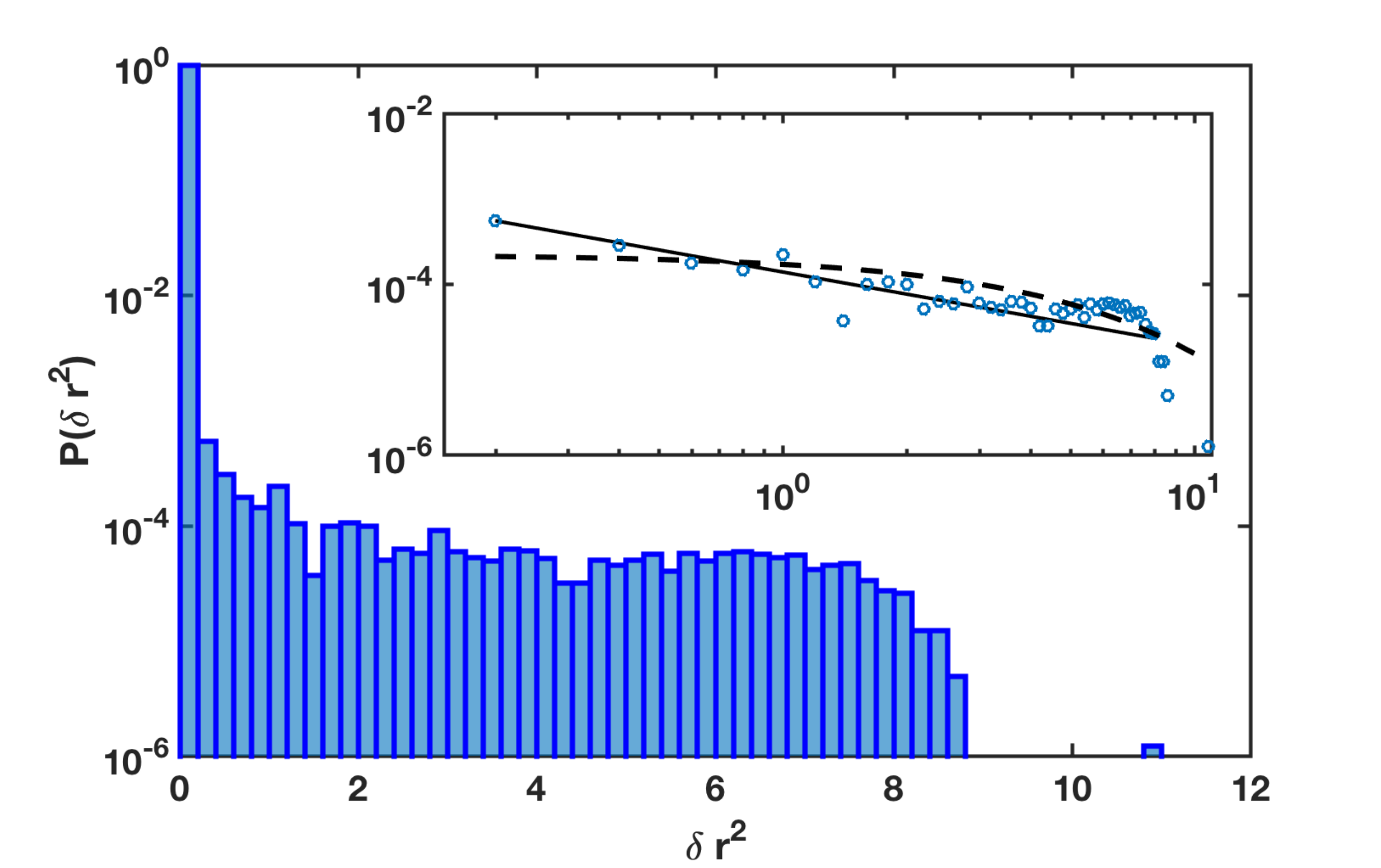} %oa
%\end{turn}
\caption{}
\label{figs1}
\end{figure}

\clearpage
\newpage
\begin{figure}
\centerline{\includegraphics[width=1.35\linewidth]{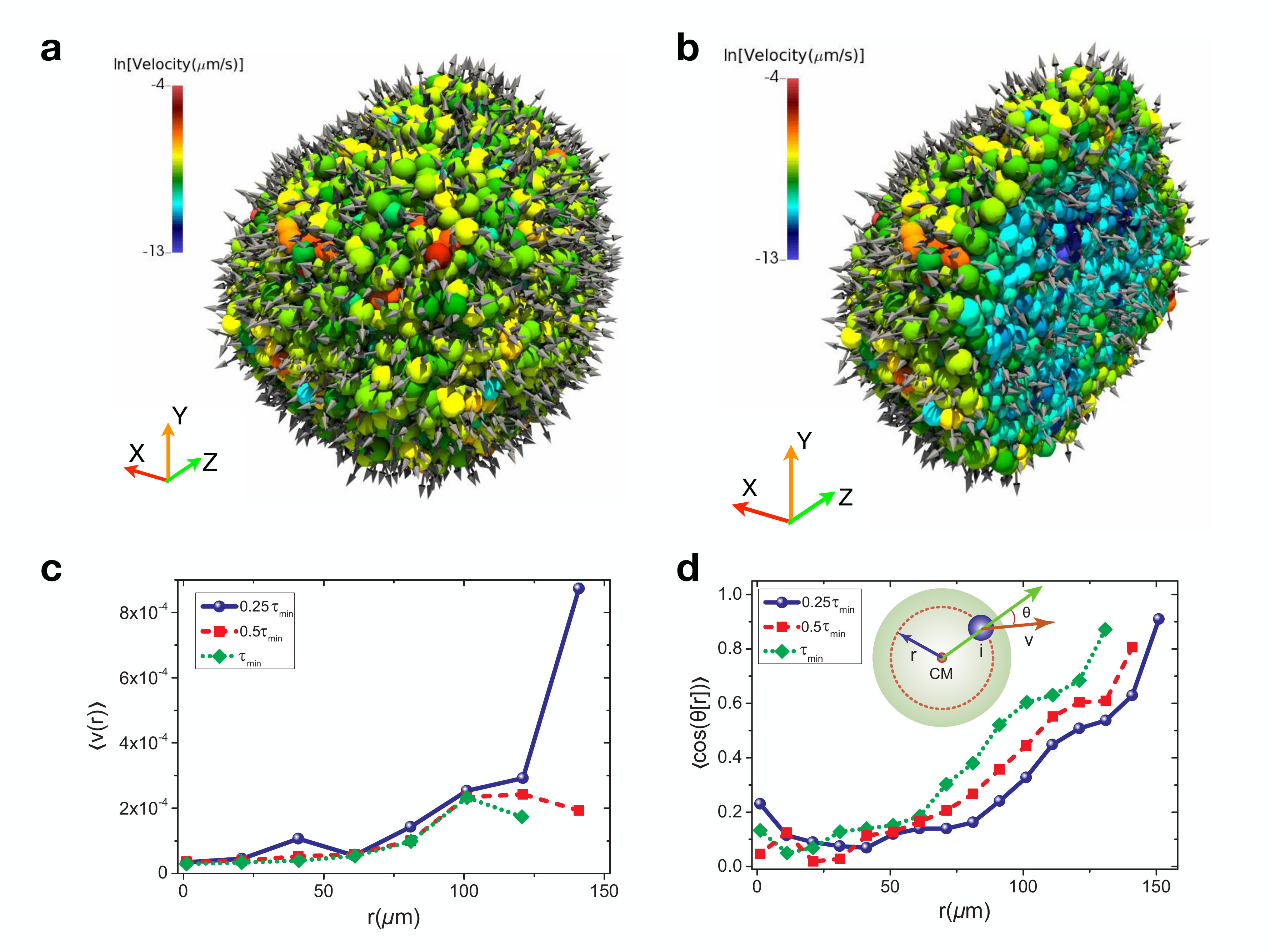}}
\caption{}
\label{figx4}
\end{figure}
%\begin{figure}
%\contcaption{}
%\end{figure}

\clearpage
\newpage
\floatsetup[figure]{style=plain,subcapbesideposition=top}
\begin{figure}
\sidesubfloat[]{\includegraphics[width=1.0\textwidth] {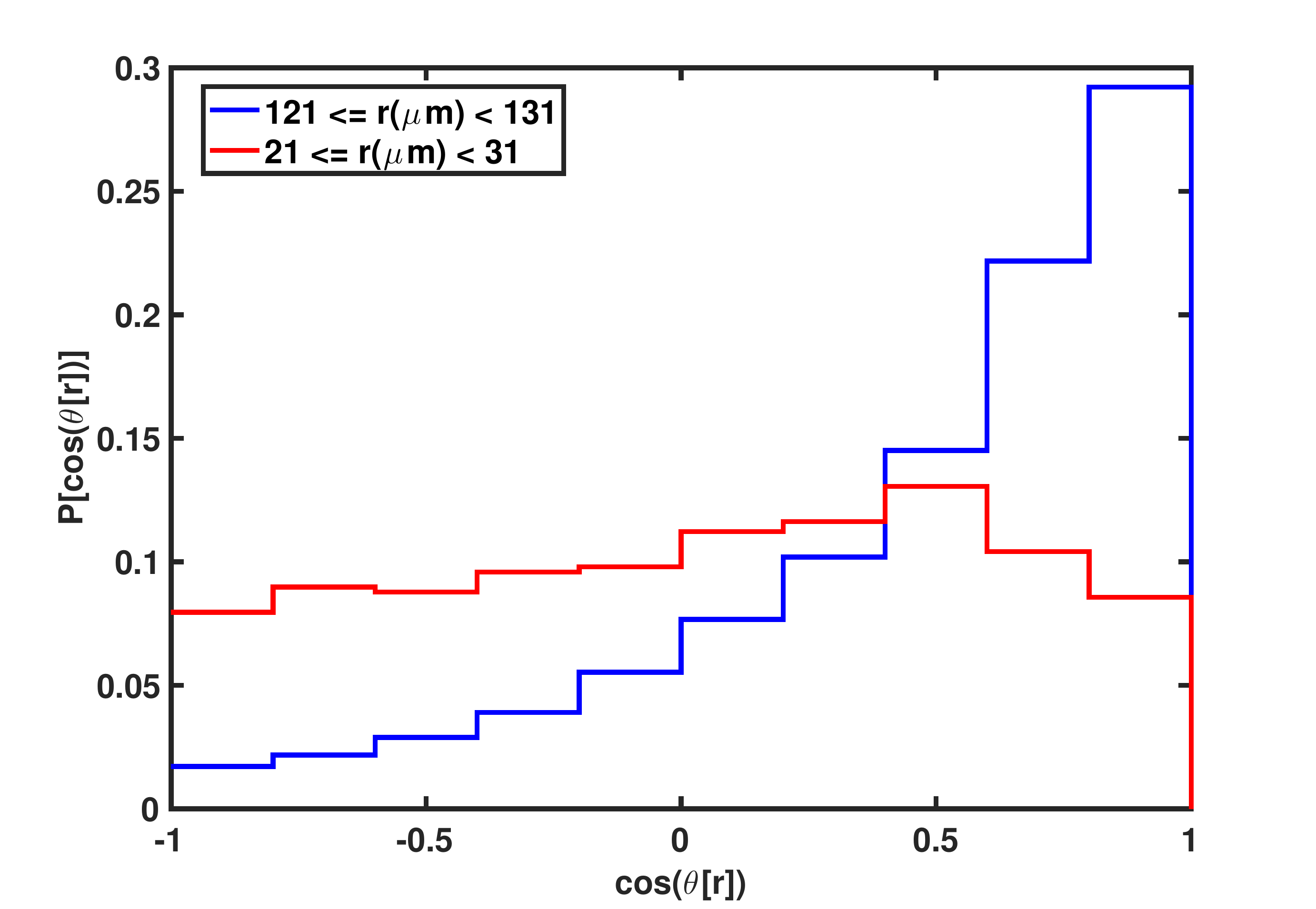} \label{figx4_2a}} \par \bigskip
\sidesubfloat[]{\includegraphics[width=1.0\textwidth] {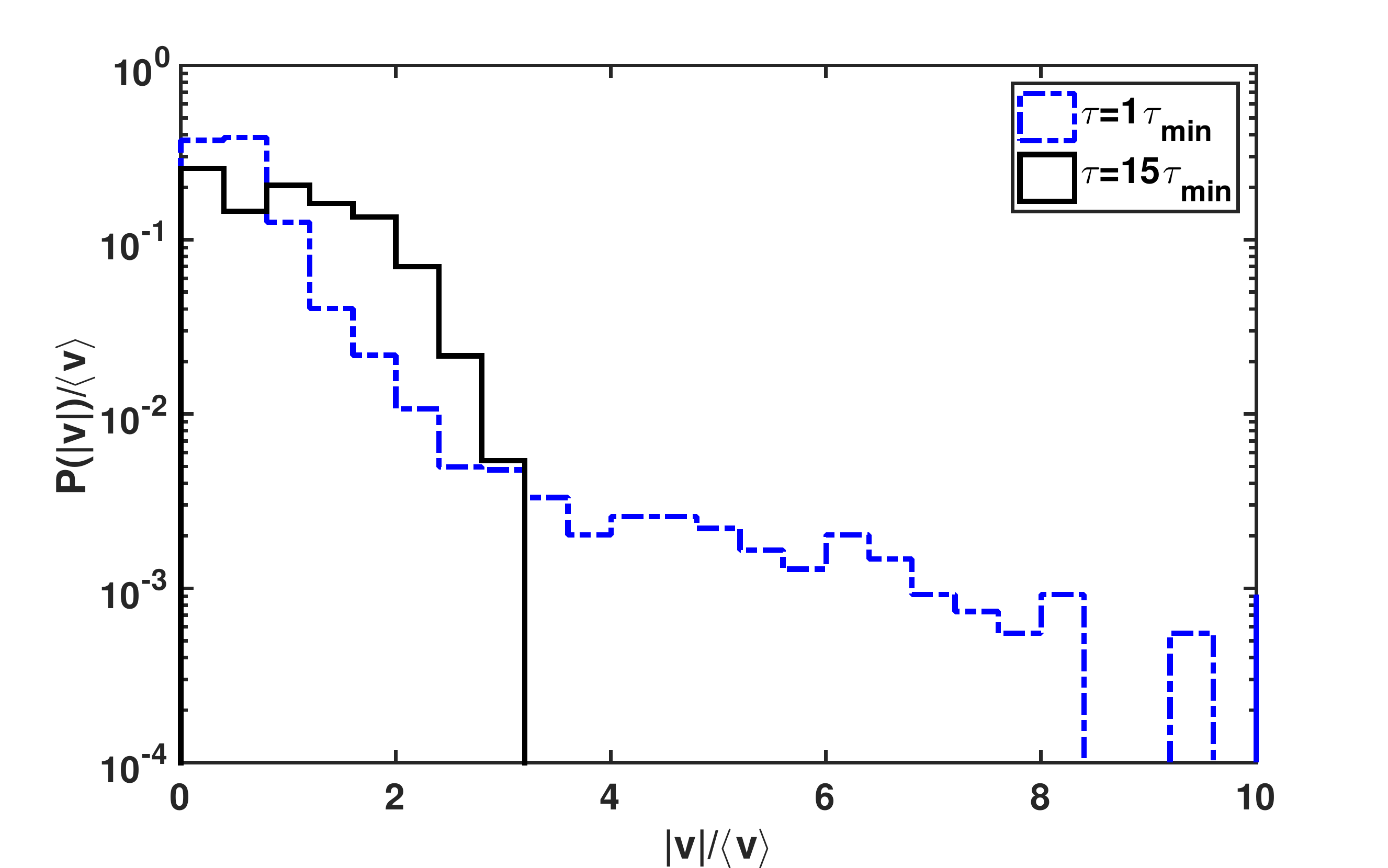} \label{figx4_2b}}
\caption{}
\label{figx4a}
\end{figure}
%\begin{figure}
%\contcaption{}
%\end{figure}

%\hrule\bigskip

\clearpage
\newpage
 \begin{figure}
\includegraphics[width=1.0\textwidth] {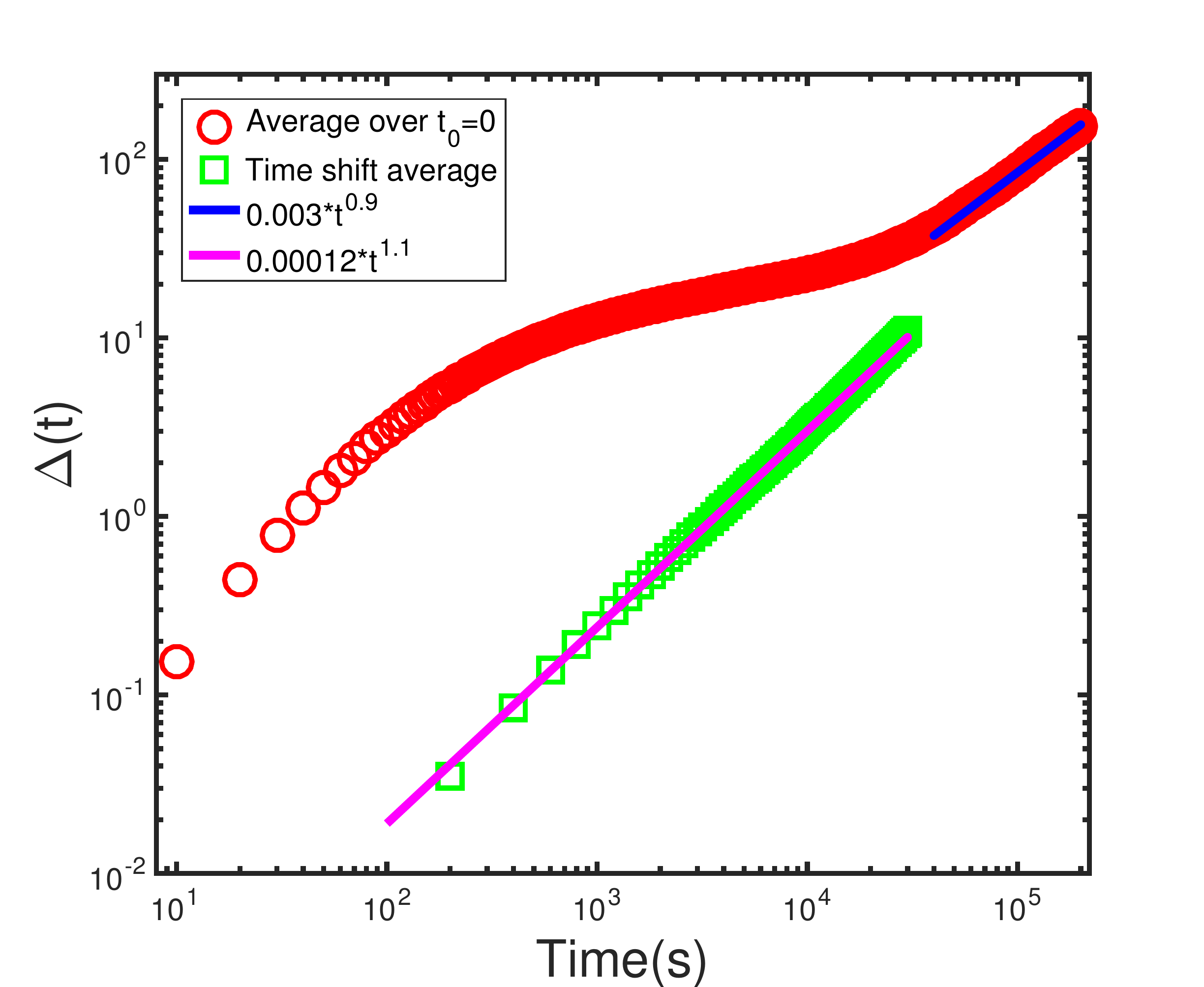}
\caption{}
 \label{msdvaried}
\end{figure}

\clearpage
\begin{figure}
\includegraphics[width=0.71\textwidth] {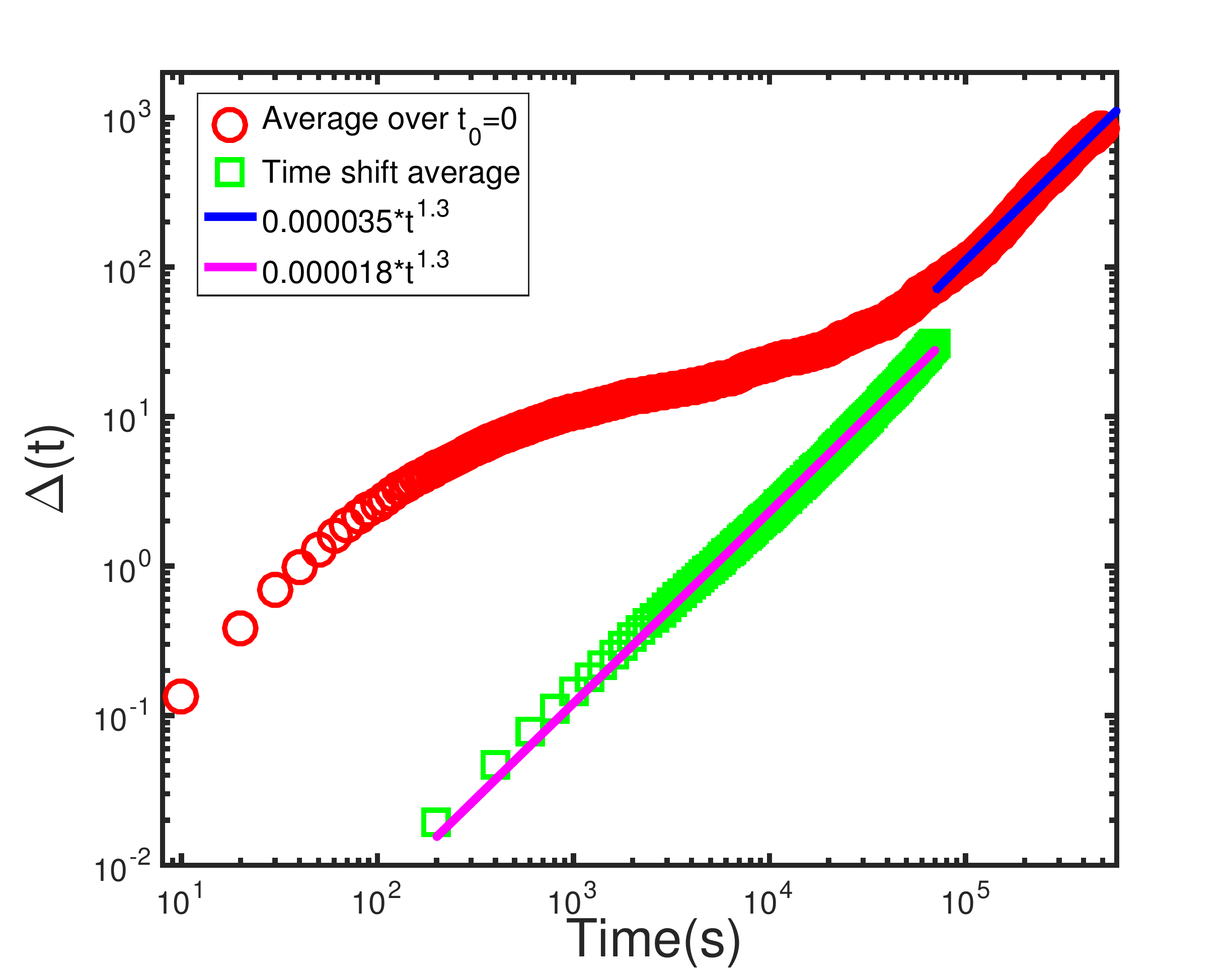} 
\caption{}
\label{msdrep7}
\end{figure}

\clearpage
\newpage
\begin{figure}
%\begin{turn}{-90}
\includegraphics[width=1.0\linewidth] {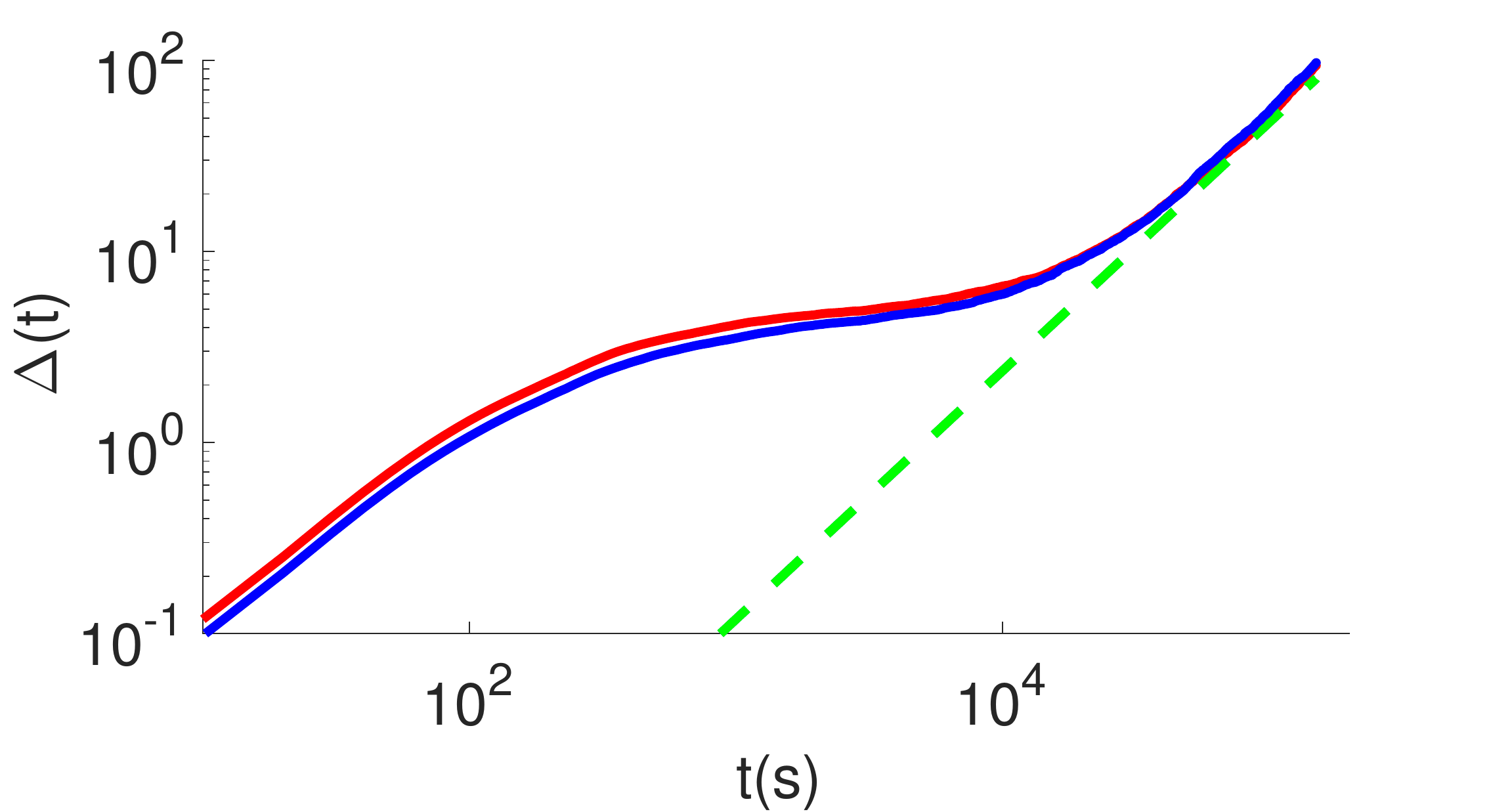} %oa
%\end{turn}
\caption{}
\label{msdnoise}
\end{figure}

\clearpage
\newpage
\begin{figure}
	\includegraphics[width=1.0\linewidth]{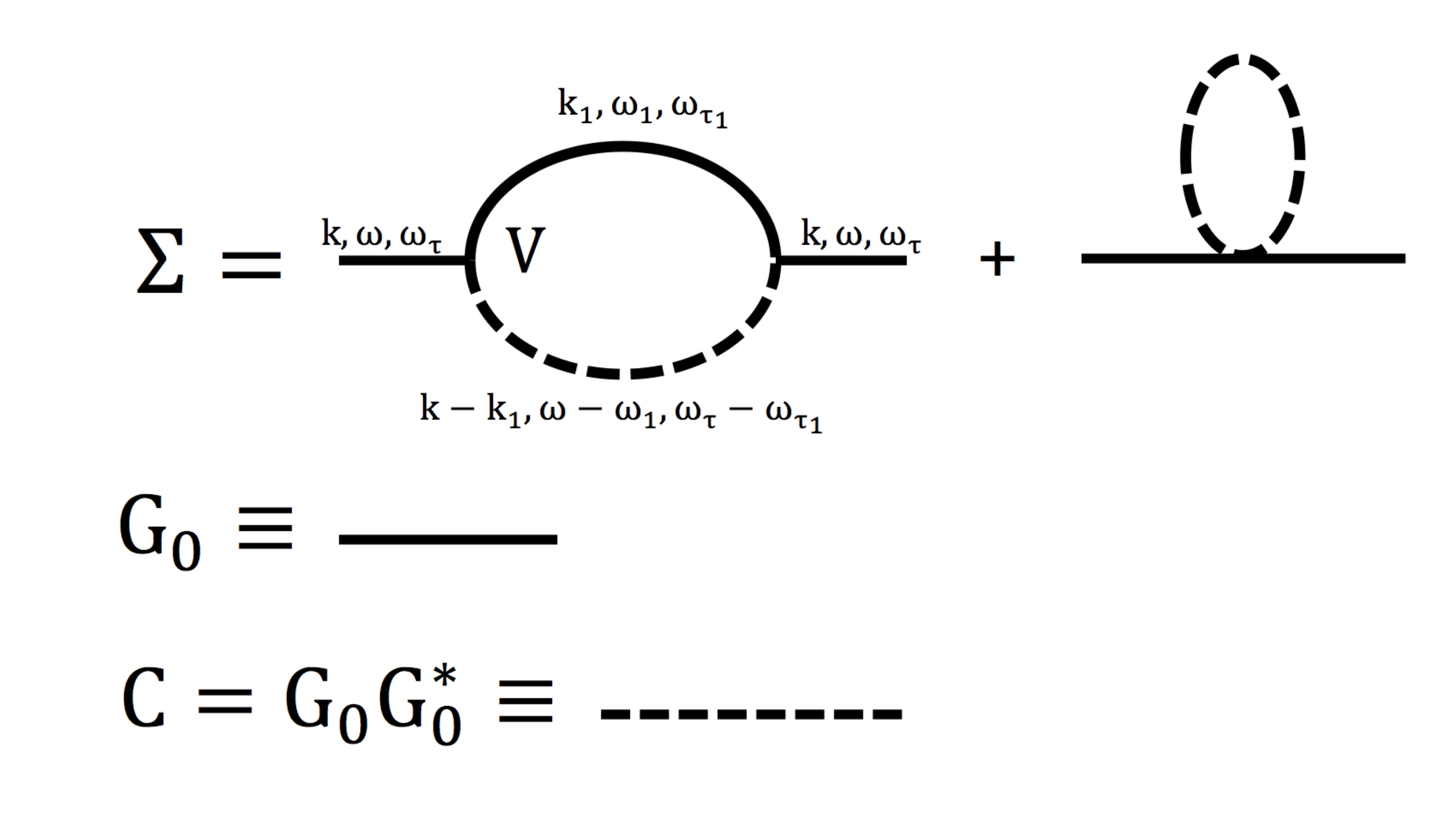}
	\caption{} 
	 	\label{fig:rg5}
\end{figure}

%\newpage
%\begin{figure}
%\includegraphics[width=1.0\linewidth] {Fadhesion.eps} %oa
%\caption{\label{fig:Fsa} The self-intermediate scattering function, $F_{s}(q,t)$, at  
%different values of $f^{ad}$. From left to right, the $f^{ad}$ values are, $5\times10^{-5}$ (red), $10^{-4}$ (orange), 
%$2\times10^{-4}$ (green), $3\times10^{-4} \mu N/\mu m^{2}$ (blue) respectively. 
%The first relaxation process depends on the strength of $f^{ad}$ while 
%the second relaxation process is not significantly affected by $f^{ad}$, as seen from the partial overlap in $F_{s}(q,t)$, at long times. }
%\label{figs4}
%\end{figure}

%\newpage
%\begin{figure}
%\includegraphics[clip,width=1.0\linewidth]{msds.eps}
%\caption{\label{fig:MSDE} Mean squared displacement, $\Delta(t)$, in units of $\mu m^2$ for different initial conditions. At $t=0$,   
%$100$ cells are randomly placed in a volume of $8000$ (blue), $64,000$ (red), and $10^{6}$ (orange) $\mu m^3$. 
%The slope of the three lines is 1.4 (purple), 1.1  (dashed green), and 1.0 (cyan), respectively at early times. In the inset, the slopes obtained from the long time limit
%are all 1.3 (solid green) irrespective of the initial condition. 
%%The universality of the exponent $\Delta(t)\propto t^{\alpha}$ at long times is explained theoretically.
%The higher exponent for tumors that are denser initially might be relevant for the virulency of cancers.}
%\label{figs5}
%\end{figure}

%\bigskip

\end{document}